\documentclass[11pt,english,fleqn]{article}
\usepackage[T1]{fontenc}
\usepackage{authblk}
\usepackage[utf8]{inputenc}
\usepackage[english]{babel}
\usepackage{graphicx}
\usepackage[sectionbib,numbers]{natbib}     
\usepackage[labelformat=simple]{subcaption}
\usepackage{tabularx}
\usepackage{multirow}
\usepackage{graphicx}
\usepackage{psfrag}
\usepackage{amsmath}
\usepackage{amssymb}
\usepackage{nicefrac}
\usepackage[]{algorithm2e}
\usepackage{fullpage}
\usepackage{listings}
\usepackage{enumerate}
\usepackage{wrapfig}
\usepackage{float}
\usepackage{parskip}
\usepackage{lipsum}
\usepackage{listings,color}
\usepackage[font=small,labelfont=normal,labelsep=default]{caption}
\usepackage[nodisplayskipstretch]{setspace}
\usepackage{epstopdf}
\AppendGraphicsExtensions{.tif}
\usepackage[colorlinks=true,urlcolor=blue,citecolor=red,linkcolor=red,bookmarks=true]{hyperref} 
\usepackage{cleveref}
\usepackage{pstricks,pstricks-add,pst-3d,pstricks-add,pst-grad,multido}

\definecolor{halfgray}{gray}{0.55} 
\definecolor{webgreen}{rgb}{0,.5,0}
\definecolor{webbrown}{rgb}{.6,0,0}
\definecolor{BlueLUH}{cmyk}{1.0,0.7,0,0}
\colorlet{LightBlue}{BlueLUH!20!white}
\colorlet{DarkBlue}{BlueLUH!80!black!20}

\title{Large Scale GPU Based Simulations of Turbulent Bubbly Flow in a Square Duct}
\author[]{Purushotam Kumar\thanks{pkumar8@illinois.edu}}
\author[]{Surya P. Vanka\thanks{spvanka@illinois.edu}}
\affil[]{Department of Mechanical Science and Engineering, University of Illinois at Urbana-Champaign}	
\date{\today}
\providecommand{\keywords}[1]{Keywords: #1}
\begin{document}
\maketitle
\onehalfspacing

\begin{abstract}	
	{\it Turbulent bubbly flows in non-circular ducts are applicable in several engineering applications, such as nuclear reactors, air conditioning and refrigeration equipment, thermal power units, electronics cooling etc. In this paper, we present the results of a numerical study of air-water turbulent bubbly flow in a periodic vertical square duct. The study is conducted using a novel numerical technique which leverages Volume of Fluid method for interface capturing and Sharp Surface Force method for accurate representation of the surface tension forces. A three-dimensional geometry construction method is employed during solution of interface equation which gives absolute conservation of mass and sharp interface between gas and liquid phases. The spurious velocity which is ubiquitous in numerical modeling of multiphase flows is substantially smaller with our technique. The entire algorithm has been implemented on a data parallel mode on multiple graphics processing units (GPU) taking advantage of the large number of available cores.
		
		We have studied the dynamics of a swarm of spherical bubbles co-flowing with the upward turbulent flow and compared results with an unladen turbulent flow. The frictional Reynolds number of the unladen $Re_{\tau}$ is 360, which is sufficient to sustain a turbulent flow. We observe the turbulence-driven secondary flows in the mean flow, with complex instantaneous turbulent vortical structures. The interaction of these secondary flows with the upwards rising bubbles is very complex and leads to significant changes in the instantaneous and time-averaged flow field. In this study, we have considered 864 bubbles rising in the square duct on a grid consisting of 32 million control volumes. The calculations have been for over 1,500,000 time-steps and took approximately 6 months of calendar time on a 4-GPU computer. 
		
		We present the results of mean void fraction distribution, mean velocities, longitudinal and transverse turbulence intensities along the wall, corner bisector, and wall bisector. A peak in the void fraction distribution near the walls is observed representing the migration of bubbles to a preferred section of the duct. The effects of turbulence-driven secondary flows and instantaneous large eddies on preferential concentration of the bubbles are discussed. The dispersed bubbles are seen to break the long elongated turbulent structures commonly observed in the unladen turbulent flow.}  
\end{abstract}

\noindent\makebox[\linewidth]{\rule{\linewidth}{0.4pt}}
\section{Introduction}
\label{ch:bubbly_flow:sec:literature_review}
%
The bubbly flow in a circular pipe or a square duct is one of the simplest fully-developed two-phase flow to envision, yet it provides important information related to the interaction between bubbles and the continuous phase. As a consequence, a significant number of experiments have been conducted to understand upward and downward bubbly flows in circular pipes. In these studies, the void fraction distribution, mean phase velocities, mean bubble size, turbulence quantities, wall shear, heat transfer from wall to fluid etc. have been explored. 
Many physical quantities of interest such as heat transfer from wall to fluid, pressure drop and local void fraction depend on the distribution of bubbles in the duct. Through experiments and numerical simulations, it has been observed that the bubbles move in the lateral direction to either accumulate near the walls, also known as wall-peaking case, or in the central region, referred to as core-peaking case. Intermediate void fraction distributions with multiple peaks along the radius are also reported in the literature. This lateral migration of bubbles is related to the lift force due to fluid-shear and mutual interaction between bubbles, which are dependent on the bubble size and drift velocity of the dispersed phase. In the upward bubbly flow, smaller bubbles $(d \le 3)$ mm concentrated in the wall region, whereas larger bubbles $(3 < d < 7)$ mm migrated to the pipe center. The change in preferential position of bubble is related to the reversal of lift force \cite{Tomiyama2002_bubble_migration} as the bubble size increases. Bubbles larger than 7 mm, initially migrate to the central region, but due to shape deformations and volumetric expansion break-up into smaller bubbles of varying sizes, and an intermediate void fraction distribution is observed. The lift direction also changes due to change in the sign of drift velocity. For example, in the downward bubbly flow, bubbles with diameter smaller than 3 mm migrate to pipe center, which is in contrast with the upward bubbly flow scenario. The works of \citet{Zun1988_transition_wall_peaking,Zun1993_space_time_evolution,Liu1989_phd_thesis,Liu1993a,Liu1993b,Tomiyama2002_bubble_migration,Colin2012} can be consulted for further information on void fraction distributions. \\
The turbulence characteristics of the surrounding flow can also plays significant role migration of the bubbles. However, apart from few experimental and numerical studies to characterize turbulence in channel and pipe flows, there is very little known about the flow field and turbulence characteristics of bubbly flow in the square ducts. In general, both mean and secondary flows are affected by bubbles in the turbulent bubbly flow. This is related to the preferential distribution of the bubble and subsequent variation of density in the cross-section. As mentioned earlier, bubbles have a tendency to concentrate either near the wall or in the central region, hence, the axial velocity depends strongly on flow conditions. For example, in the wall peaking scenario the average density near the wall decreases, therefore for the same pressure gradient the velocity near the wall increases. This results in a smoothed `M' shaped velocity profile with a flatter central portion. Similarly, for the core peaking scenario the average density in the core region decreases, hence, the mean velocity in the central region increases.\\
The turbulence generated by bubbles can also have significant impact on the structures of surrounding fluid flow, and source of this turbulence is combination of the relative motion between bubbles and fluid, their deformation, and vortex shedding behind the bubbles. Depending on the bubble Reynolds number, these vortices can be attached or detached, and their interactions with other bubbles create fluctuations that are smaller in scale than that of wall-induced turbulence. Lance and Bataille \cite{Lance1991} studied the interaction between the grid-generated isotropic turbulence and the wall-induced turbulence in a uniform bubbly flow. They showed that the turbulence remains isotropic except near the injection point of bubbles. There are two regimes of interaction between the wall-induced and bubble induced turbulence, and it depends strongly on the void fraction $\bar{\alpha}$ and $u_0^\prime/U_R$, the ratio between single phase longitudinal fluctuations and mean drift velocity. The excess turbulence kinetic energy due to bubbles first increased linearly with the void fraction for all grid-generated turbulence intensity, before increasing with a higher slope for higher grid-generated turbulence intensity. The transition between two regimes depends on the critical void fraction $\bar{\alpha} \le 1\%$; below which the hydrodynamic interactions between the bubbles were negligible, and the turbulent kinetic energy of the liquid was a summation of the grid-generated and bubble-induced turbulent kinetic energies. Above the critical void fraction, the bubbles transferred a great amount of energy to the liquid phase, primarily due to the mutual interaction between bubbles. Since, the average distance between the bubbles decreases with the increase in the void fraction, the hydrodynamic interaction is also expected to increase. Similarly, when the grid-generated turbulence intensity is increased large trajectory fluctuations and an increase in probability of mutual interaction between bubbles is observed. This is probably the reason why excess kinetic energy increases with the increase in the void fraction and turbulence intensity of the dispersed phase. \\ 
The void fraction also has a strong influence on the longitudinal velocity fluctuation correlations and the one-dimensional energy spectrum. The turbulence kinetic energy in the single phase flow is concentrated in low frequency eddies, and high frequency is described by the typical $-\frac{5}{3}$ power-law equation. The introduction of bubbles modifies the turbulence correlation and energy spectrum which can be divided in two regimes: (a) bubble-induced, and (b) wall-induced. In the first regime, the shape of correlation and energy spectrum is independent of the void fraction, but a minor deviation from single phase behavior is observed. In the second regime, the classical $-\frac{5}{3}$ power-law description of the energy spectrum is swiftly replaced by $-\frac{8}{3}$ power-law as the void fraction is increases for at a constant continuous phase flow rate. Hence, the turbulent kinetic energy in the bubbly flow is spread over a wider range of frequencies. Using the asymptotic expansion of the turbulence dissipation and production due to particles, Yuan and Michaelides \cite{Yuan1992} showed that smaller particles, whose interaction time $\tau$ is much larger than particle response time $\tau_p$, attenuated the continuous phase turbulence and the reduction was proportional to $d_p^3$. On the other hand, the larger particles ($\tau << \tau_p$) enhanced the turbulence which is related to the vortices generated by the particles. The shedding of vortices were only observed at particle Reynolds number ($Re_p$) larger than 600. The enhancement of turbulence due to particles was observed at a lower $Re_p$ which affirms role of attached vortices in turbulence enhancement. For a dilute bubbly flow ($ \bar{\alpha} \sim 10^{-4}$), Colin et al. \cite{Colin2012} showed that the turbulence intensities enhance in the upward flow and attenuate in the downward flow. In the upward flow, smaller scales possess more energy in the core region where as the larger scales possess more energy in the wall region. In the downward flow, transfer of energy from large to small scales were negligible, and bubbles always dissipated energy from the liquid.\\
Hosokawa and Tomiyama \cite{Hosokawa2013} studied bubble-induced turbulence in a laminar pipe flow (bulk Reynolds number = 900) and showed that turbulence kinetic energy, due to fluctuations introduced by bubbles, increased with the increase in the void fraction. For a core-peaking case, the turbulence intensities decreased with distance away from center. For a transition case, the peaks were observed at $r/R = 0.6$, (roughly speaking the trend of fluctuations followed the void fraction distribution). The radial and circumferential fluctuations were similar in magnitude and almost half of that in the axial direction. Alm{\'{e}}ras et al. \cite{Almeras2015} studied the bubble-induced turbulence in a large square duct with void fraction ranging from 1\% to 13\%. The turbulence intensity in the streamwise direction was larger than those in the lateral directions. It is related to the wake of bubbles as they preferentially promote the capture and transport of fluids in the streamwise direction than in the lateral directions. They injected a low-diffusive passive dye to elaborate the capture and transport process. They observed that in both streamwise and lateral directions, there are two regimes of diffusion coefficient dependence on the void fraction. At lower void fractions, where hydrodynamic interactions between bubbles are small, the diffusion coefficient increased nearly as $\bar{\alpha}^{0.4}$, whereas at higher void fractions, the diffusion coefficient were seen to be independent of $\bar{\alpha}$. The transition between two regimes occurred earlier in the lateral directions $(1\% \le \bar{\alpha}a \le 3\%)$ than in the streamwise direction $(3\% \le \bar{\alpha} \le 6\%)$.\\
In the present paper, we have studied modification of turbulence in a vertical square duct in upward turbulent bubbly flow. Despite a reasonably large number of previous studies, the bubble induced turbulence has not been fully characterized in low void fraction regimes. In particular, there are very few studies that have examined turbulence modification in square duct. A volume of fluid method with a sharp surface force method for the inclusion of surface tension forces has been developed for a multiple graphics processing units platform. The mean quantities and turbulence statistics are investigated and comparison between unladen and bubble laden turbulent flow in presented. In section \ref{sec:method} the governing equations and solution procedure used for the computations are presented. The computational details and problem description are presented in Section \ref{ch:bubbly_flow:sec:computational_details}. In Section \ref{sec:results}, we present results of the simulations performed in the study and discuss the important findings. Section \ref{ch:bubbly_flow:sec:summary} presents a summary of the present results.
%
\section{Numerical method}
\label{sec:method}
\hspace{3mm} We have developed a numerical procedure that solves the relevant governing equations on a fixed Eulerian collocated grid. The procedure consists of algorithms for interface capturing, inclusion of surface tension forces, and solution of Navier-Stokes equations. The volume of fluid (VOF) method of Hirt and Nichols \cite{Hirt1981} is used for interface capturing. The truncated interface between the gas and liquid in a control volume is represented as a oblique plane, and the interface normal and curvature are computed using a smoothed liquid volume fraction and the height function method (Rudman \cite{Rudman1998} and Cummins et al. \cite{Cummins2005}), respectively. We have incorporated the surface tension force in Navier-Stokes equations using a sharp surface force (SSF) and pressure balance methods (Francois et al.\cite{Francois2006}, Wang and Tong \cite{Wang2010}). The Navier-Stokes equations are solved using the fractional step method. The numerical procedure has been previously used to study the effects of duct confinement on dynamics of bubble rising in a square duct \cite{Kumar2015confinement,Kumar2015numerical}, the effects of magnetic field \cite{Jin2016} and variable viscosity \cite{Kumar2015non_newtonian,Kumar2019AJKFluids} on bubble dynamics.  \par 
\subsection{Governing equations}
\hspace{3mm} We assume that both the gas and liquid are isothermal, incompressible and Newtonian fluids. A single fluid approach with a method to capture the interface between gas and liquid is used. The combined governing equations for both fluids are given by following equations::
\begin{alignat}{1}
	\label{eqn:continuity}
	\nabla \cdot  \mathbf{u} = 0
\end{alignat}
\begin{equation}
	\label{eqn:momentum_equation}
	\begin{aligned}
		\frac{\partial \left(\rho \mathbf{u} \right)}{\partial t} + \nabla \cdot \left( \rho \mathbf{uu} \right) =  -\nabla p & + \nabla \cdot \left( \mu \left[ \nabla \mathbf{u} + \nabla \mathbf{u}^{T} \right] \right) \\ & + \rho \textbf{g} + \sigma \kappa \mathbf{n} \delta \left( \mathbf{x} - \mathbf{x}_f \right)
	\end{aligned}
\end{equation}
In the above equations, $\mathbf{u}$ is fluid velocity, $p$ is pressure with hydrostatic part being subtracted, $\rho$ is density, $\mu$ is dynamic viscosity, $\sigma$ is surface tension coefficient, $\kappa$ is interface curvature, $\mathbf{n}$ is interface normal, $\delta$ is the delta function, $\mathbf{x}$ is the spatial location where the equation is solved, $\mathbf{x}_f$ is the position of the interface and $\mathbf{g}$ is acceleration due to gravity. \par
\hspace{3mm} Since free surface is captured by the volume of fluid method, a time evolution of VOF function (liquid volume fraction, $\alpha$) given by
\begin{alignat}{1}
	\label{eqn:vof_eqn}
	\frac{\partial \alpha}{\partial t} + \textbf{u} \cdot \nabla \alpha = 0 
\end{alignat}
is solved. The geometry construction method (Rider and Kothe \cite{Rider1998}) coupled with a second-order operator split method (Noh and Woodward \cite{Noh1976}, Li \cite{Li1995}, Ashgriz and Poo \cite{Ashgriz1991} and Francois et al. \cite{Francois2006}) is used to solve the evolution equation. The mixture density and viscosity are calculated as linear weighting of the individual phase values, as
\begin{alignat}{2}
	\label{eqn:density_viscosity}
	\rho &= \alpha \rho_l+ (1 - \alpha)\rho_g \\
	\mu  &= \alpha \mu_l + (1 - \alpha)\mu_g 
\end{alignat}
The subscripts $l$ and $g$ denote gas and liquid phases respectively. The last term in eq. \eqref{eqn:momentum_equation} is the surface tension force, and $\sigma \kappa$ is the pressure jump at the free surface. In our current algorithm, the jump condition is enforced explicitly by representing the surface tension force as a pressure gradient given by
\begin{alignat}{1}
	\label{eqn:jump_condition}
	\sigma \kappa \mathbf{n} \delta \left( \mathbf{x} - \mathbf{x}_f \right)  = - \nabla \tilde{p} 
\end{alignat}
where $\tilde{p}$ is the pressure solely due to the surface tension force at the interface. An elliptic equation for $\tilde{p}$ is derived from continuity and momentum equations,
\begin{align}
	\label{eqn:ppe_surface_tension}
	\nabla \cdot \left( \frac{\nabla \tilde{p}}{\rho} \right) = 0
\end{align}
\hspace{3mm} The jump condition at the interface is enforced by using eq. \eqref{eqn:jump_condition} as a source term in the solution of eq. \eqref{eqn:ppe_surface_tension}. This ensures the exact difference in pressure at the interface due to the surface tension. In our work, this elliptic equation for $\tilde{p}$ is efficiently solved using a multigrid accelerated red black SOR relaxation scheme implemented on a GPU. Using this method the spurious velocities are seen to reduce to machine zero for a static bubble with exact analytical curvature and to very small values when the curvature is numerically computed. Admittedly, there are several methods to include the surface tension force in the Navier-stokes (Renardy and Renardy \cite{Renardy2002}, Sussman et al. \cite{Sussman2003,Sussman2007}, Gueyffier et al. \cite{Gueyffier1999}), however, in our experience the current method more accurate for reduction of spurious velocities and relatively easier to implement on GPUs. \par
\subsection{Solution procedure}
The following steps outline solution of the governing equations,
\begin{enumerate}
	\item Initialize the solution with initial liquid fraction ($\alpha^n$) and velocities ($\mathbf{u}^n$)
	\item Calculate the density and viscosity at $n^{th}$ time step using,
	\begin{alignat*}{1}
		\rho^n = \alpha^n \rho_l + (1 - \alpha^n)\rho_g \\
		\mu^n  = \alpha^n \mu_l  + (1 - \alpha^n)\mu_g
	\end{alignat*}
	\item Solve the interface tracking equation (using $\mathbf{u}^n$) to obtain $\alpha^{n+1}$
	\item Calculate the density ($\rho^{n+1}$) and viscosity ($\mu^{n+1}$) using $\alpha^{n+1}$
	\item Calculate the interface curvature using the height function method
	\item Solve the pressure $(\tilde{p})$ using sharp surface force method
	\begin{alignat*}{1}
		\nabla \cdot \left( \frac{\nabla \tilde{p}}{\rho}\right)^{n+1} \hspace{5mm} \text{with} ~\nabla \tilde{p} = - \left( \sigma \kappa \mathbf{n} \delta \left( \mathbf{x} - \mathbf{x}_f \right) \right)^{n+1}
	\end{alignat*}
	\item Compute the pressure gradient term $\left( \frac{\nabla \tilde{p}}{\rho}\right)^{n+1}$ at cell faces
	\item Compute the convection term $(\nabla \cdot \rho \textbf{uu})$ using geometry construction method
	\item Compute the intermediate velocities at cell centers as,
	\begin{equation*}
		\begin{aligned}
			\rho_c^{n+1}\hat{\mathbf{u}}_c = \rho_c^n \mathbf{u}_c^n &- \Delta t \sum_f \mathbf{u}_c^n \left(\sum_{k=1}^{2} \rho_k^n \alpha_k^n\right) A_f \\ 
			&+ \Delta t \sum_f \mu_f^n \left(\nabla \mathbf{u}^n + \nabla^T \mathbf{u}^n\right)_f \cdot \mathbf{n}_f + \rho_c^{n+1} \left(\frac{\nabla \tilde{p}}{\rho}\right)_{f \rightarrow c}^{n+1}
		\end{aligned}
	\end{equation*}
	\item Compute the intermediate velocities at cell faces as, 
	\begin{alignat*}{1}
		\hat{\mathbf{u}}_f = \hat{\mathbf{u}}_{c \rightarrow f}
	\end{alignat*}
	\item Compute pressure $(p)$ using,
	\begin{alignat*}{1}
		\nabla \cdot \left( \frac{\nabla p}{\rho} \right)^{n+1} = \frac{\nabla \cdot \hat{\mathbf{u}}_f}{\Delta t}
	\end{alignat*}
	\item Correct the velocity field for divergence free condition using,
	\begin{alignat*}{1}
		\mathbf{u}_c^{n+1} = \hat{\mathbf{u}}_c - \Delta t \left( \frac{\nabla p}{\rho} + \frac{\nabla \tilde{p}}{\rho} \right)_{f \rightarrow c}^{n+1}
	\end{alignat*}
\end{enumerate}
Where, $_{c \rightarrow f}$ represents the linear averaging procedure to compute face value (eg. $\rho_f$, $\mu_f$, $\mathbf{u}_f$) from its cell centered values. 

We have implemented this algorithm to run on multiple graphics processing units (GPU). We have used the CUDA-Fortran platform supported by the Portland group (PGI) Fortran compilers. Previously, we have validated the multiple GPU implementation \citep{Kumar2015numerical,Vanka2016Single,Kumar2016thesis} to study confinement effects on bubble dynamics \citep{Kumar2015confinement}, two-phase flows at T-juctions \citep{horwitz2012simulations,horwitz2013simulations,kumar2013three}, bubble dynamics in non-Newtonain fluids \citep{ Kumar2015non_newtonian,Kumar2019AJKFluids}, droplet dynamics in square duct \citep{Horwitz2014_lbm,Horwitz2019AJKFluids}, Argon bubble rising in liquid steel \citep{Vanka2015APS_DFD,Jin2016mhd_bubble} and turbulent bubbly flow \citep{Vanka2016APS_DFD}. 

%
%
%
\section{GPU Implementation}
\label{ch:gpu_implementation}
This chapter gives a brief introduction to the Graphics Processing Unit (GPU), describes its architecture and inner workings. The chapter also provides brief details of the multiple GPU implementation and speed comparison.
\subsection{Graphics Processing Units}
\label{ch:gpu_imp:sec:introduction}
The graphics processing unit (GPU) is a massively parallel device that can be used for graphics generation and computing \cite{Nickolls2010_gpu_era}. It is ubiquitously present in every personal computers,  gaming consoles and mobile phones. Until the last decade, the GPUs have been primarily used for rendering the video games and post processing high definition movies. The real-time rendering is a highly parallel operation which requires multi-threaded high efficiency power horse devices to perform the visualization operations. Along with the high-speed rendering and video editing, they are also able to do fast single precision and double precision arithmetics. With the introduction to Compute Unified Device Architecture (CUDA) \cite{Cuda_programming_guide,Nickolls2008_scalable_parallel}, a parallel computing platform and programming model the GPUs are also used in the scientific computing. Many new generation supercomputers and data centers have dedicated GPU clusters. Using CUDA, the computer programs can be written into ANSI C, ANSI C++ and Fortran. \par 
\begin{figure}[H]
	\begin{center}
		\includegraphics[width=0.5\textwidth]{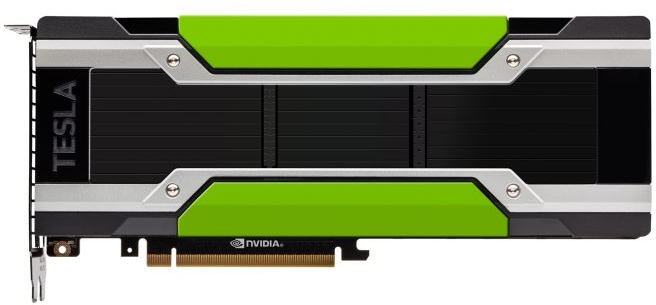}
		\caption{A Pascal P100 GPU made by NVIDIA}
		\label{fig:ch:gpu_imp:sub:architecure:main_board}
	\end{center}
\end{figure}
\Cref{fig:ch:gpu_imp:sub:architecure:main_board} shows a NVIDIA Pascal P100 GPU, which has a global memory of 16 GB and a memory bandwidth of 720 GB/s. The GPU board contains over 15 Billion transistors. It has a theoretical peak computation power of 9.3 TeraFLOPS for single precision and 4.7 TeraFLOPS for double precision computations.  
\subsection{Multiple GPU Implementation}
\label{ch:gpu_imp:sec:multigpu}
The computational performance of the GPU and adaptability of the CUDA programming environment have led to widespread acceptance of the GPUs in the scientific community. However, the exploration of fundamental sciences or large scale engineering problems require significantly larger computational resources than a single GPU can accommodate. It is true that the GPU hardware has grown significantly over the time a relatively smaller period of time. For example, the global memory per GPU has increased from 4 GB in 2009 to 24 GB in 2015. The number of CUDA core per GPU have grown from 192 in 2009 to 3584 in 2016. But it is still not sufficient to study a conduct direct numerical simulation of bubbly flow in a square duct. The obvious solution to this problem is to utilize multiple GPUs simultaneously. The multiple GPU (multi-GPU) implementations heavily borrows the programming experiences from the Massage Passing Interface (MPI) and the OpenMPI programming environments. \par 
There are three main sections of the multi-GPU implementations: 
\begin{enumerate}
	\item Domain decomposition
	\item Communication between adjacent domains
	\item Optimization of communication and computation times
\end{enumerate}
\subsubsection{Domain Decomposition}
\label{ch:gpu_imp:sub:domain_decomposition}
\begin{figure}[H]
	\begin{center}
		\includegraphics[width=0.8\textwidth]{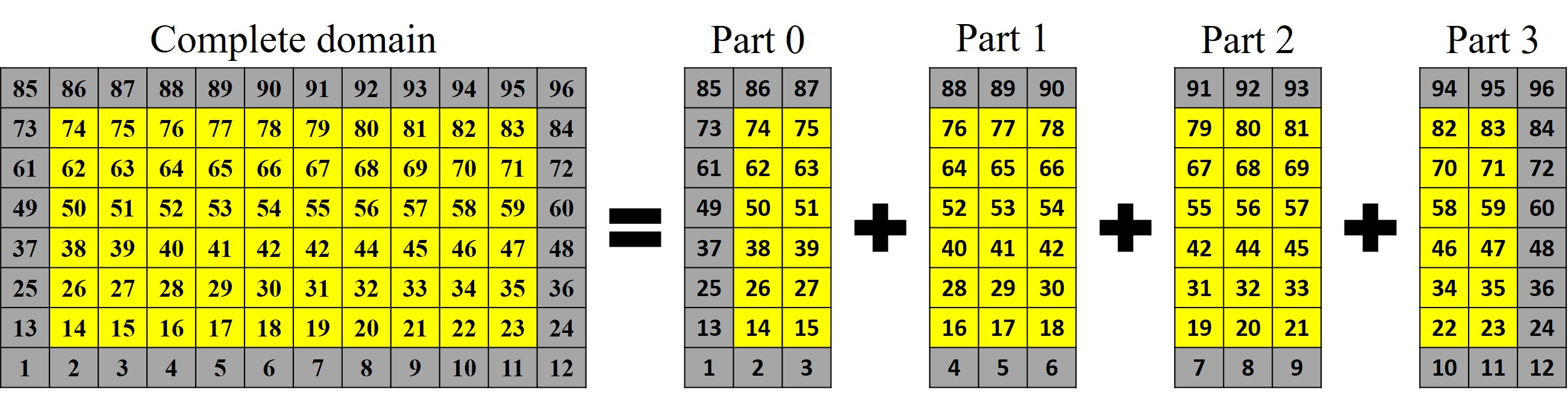}
		\caption{Illustration of Domain decomposition for 4 GPUs}
		\label{fig:ch:gpu_imp:sec:multigpu:domain_decomposition}
	\end{center}
\end{figure}

\begin{figure}[H]
	\begin{center}
		\includegraphics[width=0.8\textwidth]{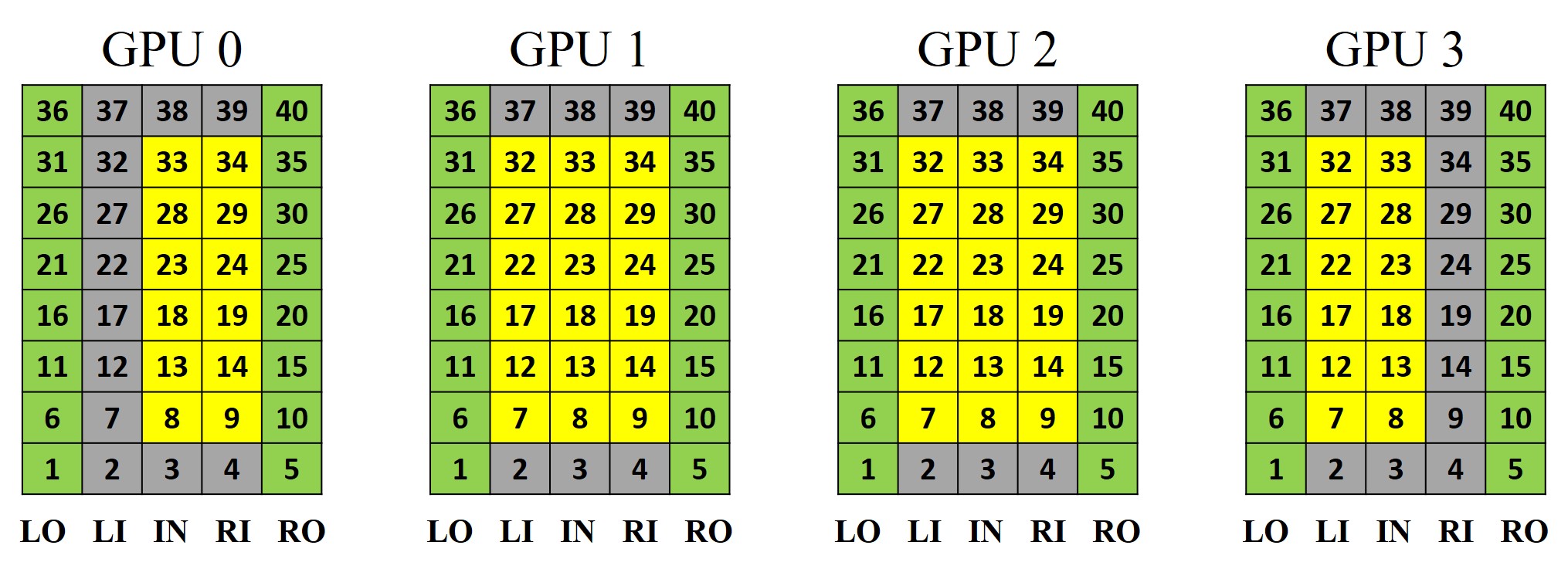}
		\caption{Additional layers at the domain boundaries}
		\label{fig:ch:gpu_imp:sec:multigpu:domain_decomposition_GPU_layout}
	\end{center}
\end{figure}

As name suggests, the domain decomposition refers to splitting of a larger computational domain into smaller subdomains which can fit on a single GPU. In case of three-dimensional Cartesian domain, the decomposition can be performed in several ways e.g. along $x$, $y$ or $z$ directions or $x$ and $y$ directions or $y$ and $z$ directions or $x$, $y$ and $z$ directions. Any of these direction can be chosen, however, a load balancing must be performed for its optimization. In our current implementation, we have chosen to split the domain along $z$ direction. This minimizes the communication time and reduces the programming difficulty. \Cref{fig:ch:gpu_imp:sec:multigpu:domain_decomposition} shows an illustration of one-directional domain decomposition along the horizontal direction of a two-dimensional computational domain. The yellow and gray colored cells represent the interior and boundary cells respectively. It can be noticed from figure that equal number of computational cells has been prescribed in each subdomain. This is enforced to ensure optimum load balancing for each GPU.
The domain decomposition alleviates the shortcomings of smaller global memory. But, it creates virtual boundaries between two neighboring domains. Hence, a mechanism to pass the information between adjacent domain/GPU is required. To accommodate the message passing between neighbors one layer is added on each side of the domain and these cells are referred as halo cells. \Cref{fig:ch:gpu_imp:sec:multigpu:domain_decomposition_GPU_layout} shows the subdomains on each GPUs and their respective data structures. The LO and RO are the left and right outer cells and used to facilitate the data transfer between neighboring domains. For example, the right outer cells of GPU0 mirrors the left interior of GPU1 and left outer cells of GPU1 mirrors the right interior of the GPU0. The LI and RI are the left and right interior cells which is useful to facilitate physical boundary conditions on GPU0 and GPU3 respectively. The left and right interior cells are also used to transfer the data between neighboring cells. Although The left and right outer cells are not needed on GPU0 and GPU3 respectively, they are added to maintain uniform data structures across the GPUs. It should be also noted that the cells numbering on subdomains are different from that on the global domain. 
\subsubsection{Communication between Neighboring Domains}
\label{ch:gpu_imp:sub:communication}
\begin{figure}[H]
	\begin{center}
		\includegraphics[width=0.95\textwidth]{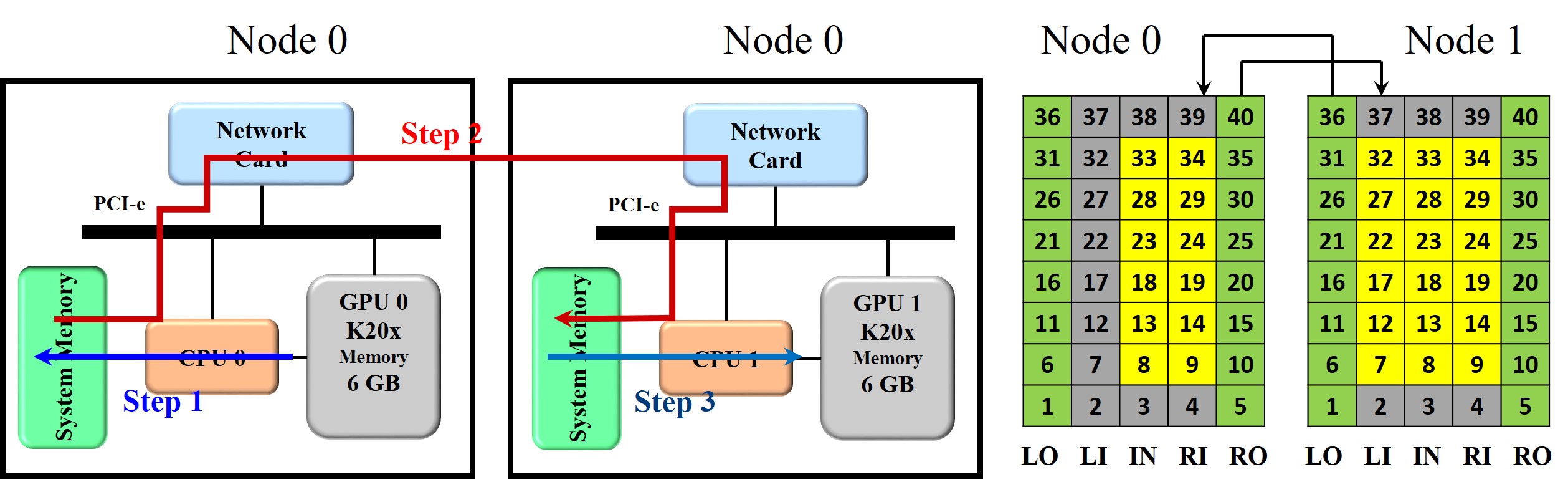}
		\caption{Data transfer process between two nodes}
		\label{fig:ch:gpu_imp:sec:multigpu:data_transfer}
	\end{center}
\end{figure}

\Cref{fig:ch:gpu_imp:sec:multigpu:data_transfer} illustrates the data transfer procedure between the two nodes connected via network. It is a three step process. In the first step, the data is copied from the GPU0 memory to the CPU0 memory on Node 0 via PCIe interface. In the second step, the data is sent from the Node 0 to Node 1 via InfiniBand \cite{InfiniBand2000}. The Message Passing Interface (MPI) \cite{Gropp1996_high_performance_MPI,Gropp1999_MPI} is used to transfer the data over the network. In the last step, the data is copied from the CPU1 memory to the GPU1 memory on Node 1. In these three steps, the data is copies from the GPU 0 to the GPU 1. This process is performed co-currently on every Node to minimize the down/waiting time. Once the transfer from left to right ($0 \rightarrow 1$, $1 \rightarrow 2$, $2 \rightarrow 3$ etc.) has taken place, the transfer from right to left ($0 \leftarrow 1$, $1 \leftarrow 2$, $2 \leftarrow 3$ etc.) is initiated. The total time taken to complete the data copying process is referred as \textbf{communication time}. 
On a shared memory machine where a multi-core CPU and multiple GPUs are present, the data copying process is less cumbersome. During the course of calculation, each GPU is assigned to a particular CPU core. The copying process is performed using same routine as explained earlier. The first and last steps of the data transfer is same to that of the distributed systems. In the second step, the data is passed from one CPU to another via pointers. Although absence of data copying over network reduces the overall communication time the reduction is relatively less prominent. This technique to balance computation time over communication time has also been applied in our Lattice Boltzmann implementation \cite{Kumar2016thesis,Horwitz2014_lbm,Horwitz2019AJKFluids}  \par 
\subsubsection{Multi-GPU Speedup} 
\label{ch:gpu_imp:sub:multi_gpu_speedup}
The performance of multi-GPU implementation has been tested on a four GPU shared memory workstation as well as on Blue Waters (BW) supercomputer which is an distributed memory cluster. It should be noted that processing speed of in-house workstation CPU is better than that of Blue Water CPU. A multi-CPU version code has also been developed for CPU vs GPU speedup comparisons. A 3D lid-driven cavity problem is selected to test the performance of CUFLOW. The governing equations are solved on a grid of $128 \times 128 \times 512$ or approximately 8 million control volumes.  
\begin{table}[!h]
	\caption{Specification of the hardwares used} \label{tab:GPUconfing}
	\centering
	\begin{tabular}{| *1{>{\centering\arraybackslash}m{1.0in}} |*2{>{\centering\arraybackslash}m{2.0in}|} @{}m{0pt}@{} }
		\cline{2-4}
		\multicolumn{1}{c|}{ } & In-house workstation & Blue waters & \\  [1ex]
		\hline
		\# of nodes & 1 & 4224 & \\
		\hline
		\multirow{2}{*}{Node CPU} & Xeon E5-2650v2 & AMD 6276 & \\  [1ex]
		{} & 2.60 GHz, 8 cores & 2.3 GHz, 16 cores & \\
		\hline
		\multirow{2}{*}{GPU/Node} & 4 $\times$ NVIDIA Tesla C2075 & 1 $\times$ NVIDIA Tesla K20x & \\   [1ex]
		{} & 5 GB each, 575 MHz & 6 GB each, 732 MHz & \\
		\hline
	\end{tabular}
\end{table}
\begin{figure}[H]
	\begin{center}
		\includegraphics[width=0.5\textwidth]{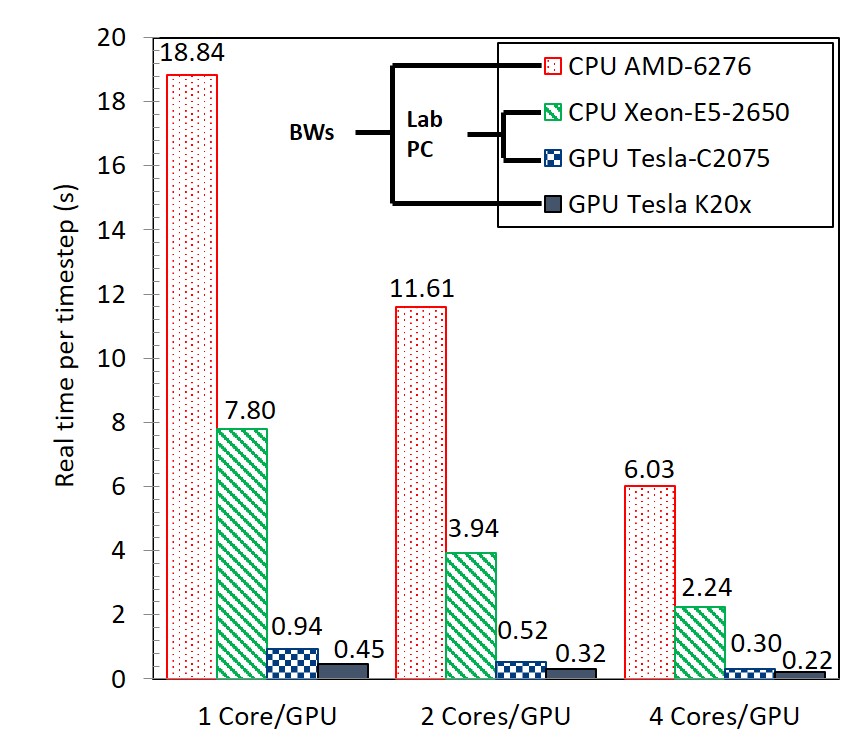}
		\caption{Time taken by multi-CPU and multi-GPU CUFLOW}
		\label{fig:ch:gpu_imp:sec:multigpu:speedups}
	\end{center}
\end{figure}
\Cref{fig:ch:gpu_imp:sec:multigpu:speedups} shows the time taken to complete one time step of the simulation. It can noticed that the CPU version of the CUFLOW takes longer time on the Blue Water supercomputer than on the in-house workstation. However, the GPU version takes lesser time on Blue water. The multi-GPU code on in-house workstation reduces the computation time by a factor of 3.2 between one GPU and four GPUs whereas on the Blue Water supercomputer the factor is approximately 2. This is primarily due to increased communication time overhead. 
%
%
%
%
\section{Computational details}
\label{ch:bubbly_flow:sec:computational_details}
%
The simulations presented here are performed for an upward bubbly flow in a square duct. In the $x ~\text{and}~ y$ directions, the flow is wall-bounded where no-slip and no-penetration conditions are applied. In the $z$ or streamwise direction the flow is considered periodic. The gravitational acceleration acts in the downward or negative $z$ direction, and the contribution due to hydrostatic pressure on the continuous phase is removed with the addition of $(\rho_0 \mathbf{g})$, where $\rho_0 = (1-\bar{\phi})\rho_l + \bar{\phi}\rho_l$ $(\bar{\phi}: \text{average void fraction})$ to Navier-stokes equations. This was employed in order to prevent a uniform downward acceleration of the continuous phase. A constant mean pressure gradient based on bulk Reynolds number $Re_b = \rho_l w_b W \mu_l^{-1}$ of unladen flow has been applied along positive $z-$direction. It has been kept constant between unladen and laden flows, hence the bulk velocity of laden flow is expected to increase as the average density within the domain will decrease. The bulk Reynolds number of unladen flow simulation is set to 5300, so that sustainable turbulence is realized. The corresponding frictional Reynolds number $Re_{\tau} = \rho_l u_{\tau} W \mu_l^{-1}$ is approximately 360. \par
Two simulations of turbulent flow: (a) unladen flow, and (b) bubble laden flow are performed in the present study. Both simulations are performed on a Cartesian grid of $192 \times 192 \times 768$ control volumes with a uniform grid spacing in $x$, $y$, and $z$ directions. The extent of a square duct in $x$, $y$, and $z$ directions is $48 mm \times 48 mm \times 192 mm$. We note that the canonical height $(H)$ used to understand unladen turbulence flow in ducts and channels is $H = 2\pi W$ \cite{Kim1987_turbulence_stat,Madabhushi1991} or higher, however, for bubble laden flow, it has been observed that the correlation length of two-phase flow is smaller than that of single phase flow \cite{Lance1991,Panidis2000,Hosokawa2013,Santarelli2015}. Hence, for computational efficiency we have selected a smaller duct length (four times the duct width). Admittedly, turbulence quantities of unladen flow are not expected to match exactly with results available in literature \cite{Madabhushi1991}. However, since the primary focus of this work is on understanding the mechanisms for capturing and redistribution of the bubble by turbulence and modification of the velocity structures, root mean square velocities and Reynolds stresses by the dispersed phase this domain size is adequate. \par
\hspace{3mm} The unladen flow is started with a perturbed solenoidal velocity and integrated till a stationary flow field is realized. The initial bulk velocity and fluctuations are given by,
\begin{equation}
	\label{eq:ch:bubbly_flow:sec:computational_details:solenoidal}
	\begin{aligned}
		w(x,y,z) =  
		\begin{cases}
			w_b \left(\frac{2y}{W}\right)^{1/7}      & y/W \in [0,0.5] \\
			w_b \left(2 - \frac{2y}{W}\right)^{1/7}  & y/W \in [1,0.5] 
		\end{cases}
	\end{aligned}  
\end{equation}
\begin{equation}
	\label{eq:ch:bubbly_flow:sec:computational_details:purturbed}
	\begin{aligned}
		& u^\prime(x,y,z) = 0.08 w_b \cdot \text{cos}\left(\frac{n \pi x}{W}\right) \frac{W}{H} \left[\text{cos}\left(\frac{n \pi z}{H}\right)  \text{sin}\left(\frac{n \pi y}{W}\right) + (\text{sin}\left(\frac{n \pi z}{H}\right)  \text{cos}\left(\frac{n \pi y}{W}\right) \right] \\
		& v^\prime(x,y,z) = 0.08 w_b \cdot \text{sin}\left(\frac{n \pi z}{H}\right) \text{sin}\left(\frac{n \pi y}{W}\right) \text{sin}\left(\frac{n \pi x}{W}\right) \\
		& w^\prime(x,y,z) = 0.08 w_b \cdot \text{sin}\left(\frac{n \pi z}{H}\right) \text{sin}\left(\frac{n \pi y}{W}\right) \text{sin}\left(\frac{n \pi x}{W}\right) 
	\end{aligned}
\end{equation}
The bulk velocity $w_b$ of unladen flow is calculated from the bulk Reynolds number as $w_b= Re_b \mu_l \rho_l^{-1} W^{-1}$. The one-seventh law of turbulent channel flow with $8\%$ divergence free velocity fluctuations is prescribed as the initial condition for velocity field. Then the governing equations are integrated with the initial conditions and after a sufficient time, $t = 60T_b$, where $T_b = H/w_b$ is the bulk time, the effects of initial condition are seen to diminish and a stationary turbulence field is realized. At this point, we initiate the calculation of mean variables and collect them until fully developed turbulence is realized and a stationary mean velocity profile is achieved. The fully developed turbulence (or stationary mean flow) is achieved by approximately $t = 145T_b$. Turbulence statistics such as rms velocities $(\sqrt{\left\langle u^\prime u^\prime\right\rangle},\sqrt{\left\langle v^\prime v^\prime\right\rangle}, \sqrt{\left\langle w^\prime w^\prime\right\rangle})$ and Reynolds stresses $(\left\langle u^\prime u^\prime\right\rangle, \left\langle v^\prime v^\prime\right\rangle, \left\langle w^\prime w^\prime\right\rangle, \left\langle u^\prime v^\prime\right\rangle, \left\langle u^\prime w^\prime\right\rangle)$ are collected for an additional 55 bulk times (or until $t = 200T_b$). The results of single phase calculations are presented in \cref{ch:bubbly_flow:sec:unladen_validation}.\par
The turbulent flow field from the solution of unladen flow is taken as the initial condition for laden flow calculation. At $t = 200T_b$, 864 mono-dispersed bubbles are uniformly placed in the computational domain. Initial velocities inside the bubbles are set to that of the liquid velocities at respective spatial locations. Based on the total volume of bubbles, the void fraction of the laden flow is approximately $\bar{\phi} = 0.82\%$. Since void fraction is less than $2\%$, the mixture can be considered dilute and hydrodynamic interactions among bubbles are expected to be negligible. The bubbles are arranged in a regular array of $6 \times 6 \times 24$ bubbles in $x, y, ~\text{and} ~z$ directions respectively. Although they are placed in a regular array, their initial configuration is expected to have no effect on the stationary-state results. Bunner and Tryggvason \cite{Bunner2002,Bunner2003} have explored the effects of the initial configuration of bubbles and found negligible effect on the final solution. \par
\begin{figure}[H]
	\begin{center}
		\includegraphics[height=0.4\textwidth]{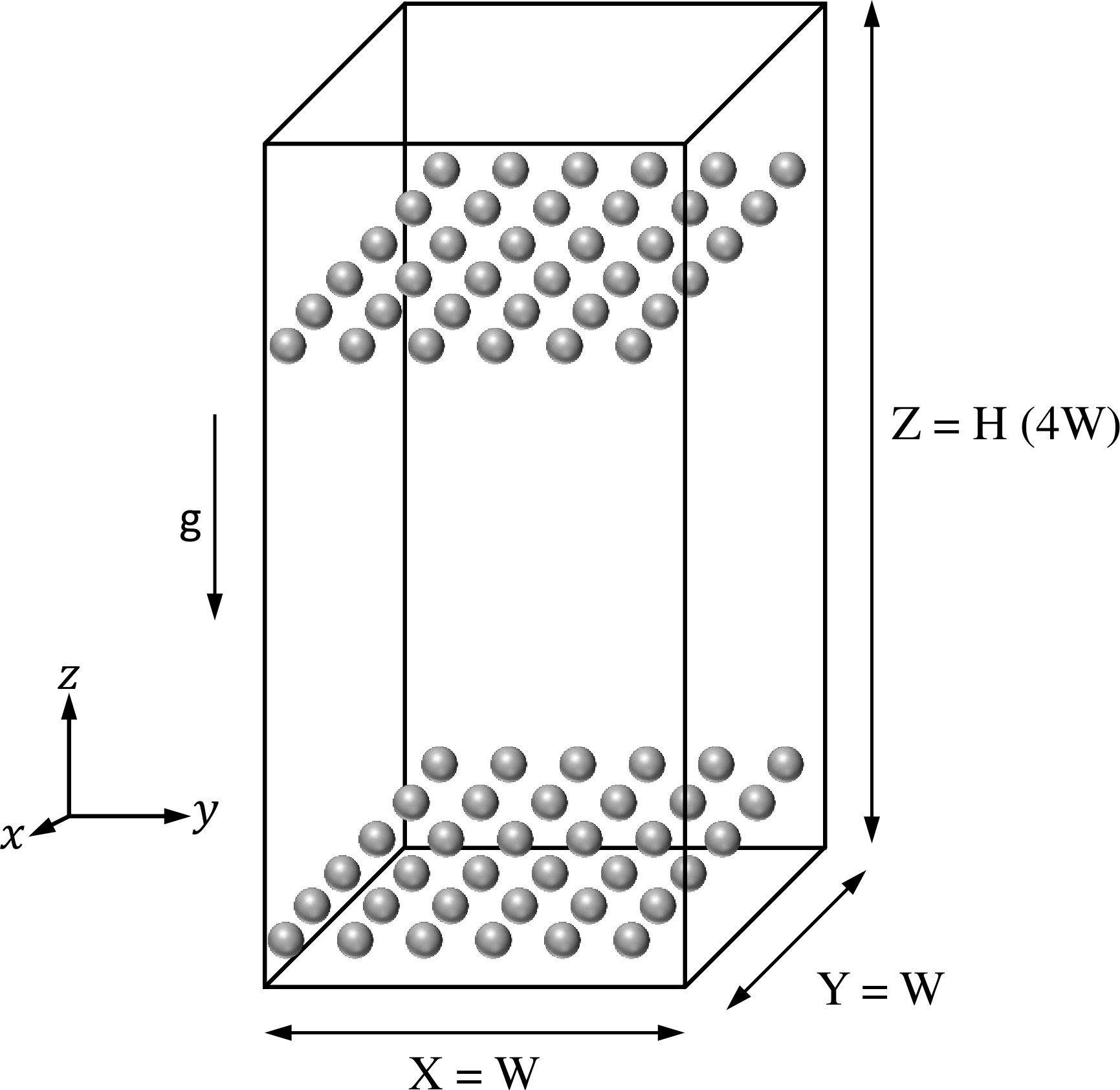}
	\end{center}
	\caption{Computational domain and initial bubble locations}
	\label{fig:problem_statement:ch:bubbly_flow}
\end{figure}

\Cref{fig:problem_statement:ch:bubbly_flow} shows the illustration of computational domain and the initial configuration of the top and bottom layers of 36 bubbles (the other 22 layers are not shown in the picture, but they are present in the simulations). The liquid and gas densities are 1000 and 1.2 kg/m$^3$ respectively, which results in a density ratio of $\rho_l/\rho_g = 833$. The dynamic viscosity of the liquid and gas are $0.001$ and $1.8 \times 10^{-5}$ Pa.s, respectively. The surface tension $(\sigma)$ between liquid and gas is $0.072$ N/m. The diameter of each bubble is $2$ mm. The \textit{Bond number} $Bo = \rho_l g d^2/\sigma$ (or \textit{E\"otvos number} $Eo = \Delta \rho g d^2/\sigma$) and \textit{Morton number} $Mo = g\mu_l^4/\rho_l \sigma^3$ are approximately 0.55 and $2.6 \times 10^{-11}$, respectively. Based on the combination of \textit{Bond} and \textit{Morton} numbers, the expected terminal shape of a bubble rising in a quiescent liquid column is nearly spherical \cite{Grace1973}, and it is expected to follow a vertical straight path during its rise. The choice of smaller bubble is to explore the effects of spherical bubbles. However, due to local shear and turbulence, a small degree of deformation as well as a non-rectilinear rise path is expected. \par
%
%
%
%
\section{Result and discussion}
\label{sec:results}
\subsection{Validation: Unladen Duct Flow}
\label{ch:bubbly_flow:sec:unladen_validation}
We present results of a validation study for predictions of single phase turbulence and compare them with results in literature. The single phase turbulence predictions are verified qualitatively and quantitatively with the works of \citet{Gavrilakis1992,Huser1993,Broglia2003,Zhang2015}. The frictional Reynolds number $Re_{\tau} = 360$ of the current work is intermediate to \citet{Gavrilakis1992,Huser1993,Zhang2015}, and it is same as Broglia et al. \cite{Broglia2003}, a large eddy simulation, hence, only qualitative comparisons can be made for some turbulence quantities. The algorithm used for predicting turbulent flows in square ducts has also been validated by \citet{Winkler2004_preferential,Winkler2006_wall_deposition,Chaudhary2009,Chaudhary2010_DNS_MHD,Chaudhary2011_DNS_MHD,Shinn2011}. The presented quantities are averaged over four statistically invariant transformations: time, $z$-direction, to account for periodicity, and two central planes: $x$ and $y$, to enforce quadrant symmetry.  
\begin{figure}[H]
	\begin{center}
		\includegraphics[width=0.6\textwidth]{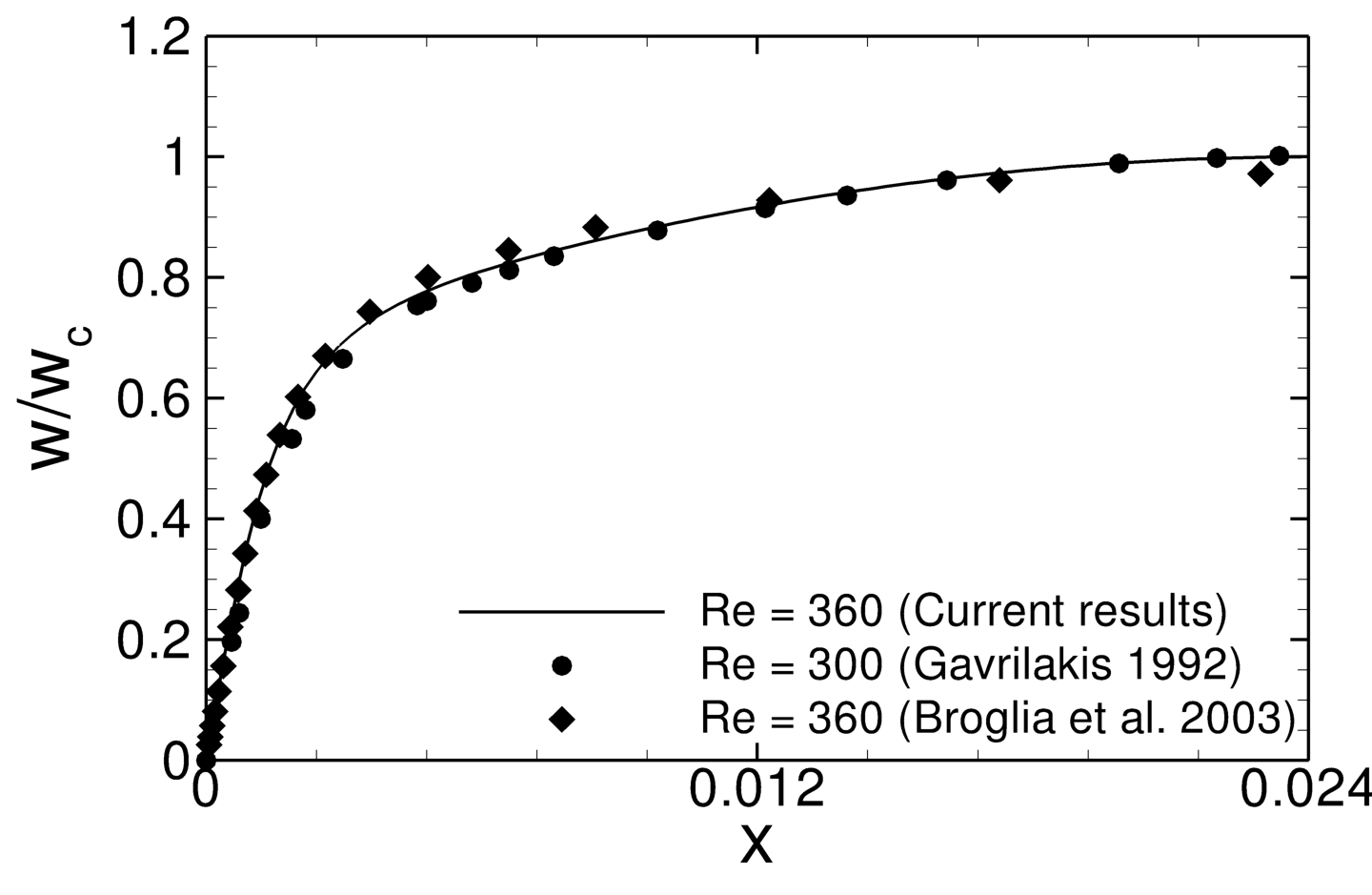}
	\end{center}
	\caption{Averaged streamwise velocity $(w)$ y-wall bisector}
	\label{fig:validation_single_phase_stat_mean:ch:bubbly_flow:sec:results}
\end{figure}        
\Cref{fig:validation_single_phase_stat_mean:ch:bubbly_flow:sec:results} shows the mean streamwise velocity distribution along the wall bisector. It has been scaled by the centerline streamwise velocity $(w_c)$ which is $w_c = 0.1464$. It can be noticed that the current prediction of streamwise velocity profile is very close to the predictions made by Gavrilakis \cite{Gavrilakis1992} for $x > 0.0072~mm$ (or $x > 0.3W)$. For $x < 0.3W$, the predicted mean velocity is approximately 1-2\% higher. It should be noted that they used a stretched grid of $127 \times 127$ nodes in the cross-section and their first grid node was at $\Delta x^{+} = 0.45$, whereas the current results are obtained using a uniform grid of $192 \times 192$ nodes with the first node at $\Delta x^{+} = 1.875$. Hence, a small difference in the near-wall statistics is expected between these two studies. A similar qualitative agreement can also be observed between current and Broglia et al. \cite{Broglia2003} predictions. The current results are nearly identical for $x \le 0.1W$ and are within 1-2\% for $x > 0.1W$. It is also noteworthy that point-wise data for comparison purposes were extracted from the original articles using Matlab imaging tool and manual hand picking, hence a small human error can also be expected. \par
\begin{figure}[H]
	\begin{center}
		\begin{subfigure}[b]{0.48\textwidth}
			\includegraphics[width=1\textwidth]{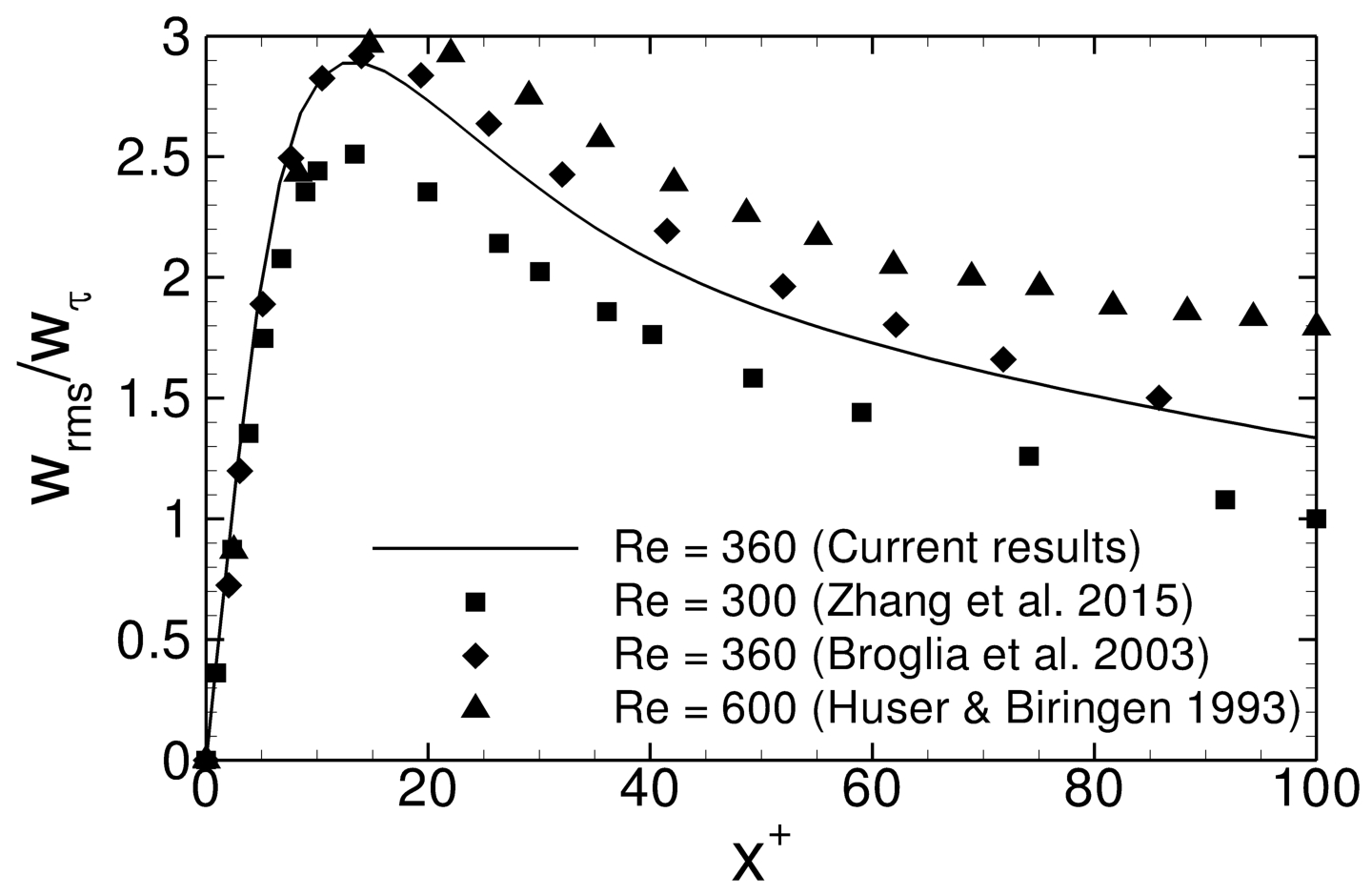}
			\caption{}
			\label{fig:validation_wrms_wtau:ch:bubbly_flow:sec:results}
		\end{subfigure} %
		\hspace{3mm}
		\begin{subfigure}[b]{0.48\textwidth}
			\includegraphics[width=1\textwidth]{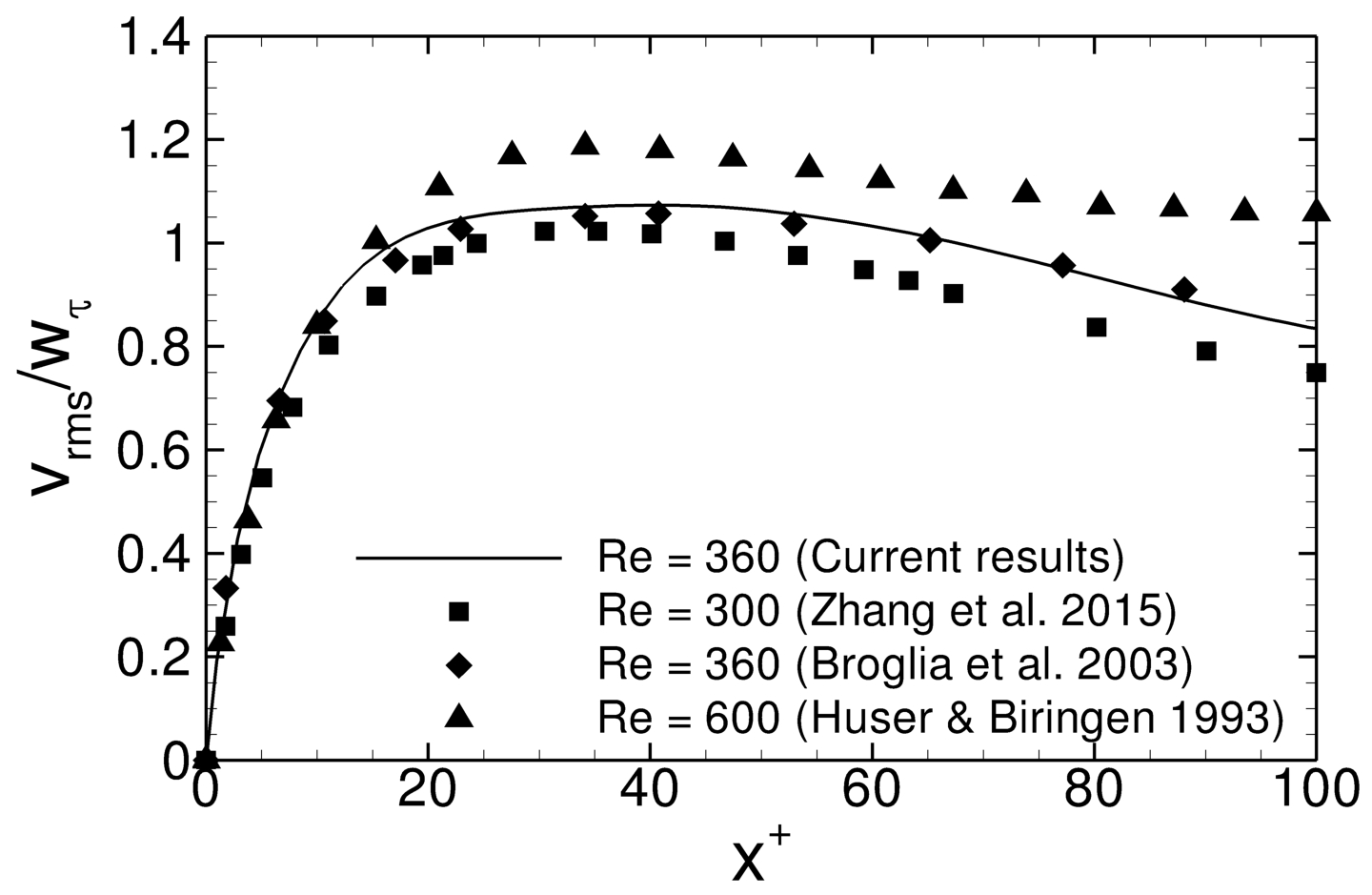}
			\caption{}
			\label{fig:validation_vrms_wtau:ch:bubbly_flow:sec:results}
		\end{subfigure} %
		\caption{Root mean square velocity fluctuations along the wall bisector: (a) $w_{rms}$ and (b) $v_{rms}$}
		\label{fig:validation_single_phase_stat_rms:ch:bubbly_flow:sec:results}
	\end{center}
\end{figure}

\Cref{fig:validation_single_phase_stat_rms:ch:bubbly_flow:sec:results} shows the streamwise $(w_{rms})$ and spanwise $(v_{rms})$ velocity fluctuations along the y-wall bisector for $x^+ \le 100$. A qualitative agreement between current predictions and Zhang et al. \cite{Zhang2015} prediction at $Re_{\tau} = 300$ and Huser and Biringen \cite{Huser1993} at $Re_{\tau} = 600$ can be noticed for both $(w_{rms})$ and $(v_{rms})$ from the \cref{fig:validation_wrms_wtau:ch:bubbly_flow:sec:results,fig:validation_vrms_wtau:ch:bubbly_flow:sec:results}, respectively. In direct comparison with $Re_{\tau} = 360$, it is evident that $v_{rms}$ from our prediction is in close agreement with that of Broglia et al. \cite{Broglia2003}, whereas the streamwise rms velocity, $w_{rms}$, is identical to current work for $x^{+} \le 18$ and higher for $x^{+} > 18$. The higher value of $w_{rms}$ in Broglia et al. \cite{Broglia2003} is the artifact of a larger streamwise grid spacing. A similar effect of grid resolution has also been observed by Zhang et al. \cite{Zhang2015}.
Another way to measure the solution's accuracy is to compare the ratio of centerline streamwise and bulk velocities, $(w_c/w_b)$, and the ratio of frictional and bulk velocities, $(w_{\tau}/w_b)$ to those reported in the literature.  The values of $(w_c/w_b)$ and $(w_{\tau}/w_b)$ are 1.33 and 0.678, respectively which are within 0.3\% of that reported in the literature \cite{Madabhushi1991,Broglia2003}. These simulations validate the grid used and provide data for comparisons with bubble laden flow. 
\subsection{Bubble Laden Flow}
\label{ch:bubbly_flow:sec:results}
\begin{figure}[H]
	\begin{center}
		\includegraphics[width=0.5\textwidth]{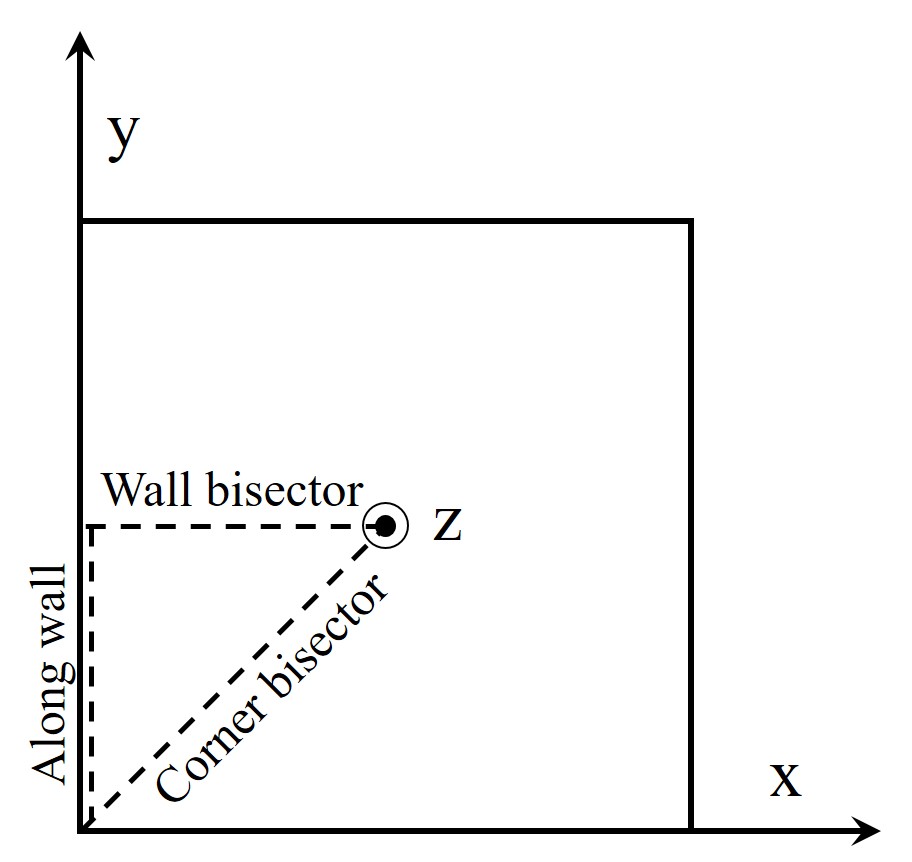}
	\end{center}
	\caption{Locations along which the quantities are presented}
	\label{fig:presentation_direction:ch:bubbly_flow:sec:results}
\end{figure}

Now, we present the results of study to analyze the effects of bubbles on liquid phase turbulence and discuss interaction between bubbles and turbulence. We discuss the findings of unladen flow and compare them with the bubble laden flow results. The instantaneous flow discussions are presented first, followed by a discussion on the turbulence statistics. Most of the turbulence quantities are calculated and discussed near the wall region, along the corner bisector and along the wall bisector. These locations are shown in \cref{fig:presentation_direction:ch:bubbly_flow:sec:results}. 
\subsubsection{Bubble Distribution}
\begin{figure}[H]
	\begin{center}
		\begin{subfigure}[b]{0.29\textwidth}
			\includegraphics[width=1\textwidth]{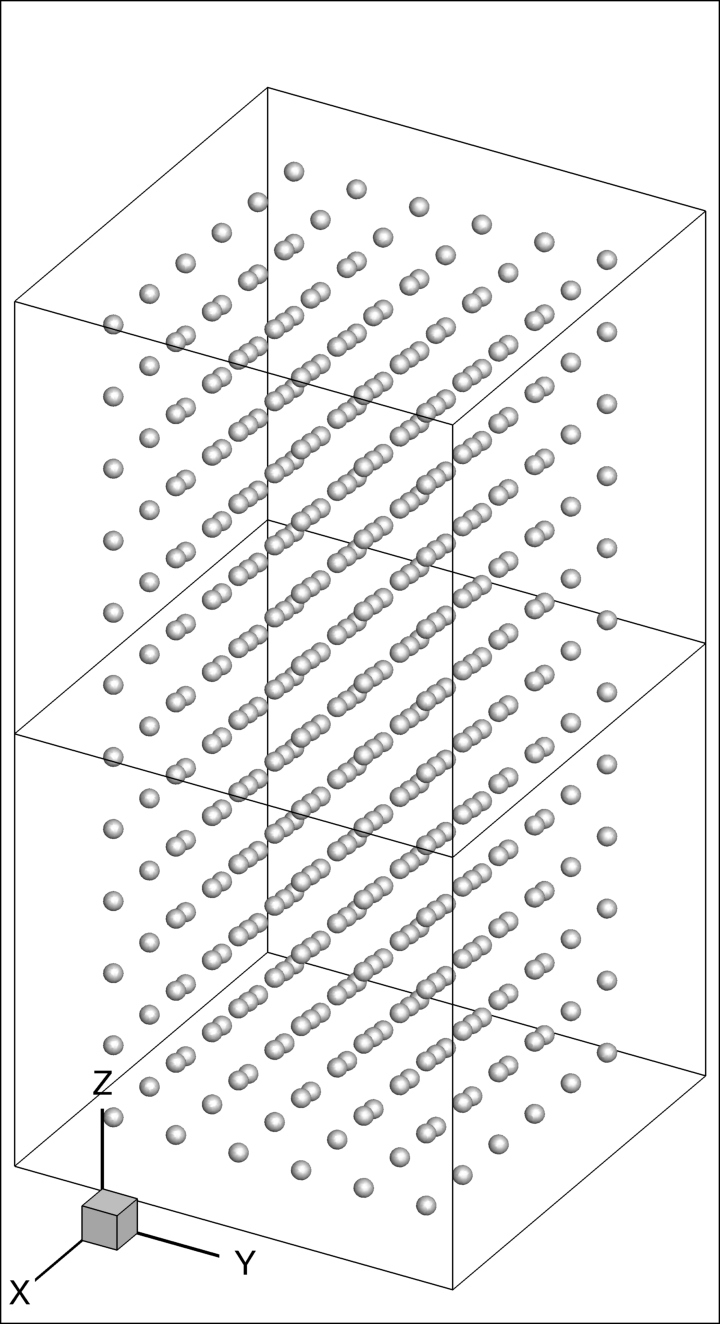}
			\caption{Isometric view}
			\label{fig:bubble_initial_iso_surface_isometric:ch:bubbly_flow:sec:results}
		\end{subfigure} %
		\hspace{1mm}
		\begin{subfigure}[b]{0.29\textwidth}
			\includegraphics[width=1\textwidth]{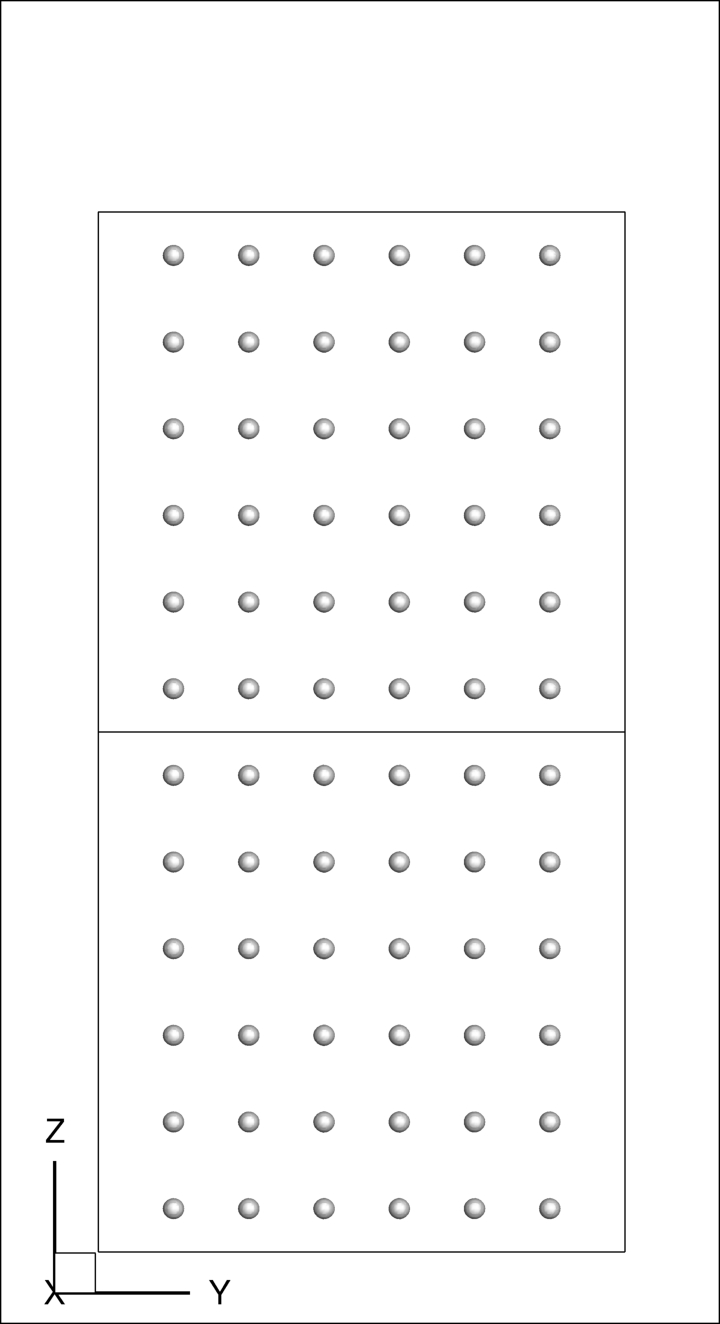}
			\caption{Front view}
			\label{fig:bubble_initial_iso_surface_front:ch:bubbly_flow:sec:results}
		\end{subfigure} %
		\hspace{1mm}
		\begin{subfigure}[b]{0.29\textwidth}
			\includegraphics[width=1\textwidth]{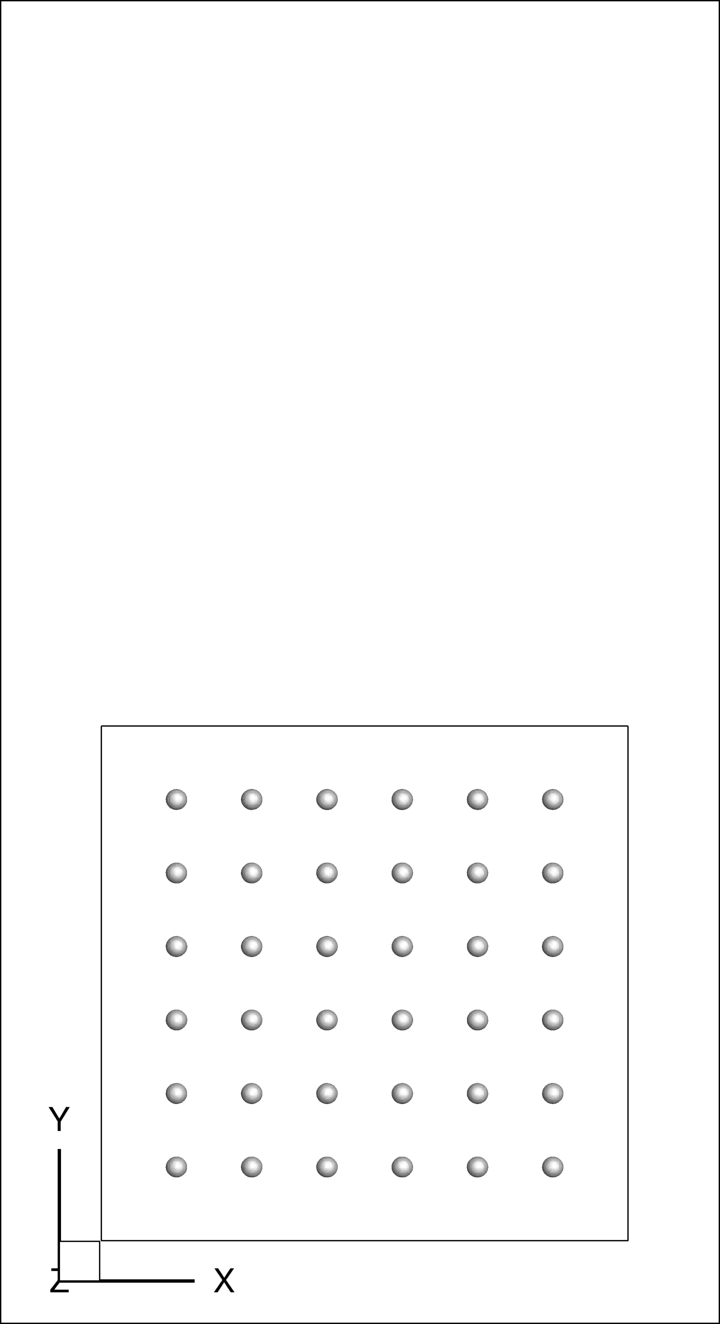}
			\caption{Top view}
			\label{fig:bubble_initial_iso_surface_top:ch:bubbly_flow:sec:results}
		\end{subfigure} %
		\caption{Initial configuration of bubbles in the periodic domain}
		\label{fig:bubble_initial_iso_surface:ch:bubbly_flow:sec:results}
	\end{center}
\end{figure}

\begin{figure}[H]
	\begin{center}
		\begin{subfigure}[b]{0.29\textwidth}
			\includegraphics[width=1\textwidth]{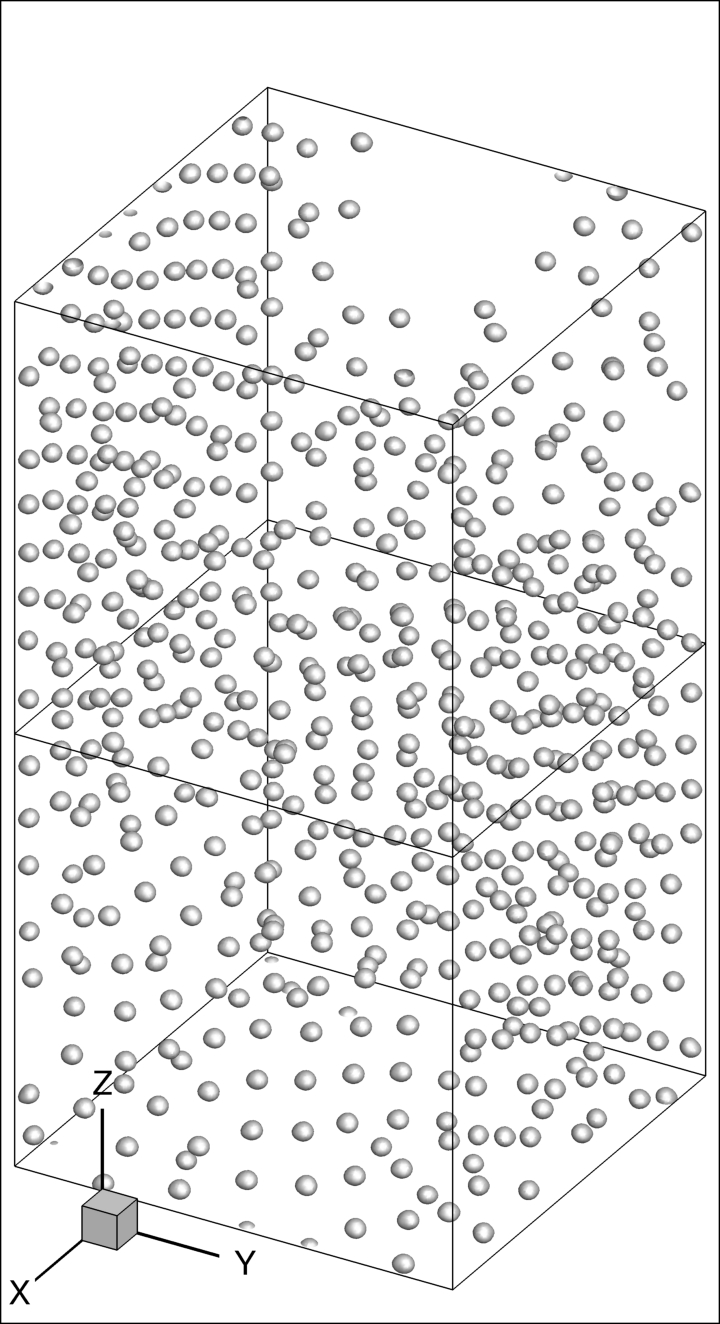}
			\caption{Isometric view}
			\label{fig:bubble_iso_surface_isometric:ch:bubbly_flow:sec:results}
		\end{subfigure} %
		\hspace{1mm}
		\begin{subfigure}[b]{0.29\textwidth}
			\includegraphics[width=1\textwidth]{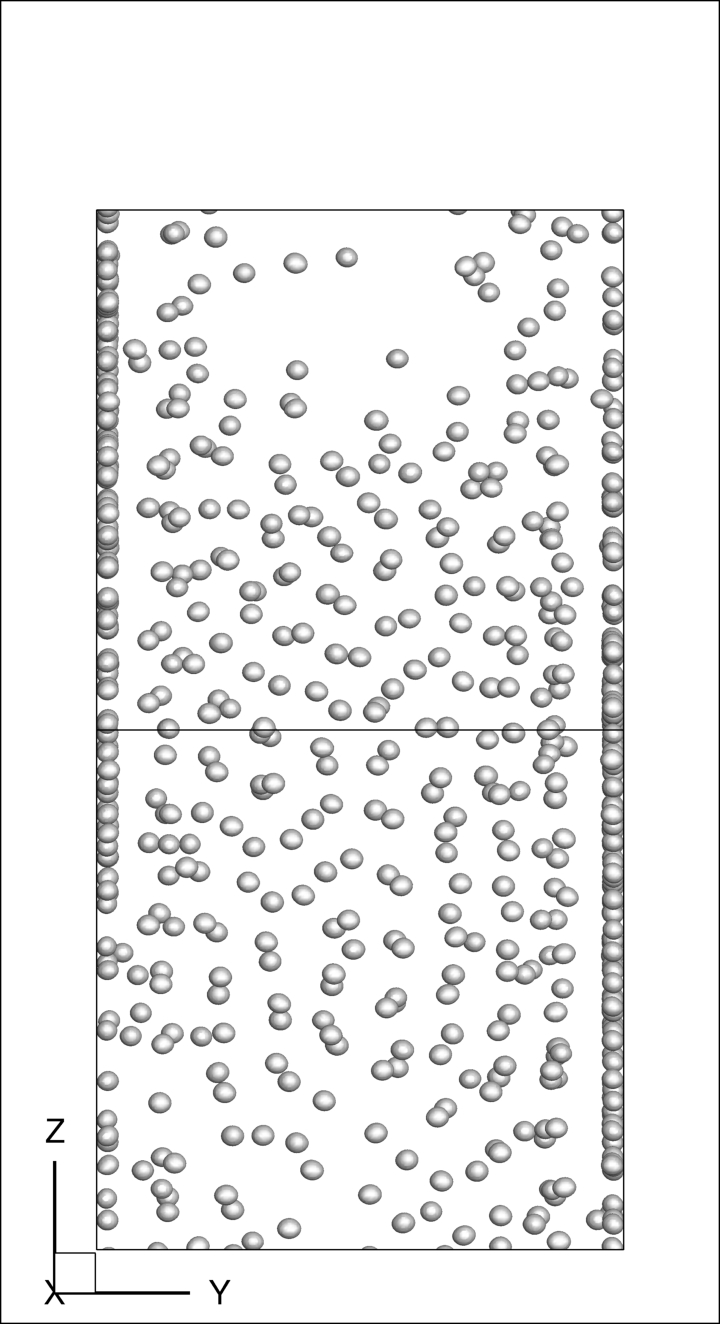}
			\caption{Front view}
			\label{fig:bubble_iso_surface_front:ch:bubbly_flow:sec:results}
		\end{subfigure} %
		\hspace{1mm}
		\begin{subfigure}[b]{0.29\textwidth}
			\includegraphics[width=1\textwidth]{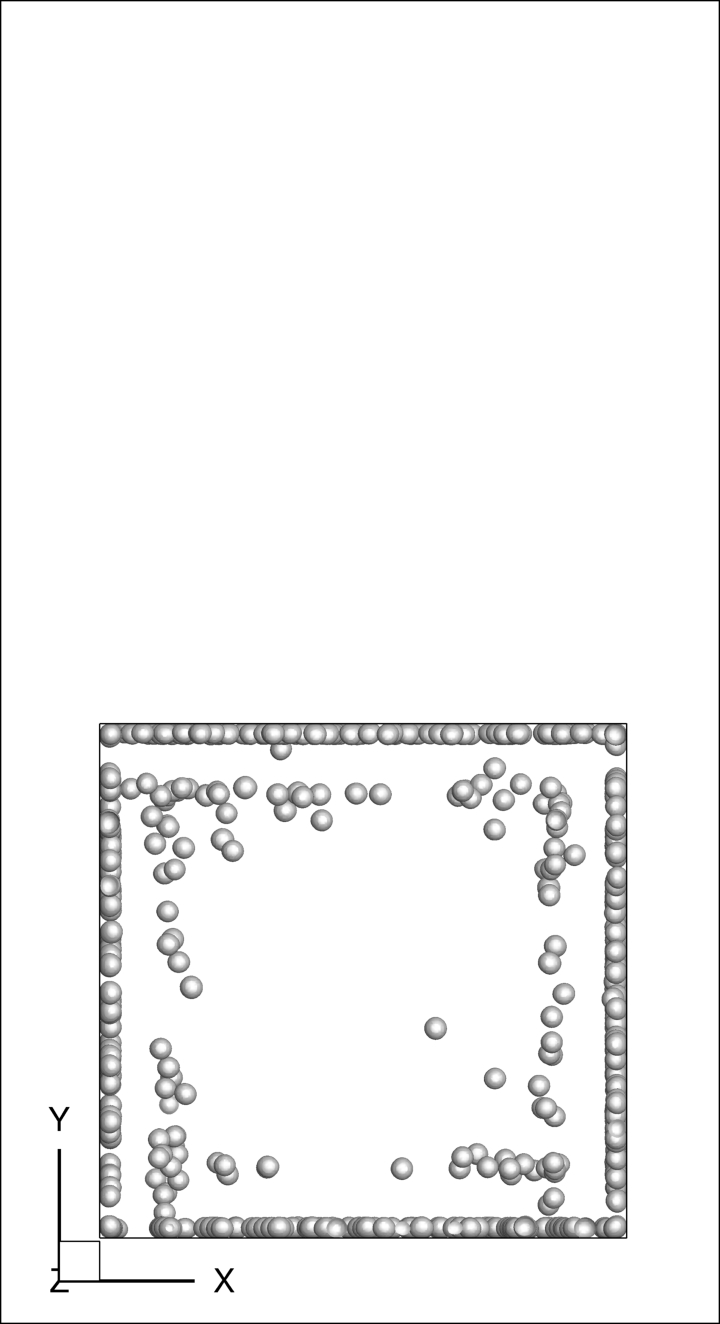}
			\caption{Top view}
			\label{fig:bubble_iso_surface_top:ch:bubbly_flow:sec:results}
		\end{subfigure} %
		\caption{Instantaneous configuration of bubbles as seen in (a) isometric, (b) front and (c) top views at $t = 274$ s or after 40 bulk time units after introduction of bubbles in the unladen flow}
		\label{fig:bubble_iso_surface:ch:bubbly_flow:sec:results}
	\end{center}
\end{figure}

\Cref{fig:bubble_initial_iso_surface:ch:bubbly_flow:sec:results} shows the initial placement of bubbles in the unladen flow. The isometric, front and top views are shown in \Cref{fig:bubble_initial_iso_surface_isometric:ch:bubbly_flow:sec:results,fig:bubble_initial_iso_surface_front:ch:bubbly_flow:sec:results,fig:bubble_initial_iso_surface_top:ch:bubbly_flow:sec:results}. Furthermore, in order to maintain a better aspect ratio of the figures, only half of the computational domain is shown in the \cref{fig:bubble_initial_iso_surface_isometric:ch:bubbly_flow:sec:results,fig:bubble_iso_surface_isometric:ch:bubbly_flow:sec:results}.
\Cref{fig:bubble_iso_surface:ch:bubbly_flow:sec:results} shows the transient distribution of bubbles at $t_{laden} = 40T_b$ where $t_{laden}$ is the total time for which the bubble laden flow calculation has been conducted. The isometric view given in \cref{fig:bubble_iso_surface_isometric:ch:bubbly_flow:sec:results} shows that the bubble are randomly distributed in the domain. However, a clearer  idea of bubble distribution is gathered from the front and top views shown in \cref{fig:bubble_iso_surface_front:ch:bubbly_flow:sec:results,fig:bubble_iso_surface_top:ch:bubbly_flow:sec:results} respectively.

Comparing \cref{fig:bubble_initial_iso_surface:ch:bubbly_flow:sec:results,fig:bubble_iso_surface:ch:bubbly_flow:sec:results}, we can state that the bubbles are significantly affected by the flow as they have moved away from their initial positions. From the top view it can be noticed that a majority of bubbles have migrated towards the wall and formed a layer. These layers of bubbles are very close (less than a control volume) to the wall\footnote{We have used a complete wetting boundary condition for the sidewalls which ensure a thin layer of liquid at walls. Due to this the bubbles are able to get very close the wall without making a contact with them.} as seen in both the front and top views. Away from the first layer near the wall, a cluster of bubbles are observed, and they are located in the regions between the wall and the half distance from the duct center. Furthermore, the cluster is sparse and skewed i.e. a larger number of bubbles are present near the corner region than that near the wall bisector region. The presence of a bubble layer near the wall and a cluster of bubbles away from the wall are clear indications that the current scenario is wall wall-peaking with a secondary intermediate-peak. 
The ensemble-averaged distributions of the void fraction $\left\langle \phi \right\rangle$ along wall bisector, wall and corner bisector are shown in \cref{fig:mean_void_fraction_wall_bisector:ch:bubbly_flow:sec:results,fig:mean_void_fraction_wall_vicinity:ch:bubbly_flow:sec:results,fig:mean_void_fraction_corner_bisector:ch:bubbly_flow:sec:results}. They are symmetric about duct center and wall bisector therefore only half of the domain are shown in the figures. It can be noticed from \cref{fig:mean_void_fraction_wall_bisector:ch:bubbly_flow:sec:results,fig:mean_void_fraction_corner_bisector:ch:bubbly_flow:sec:results} that there are two peaks along the wall and corner bisectors. The observation is in the direction as that observed in \cref{fig:bubble_iso_surface:ch:bubbly_flow:sec:results}. The thickness of the peak at these locations are approximately 2 mm, or one bubble diameter. Furthermore, the peak value along the corner bisector is higher that along the wall; therefore, the bubble density near the corner is larger than that near the wall bisector. 

In the wall direction there are multiple peaks and valleys which indicate that there are a large number of bubbles close to the wall, and there is a significant gap between these bubbles. The height and width of the first peak is similar to that along the corner bisector. Rest of peaks are smaller and wider than the first peak, hence, the bubble density along the wall is highest in the corner region, and it is nearly constant away from the corner region.
\begin{figure}[H]
	\begin{center}
		\begin{subfigure}[b]{0.49\textwidth}
			\includegraphics[width=1\textwidth]{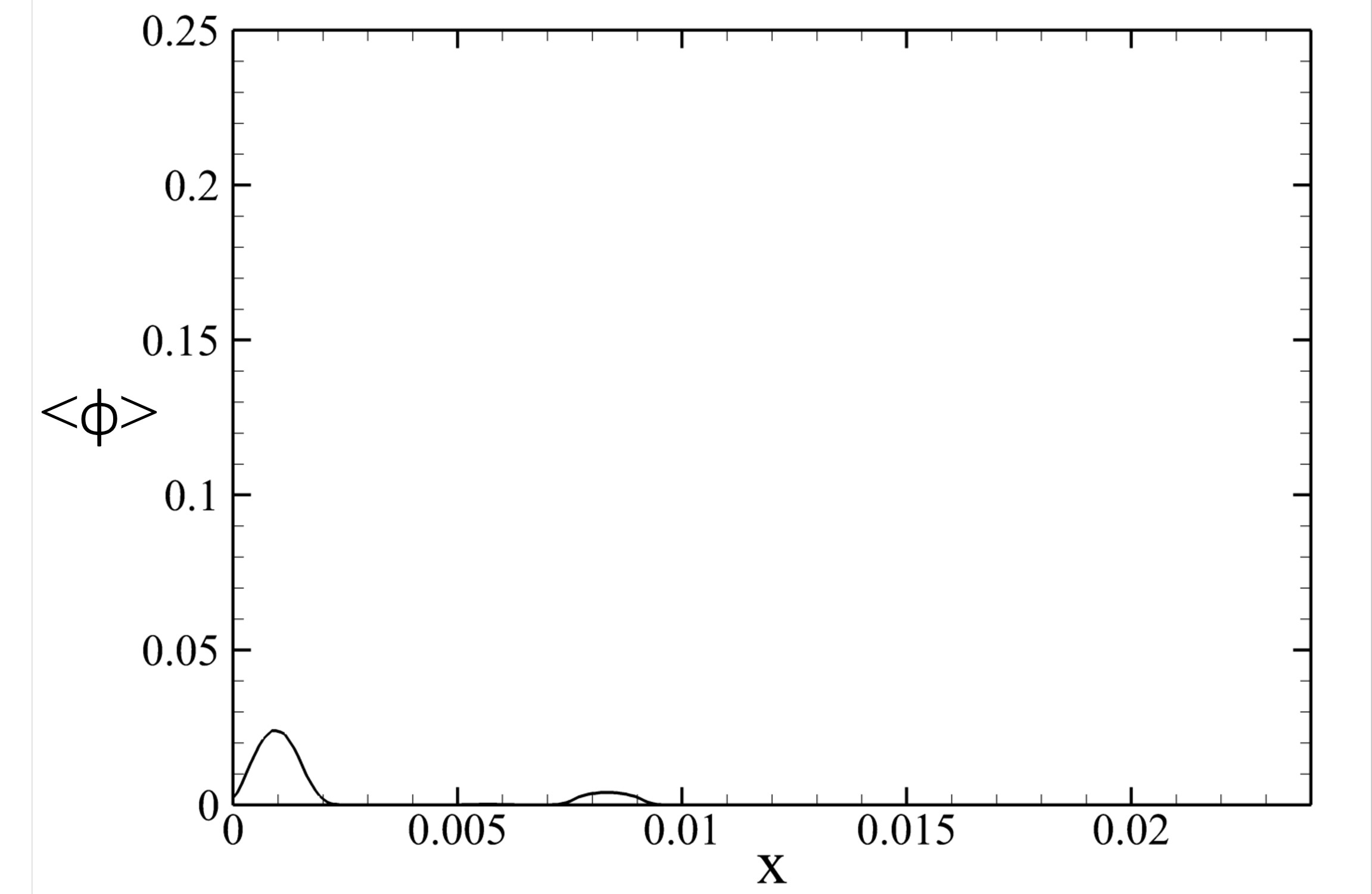}
			\caption{Along wall bisector}
			\label{fig:mean_void_fraction_wall_bisector:ch:bubbly_flow:sec:results}
		\end{subfigure} %
		\hspace{1mm}
		\begin{subfigure}[b]{0.49\textwidth}
			\includegraphics[width=1\textwidth]{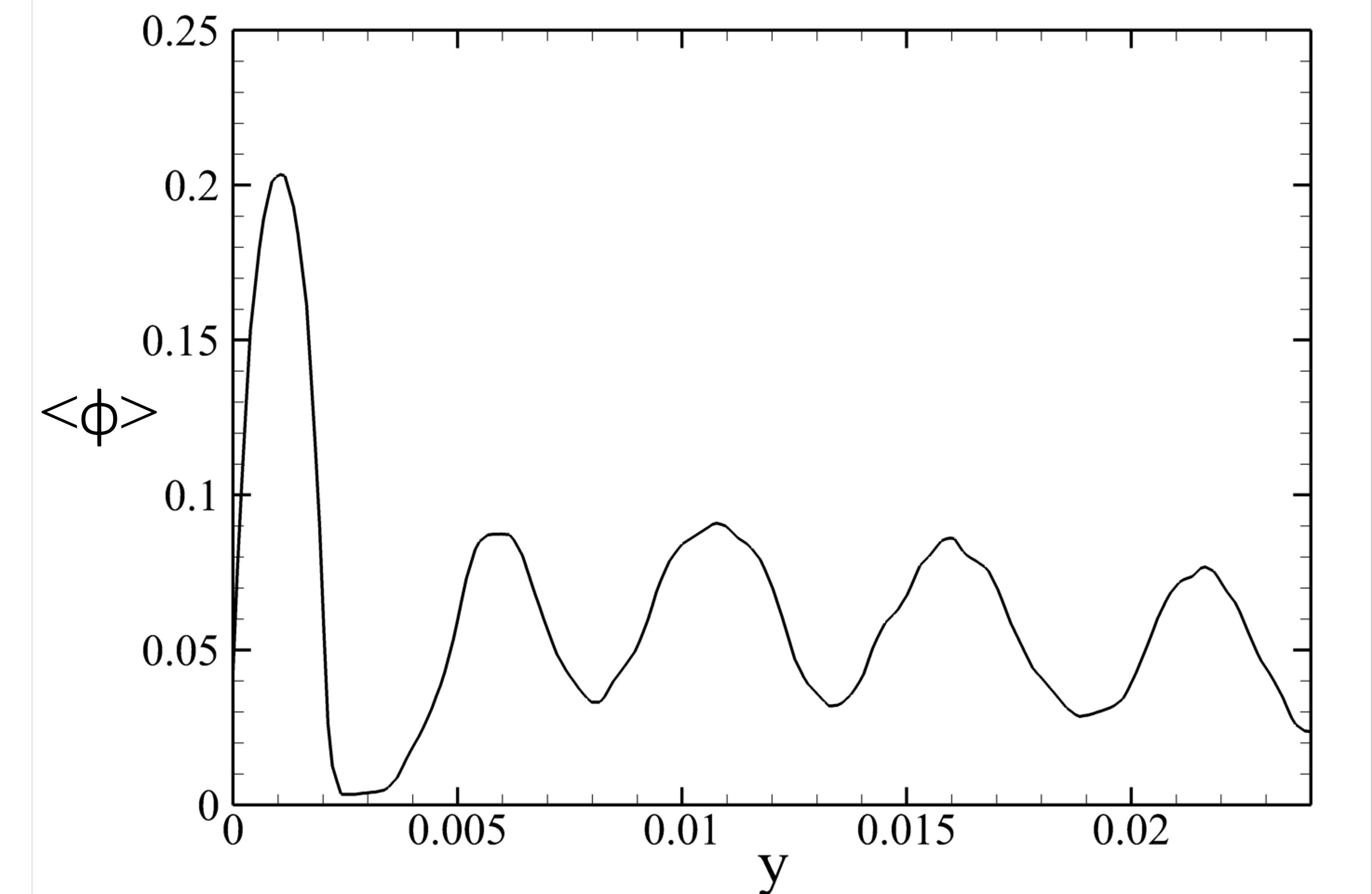}
			\caption{Along wall}
			\label{fig:mean_void_fraction_wall_vicinity:ch:bubbly_flow:sec:results}
		\end{subfigure} %
	\end{center}
\end{figure}	
\begin{figure}[H]
	\begin{center}
		\ContinuedFloat
		\hspace{1mm}
		\begin{subfigure}[b]{0.49\textwidth}
			\includegraphics[width=1\textwidth]{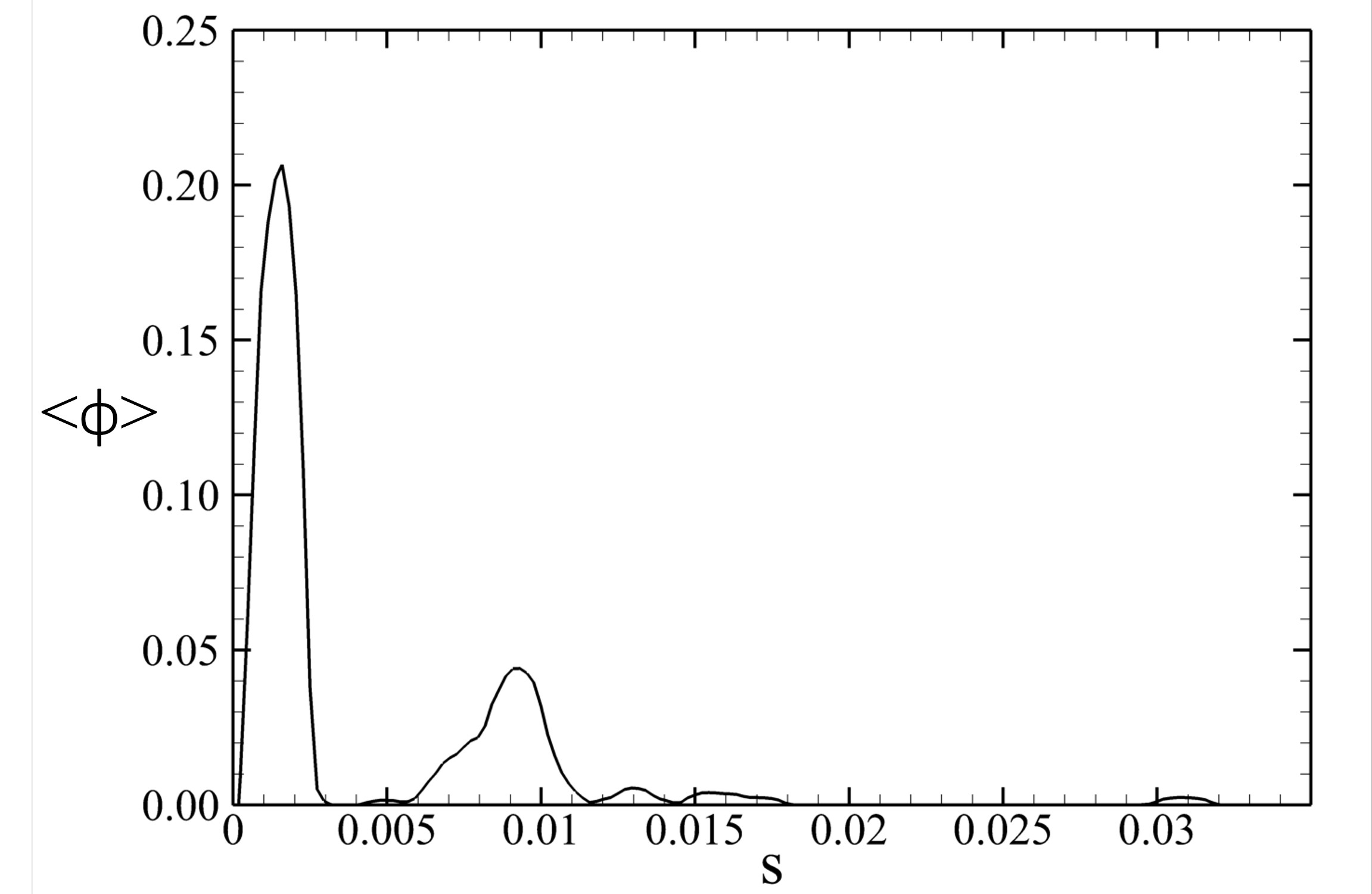}
			\caption{Along corner bisector}
			\label{fig:mean_void_fraction_corner_bisector:ch:bubbly_flow:sec:results}
		\end{subfigure} %
		\caption{Ensemble-averaged void fraction profiles along (a) wall bisector, (b) wall and (c) corner bisector at $t = 274$ s}
		\label{fig:mean_void_fraction:ch:bubbly_flow:sec:results}
	\end{center}
\end{figure}
\subsubsection{Instantaneous Flow Structures}
\label{ch:bubbly_flow:sec:results:sub:laden:subsub:flow_struct}
\begin{figure}[H]
	\begin{center}
		\begin{subfigure}[b]{0.3\textwidth}
			\includegraphics[width=1.0\textwidth]{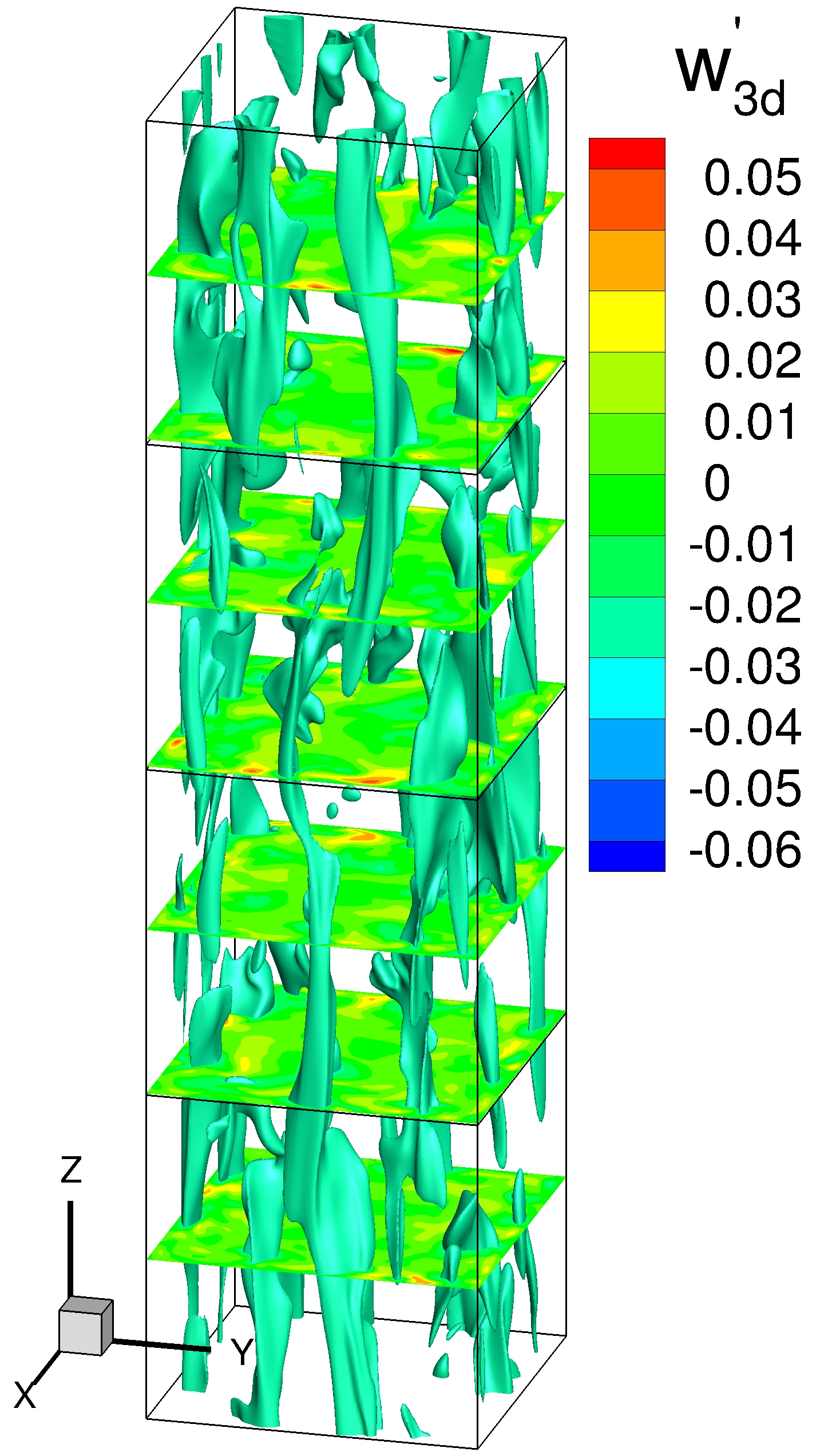}
			\caption{}
			\label{fig:wprime_3d_unladen:ch:bubbly_flow}
		\end{subfigure} %
		\hspace{1mm}
		\begin{subfigure}[b]{0.3\textwidth}
			\includegraphics[width=1.0\textwidth]{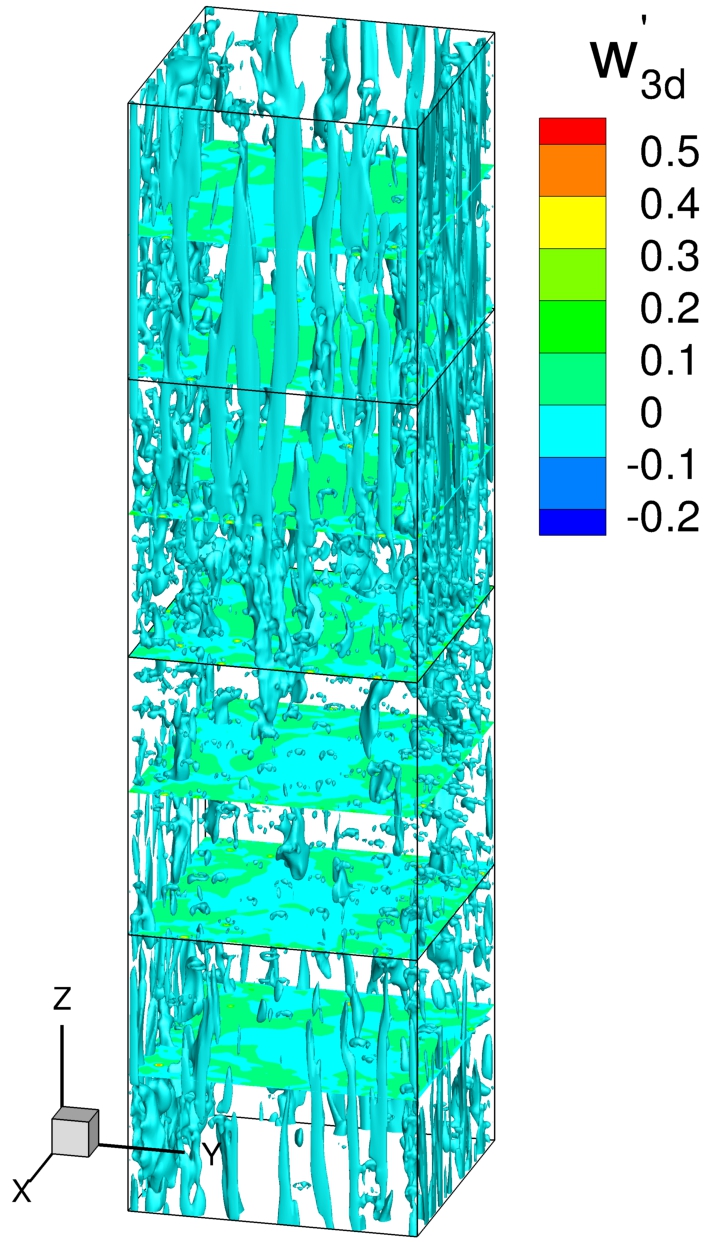}
			\caption{}
			\label{fig:wprime_3d_laden:ch:bubbly_flow}
		\end{subfigure} %
		\caption[Instantaneous three-dimensional velocity fluctuations for (a) unladen flow and (b) bubble laden flow. Displayed are the iso-surface of $w^{\prime}_{3d}/w_c = 0.15$, the vertical distance between planes used for contour plots is $W/2$.]{Instantaneous three-dimensional velocity fluctuations for (a) unladen flow and (b) bubble laden flow. Displayed are the iso-surface\footnotemark of $w^{\prime}_{3d}/w_c = 0.15$, the vertical distance between planes used for contour plots is $W/2$.}
		\label{fig:wprime_3d:ch:bubbly_flow}
	\end{center}
\end{figure}

The three-dimensional fluctuation of the instantaneous streamwise velocity $(w^{\prime}_{3d})$ can give a qualitative picture of the flow field. The expression for $w^{\prime}_{3d}$ is given by
\begin{equation}
	w^{\prime}_{3d}(x,y,z,t) = w(x,y,z,t) - \left\langle w \right\rangle_z(x,y,t)
\end{equation}
The instantaneous streamwise velocity, $w(x,y,z,t)$, is averaged along the periodicity direction to get $\left\langle w \right\rangle_z(x,y,t)$. The average is performed only in the regions where liquid is present. The iso-surfaces and contours of instantaneous streamwise velocity fluctuations $w^{\prime}_{3d}$ are shown in \cref{fig:wprime_3d:ch:bubbly_flow}. 

\Cref{fig:wprime_3d_unladen:ch:bubbly_flow} shows the instantaneous streamwise flow structures in the unladen flow. It can be noticed that the instantaneous flow is dominated by elongated structures in the streamwise direction, and they are mostly located in the near wall regions and extend nearly half of duct in the streamwise direction. The magnitude of instantaneous fluctuations $w^{\prime}_{3d}$ is approximately half of bulk velocity. 

In the case of laden flow, the instantaneous flow consists of both positive and negative fluctuations, but the long elongated structures observed in unladen flow split into a set of smaller and slimmer structures as shown in \cref{fig:wprime_3d_laden:ch:bubbly_flow}. The wall region is dominated by relatively longer instantaneous structures, whereas the core region is dominated by smaller structures. The smallest size of structures in the core region is comparable to the bubble size. These structures, in both core and wall regions, are more oscillatory than the ones observed in the unladen flow. The magnitude of fluctuations are approximately ten times larger than of unladen flow.
\footnotetext{It should be noted that the centerline velocity of unladen flow is approximately 2.5 times smaller than that of bubble laden flow. Hence, the value of iso-surfaces in \cref{fig:wprime_3d_unladen:ch:bubbly_flow,fig:wprime_3d_laden:ch:bubbly_flow} are not same.}
\subsubsection{Instantaneous Streamwise Velocity}
\label{ch:unladen_flow:sec:results:sub:transient_vel}
\begin{figure}[H]
	\begin{center}
		\begin{subfigure}[b]{0.48\textwidth}
			\includegraphics[width=1\textwidth]{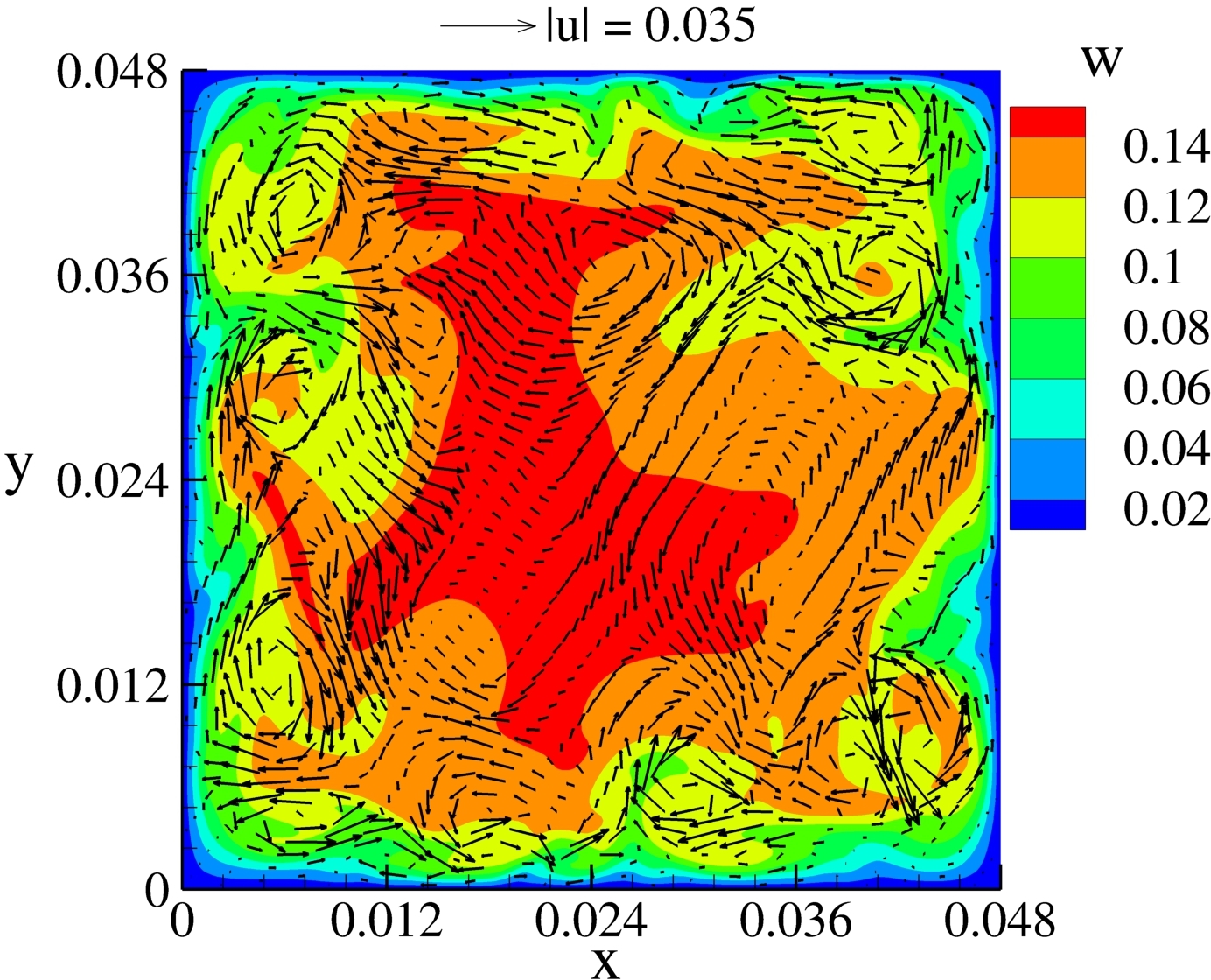}
			\caption{}
			\label{fig:transient_w_vel_unladen:ch:bubbly_flow:sec:results}
		\end{subfigure} %
		\hspace{3mm}
		\begin{subfigure}[b]{0.485549133\textwidth}
			\includegraphics[width=1\textwidth]{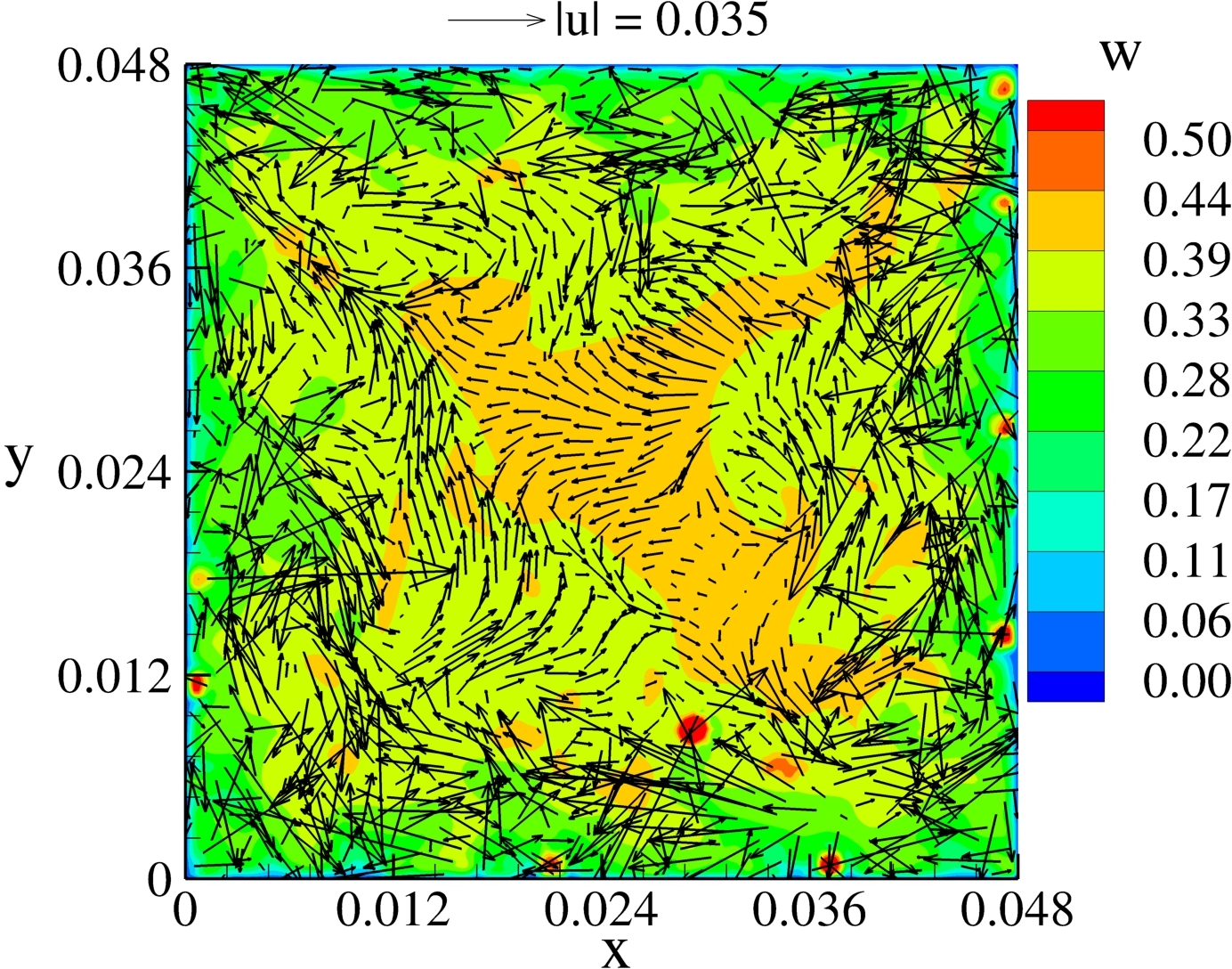}
			\caption{}
			\label{fig:transient_w_vel_laden:ch:bubbly_flow:sec:results}
		\end{subfigure} %
		\caption{Contours of instantaneous streamwise velocity overlaid with cross-stream velocity vectors for (a) unladen flow and (b) bubble laden flow}
		\label{fig:transient_w_vel:ch:bubbly_flow:sec:results}
	\end{center}
\end{figure}

Instantaneous flow field of unladen flow in the duct cross-section is shown in \cref{fig:transient_w_vel_unladen:ch:bubbly_flow:sec:results}. The velocities are averaged along the streamwise direction to increase the sample size. The figure shows the contour of instantaneous streamwise velocity overlaid with instantaneous in-plane velocity vectors. Significant fluctuations in the velocity can be observed from the contours of the velocity, which is indicative of the turbulent flow. The presence of the in-plane velocity vectors are also a sign of turbulent flow. It can be noticed that there is a large variation in the magnitude of the in-plane velocities, indicated by the length of vectors, at similar locations in different quadrants. The phenomenon is due to the dynamic nature of the turbulent flow. \par 
\Cref{fig:transient_w_vel_laden:ch:bubbly_flow:sec:results} shows the instantaneous streamwise velocity of the bubble laden flow. It can be noticed that the streamwise velocity fluctuations are much larger compared to the unladen flow, and they are as high as bulk velocity. Furthermore, the maximum fluctuations are observed in the corner and wall regions where bubbles are located. The streamwise velocity fluctuations in the vicinity of the bubbles are comparable to the terminal velocity of the bubble. The bulk velocity of the bubble laden flow is close to 0.322 m/s which is nearly three times higher than that of unladen flow. The increase of bulk velocity is a combined result of the upward motion of bubbles (and subsequent dragging of the surrounding liquid) and significant reduction in viscosity near the walls due to the preferential accumulation of bubbles. The cross-stream velocities are also higher than their unladen flow counterparts, and they are located near wall and corner regions. 
\subsubsection{Ensemble-averaged Distribution of Streamwise and Spanwise Velocity}
\label{ch:unladen_flow:sec:results:sub:mean_vel}
\begin{figure}[H]
	\begin{center}
		\begin{subfigure}[b]{0.48\textwidth}
			\includegraphics[width=1\textwidth]{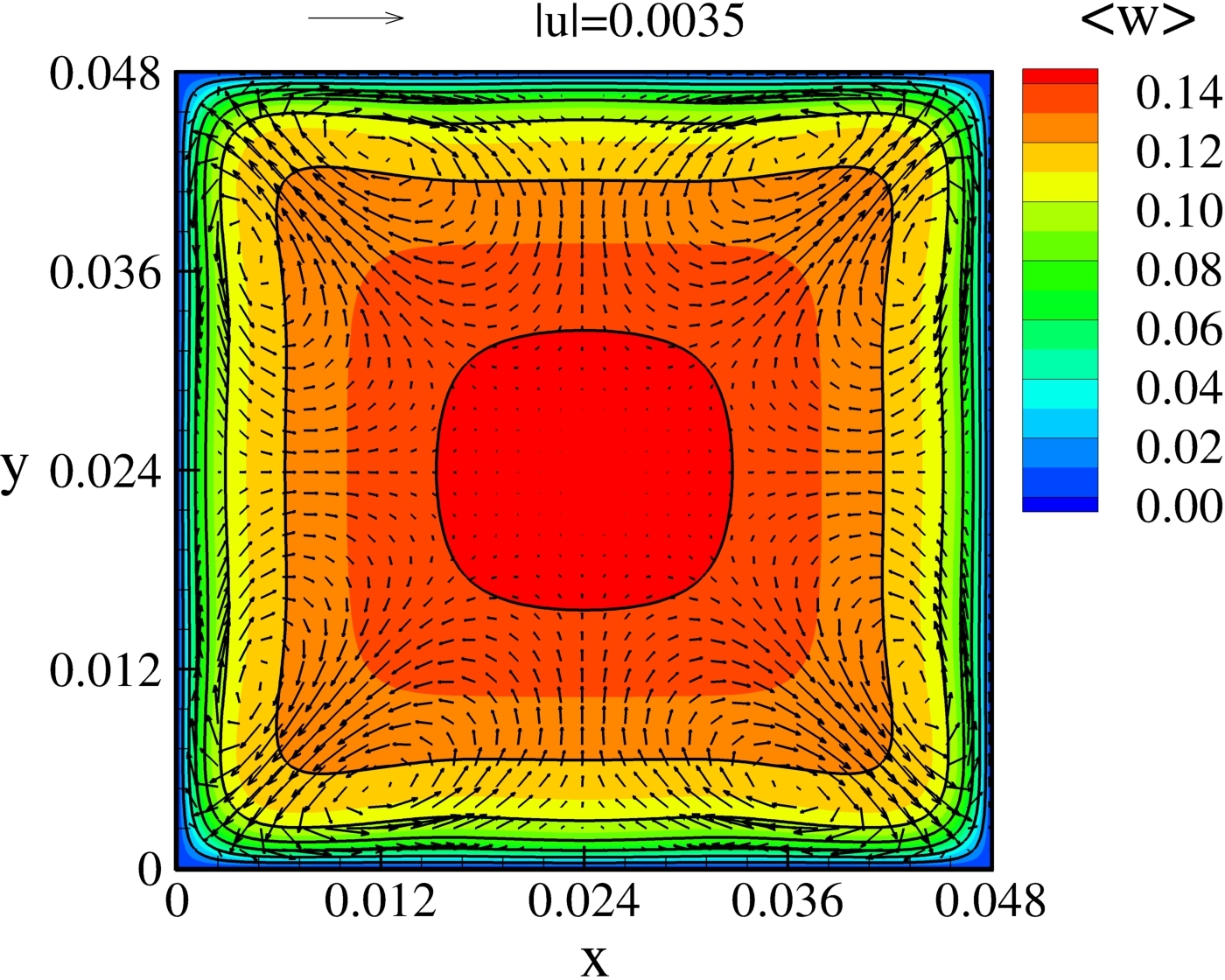}
			\caption{}
			\label{fig:mean_w_vel_unladen_vectors:ch:bubbly_flow:sec:results}
		\end{subfigure} %
		\hspace{3mm}
		\begin{subfigure}[b]{0.48\textwidth}
			\includegraphics[width=1\textwidth]{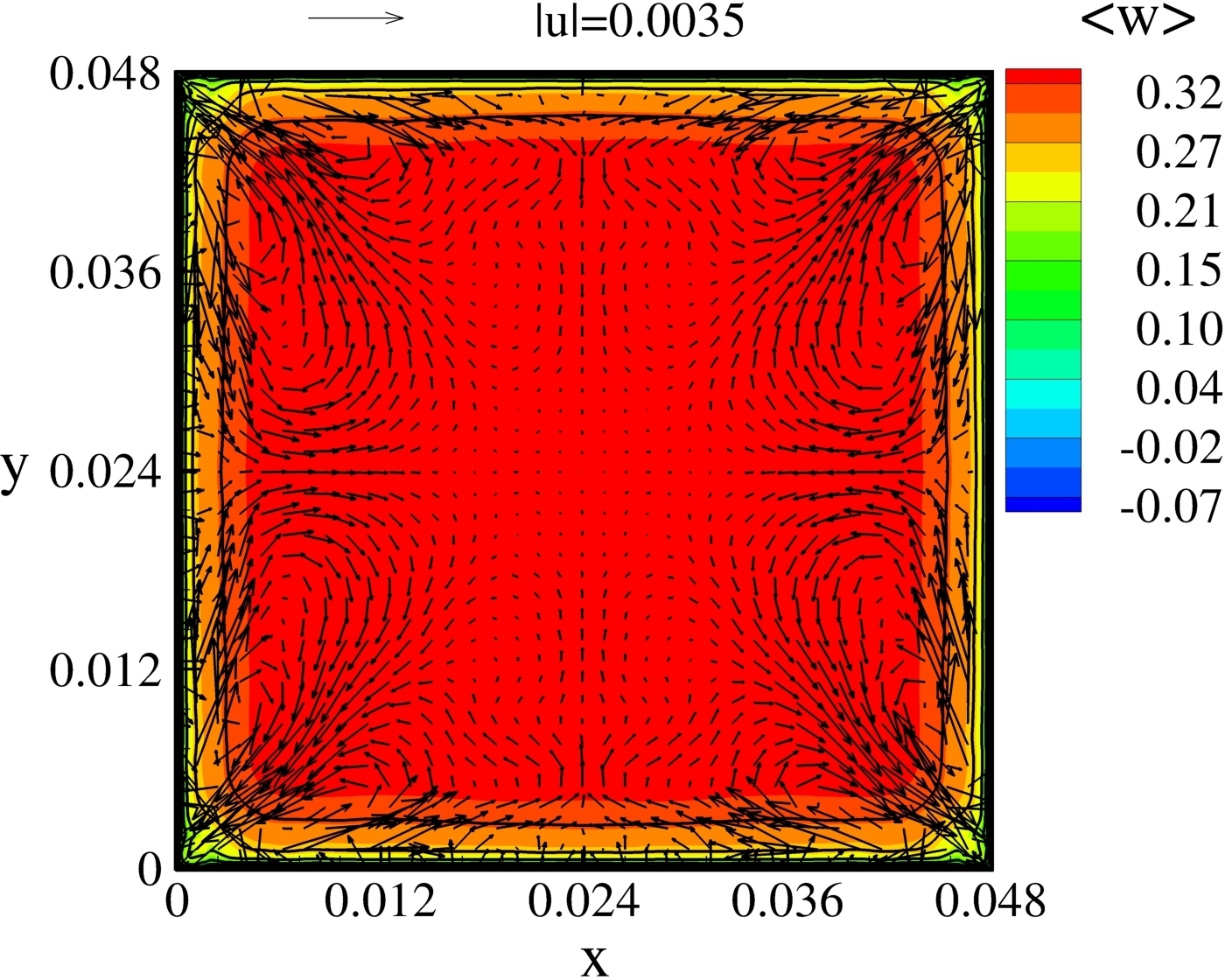}
			\caption{}
			\label{fig:mean_w_vel_laden_vectors:ch:bubbly_flow:sec:results}
		\end{subfigure} %
		\caption{Contours of the mean streamwise velocity overlaid with mean cross-stream velocity vectors for (a) unladen flow and (b) bubble laden flow}
		\label{fig:mean_w_vel_vectors:ch:bubbly_flow:sec:results}
	\end{center}
\end{figure}

The contours of ensemble-averaged\footnote{The ensemble average of r.m.s and turbulence quantities ($\left\langle u \right\rangle$,$\left\langle v \right\rangle$,$\left\langle w \right\rangle$, $\left\langle u^{\prime}u^{\prime} \right\rangle$, $\left\langle v^{\prime}v^{\prime} \right\rangle$, $\left\langle w^{\prime}w^{\prime} \right\rangle$, $\left\langle u^{\prime}v^{\prime} \right\rangle$, $\left\langle u^{\prime}w^{\prime} \right\rangle$, $\left\langle \phi \right\rangle$) are performed over time, $z$-direction, and $x$ and $y$ central planes. Then the averaged values are reflected back to other three quadrants and the appropriate sign of the quantities are maintained. Hence, the quantities are expected to be quadrant symmetric.} streamwise velocity of unladen and laden flows in the duct cross-section are given in \cref{fig:mean_w_vel_unladen_vectors:ch:bubbly_flow:sec:results,fig:mean_w_vel_laden_vectors:ch:bubbly_flow:sec:results}. The contours are overlaid with the ensemble-averaged in-plane velocity ($\left\langle u \right\rangle$,$\left\langle v \right\rangle$) vectors. It can be noticed that there are two counter-rotating vortices in each quadrant and a total of eight vortices in the duct cross-section. They are the results of the mean secondary flow generated by anisotropy of turbulent stresses. They are responsible for the transfer of high energy eddies from the duct center to the walls. The mean secondary flow transports the fluid momentum from the bulk region to the corners along their bisectors, and then back to the bulk region along wall bisectors. The magnitude of mean secondary velocities reach a maximum along the diagonal of the duct. \par 
Comparing unladen and bubble laden flows, we notice that the streamwise velocity increases with the introduction of bubbles. The centerline velocity, $w_c$, increases from $0.146$ m/s for unladen flow to $0.325$ m/s for bubble laden flow. The velocity increases everywhere in the cross-section and not just at the centerline. It can also be observed that the streamwise velocity of bubble laden flow is flatter than that of unladen flow and the velocity contour lines in the corner region are rounder in laden flow case than that in unladen flow. \par 
\begin{figure}[H]
	\begin{center}
		\begin{subfigure}[b]{0.48\textwidth}    
			\includegraphics[width=1\textwidth]{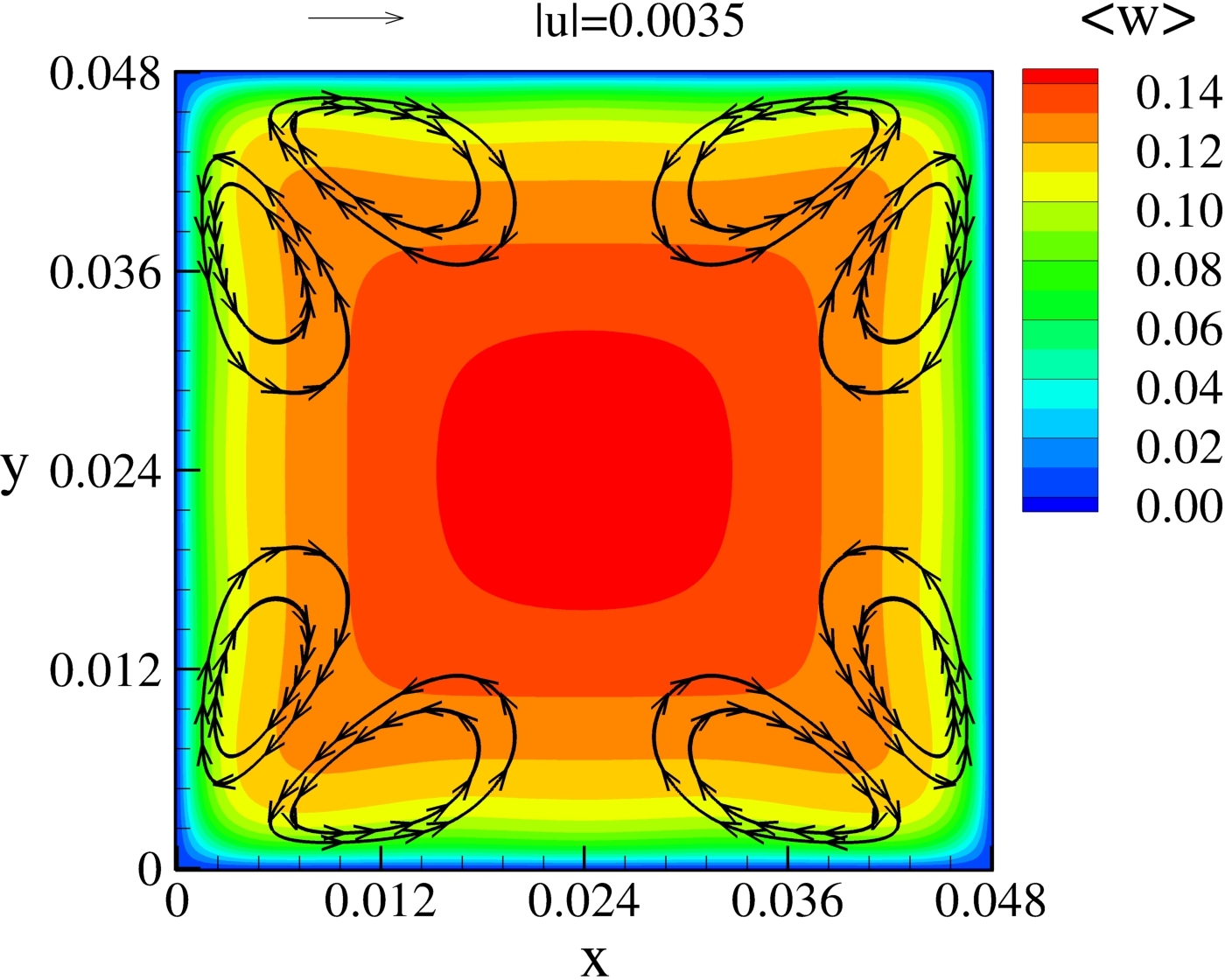}
			\caption{}
			\label{fig:mean_w_vel_streamlines_unladen:ch:bubbly_flow:sec:results}
		\end{subfigure}
		\hspace{3mm}
		\begin{subfigure}[b]{0.48\textwidth}    
			\includegraphics[width=1\textwidth]{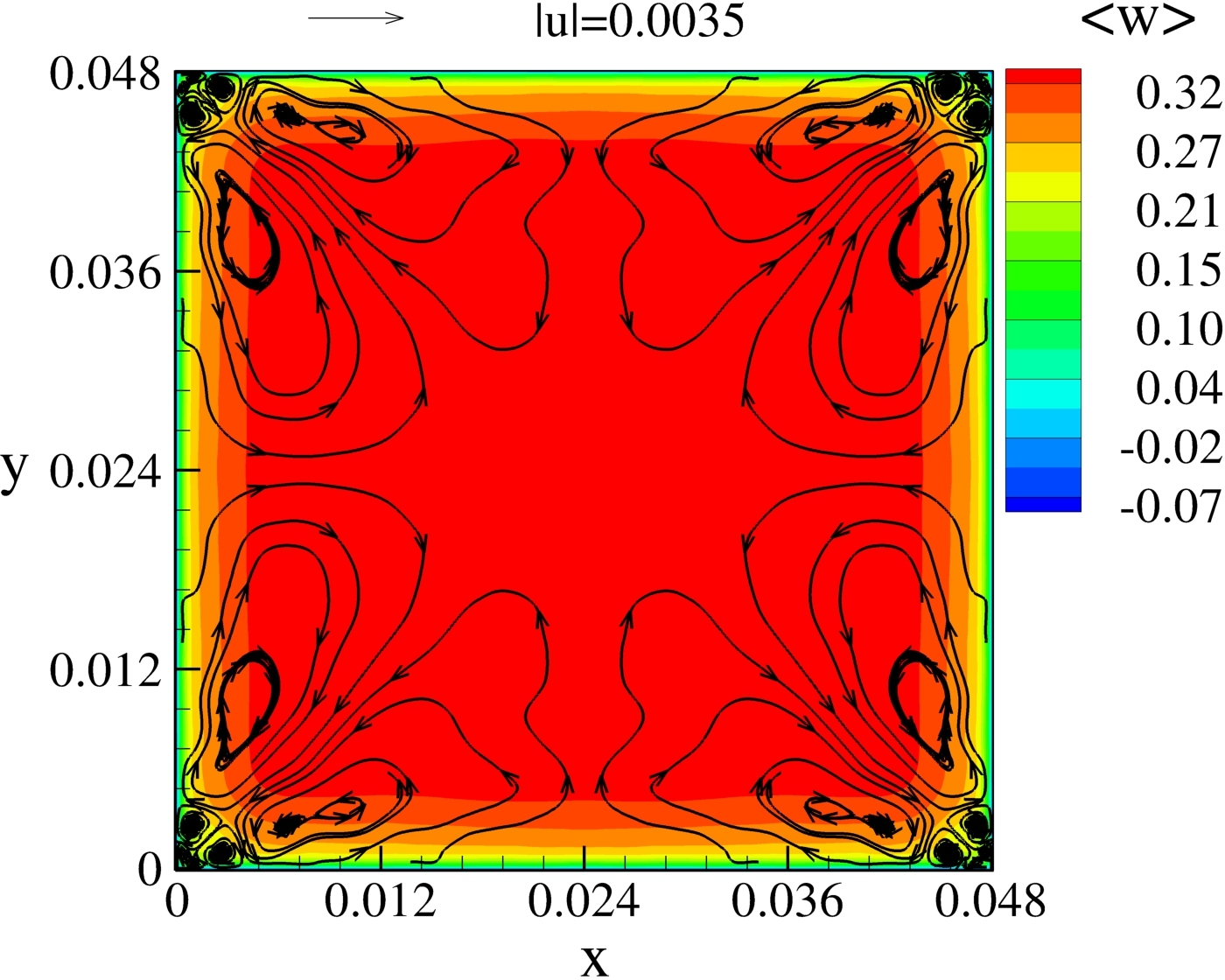}
			\caption{}
			\label{fig:mean_w_vel_streamlines_laden:ch:bubbly_flow:sec:results}
		\end{subfigure}
		\caption{Contours of mean streamwise velocity overlaid with mean cross-stream stream functions for (a) unladen flow and (b) bubble laden flow}
		\label{fig:mean_w_vel_streamlines:ch:bubbly_flow:sec:results}
	\end{center}
\end{figure}

\Cref{fig:mean_w_vel_streamlines_unladen:ch:bubbly_flow:sec:results,fig:mean_w_vel_streamlines_laden:ch:bubbly_flow:sec:results} show the mean in-plane streamlines in the cross-sectional plane. It can be noticed that the introduction of bubbles in unladen flow does not change direction of the secondary flows, i.e. fluid from the bulk region to corners are transferred along the corner bisectors and then back to the bulk region along wall bisectors. However, the introduction of bubbles makes the flow a bit more complex and introduces additional counter-rotating vortices near the corner bisectors. They are absent in the unladen flow, and streamlines reach the wall before they bend. The additional vortices in the laden flow affect the streamline velocity $w$ distribution in the corner region, as it is reflected in the form of a pointed counter line in \cref{fig:mean_w_vel_laden_vectors:ch:bubbly_flow:sec:results}. In the unladen flow, counter-rotating vortices are nearly symmetric about the corner bisectors, whereas in the laden flow, they are clearly asymmetric. It can also be observed that the secondary flow streamlines in the absence of bubble are smooth and free of oscillations, whereas they are tortuous and twisted when bubbles are introduced. This may be because the bubble distribution is not fully symmetric and/or the flow can never reach full symmetry. It is also possible that the flow has not reached a stationary state and more sampling of the results are required. 
\begin{figure}[H]
	\begin{center}
		\begin{subfigure}[b]{0.48\textwidth}
			\includegraphics[width=1\textwidth]{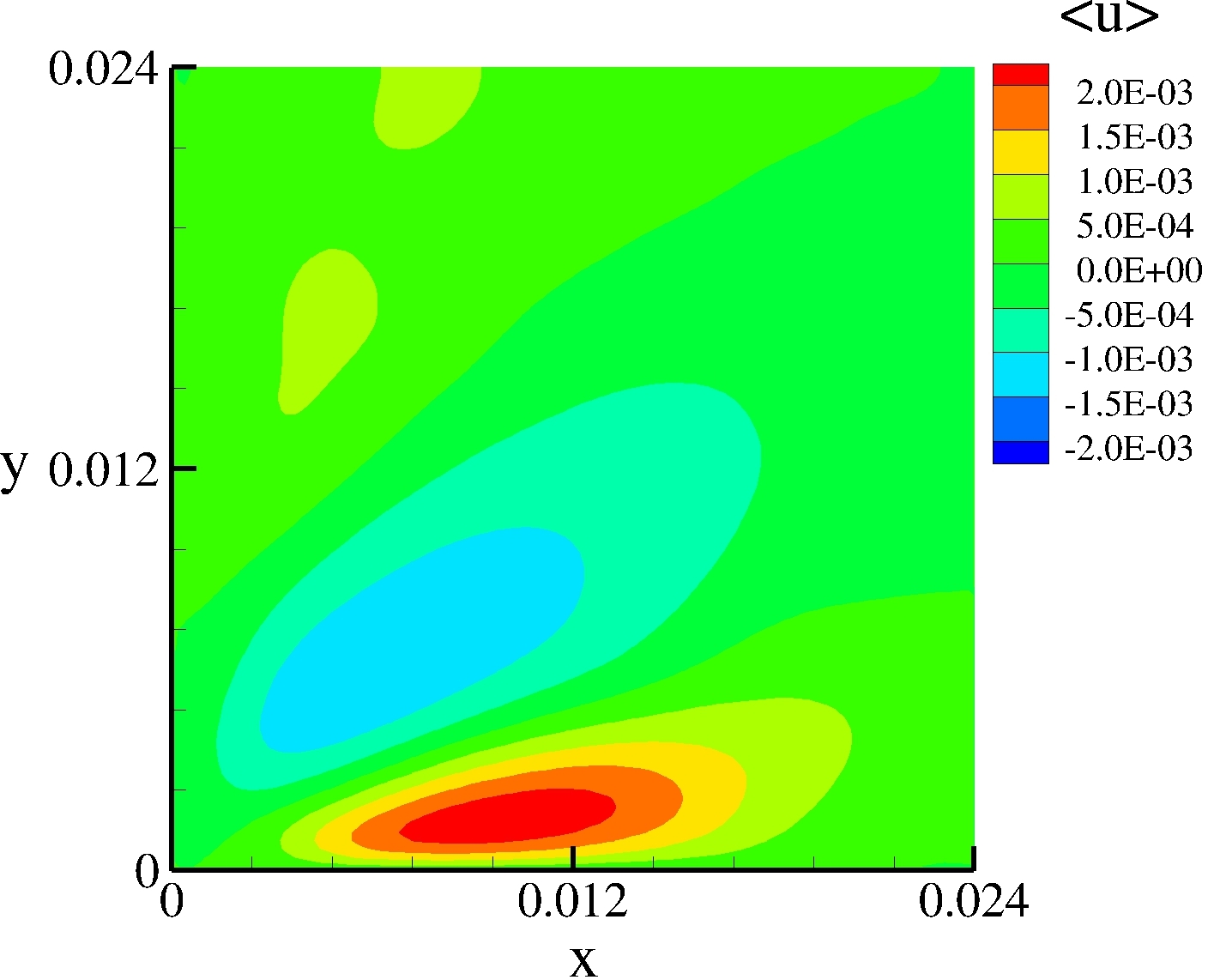}
			\caption{}
			\label{fig:mean_u_vel_unladen:ch:bubbly_flow:sec:results}
		\end{subfigure} %
		\hspace{3mm}
		\begin{subfigure}[b]{0.48\textwidth}
			\includegraphics[width=1\textwidth]{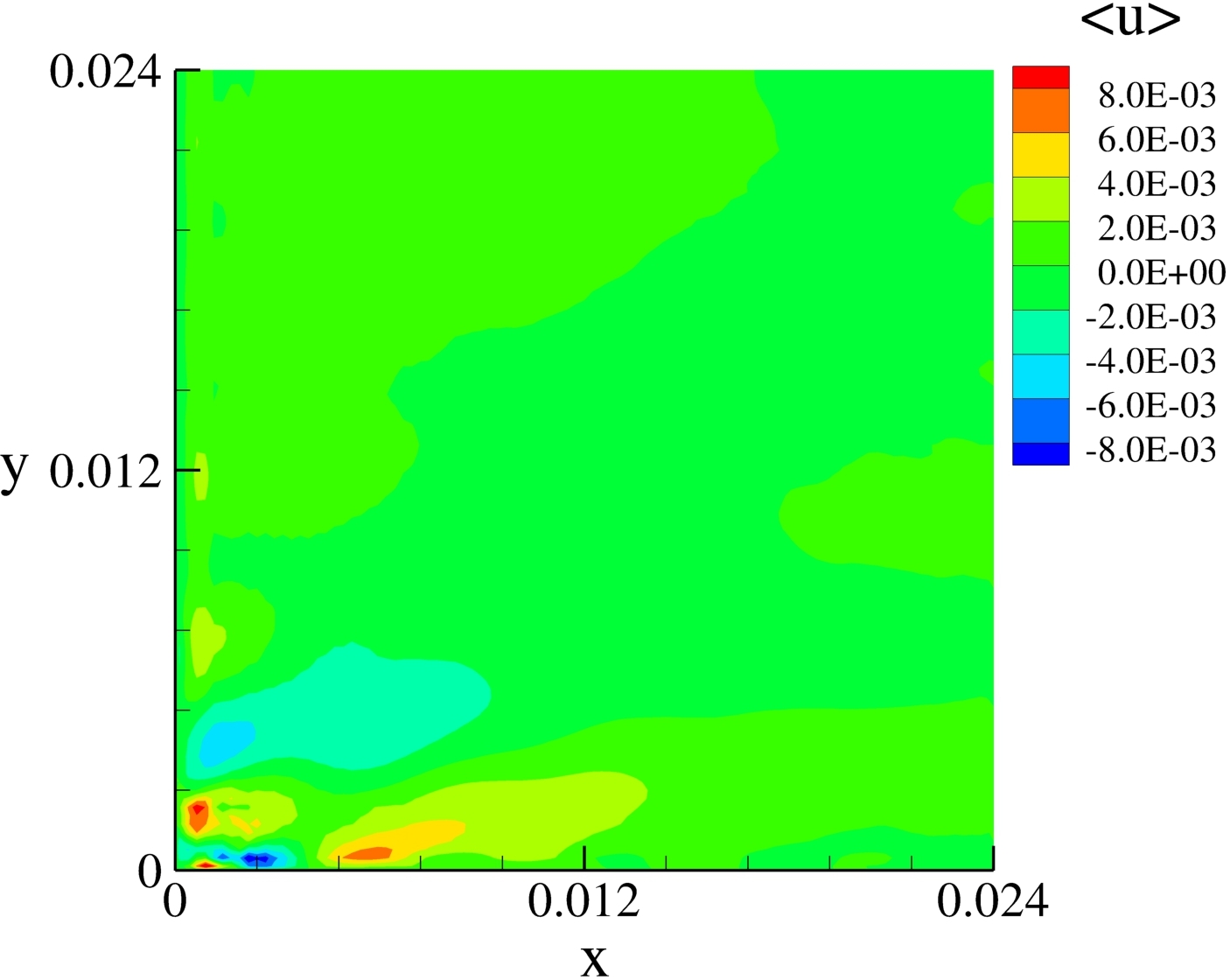}
			\caption{}
			\label{fig:mean_u_vel_laden:ch:bubbly_flow:sec:results}
		\end{subfigure} %
		\caption{Contours of mean $x$-direction velocity in the lower left cross-section of the duct: (a) unladen flow and (b) bubble laden flow}
		\label{fig:mean_u_vel:ch:bubbly_flow:sec:results}
	\end{center}
\end{figure}
Contours of $x$-direction spanwise velocity of unladen and laden flows are shown in \cref{fig:mean_u_vel_unladen:ch:bubbly_flow:sec:results,fig:mean_u_vel_laden:ch:bubbly_flow:sec:results}. The maximum magnitude of $\langle u \rangle$ is approximately 2\% of streamwise bulk velocity $(w_b)$ for both unladen and laden flows. The unladen flow prediction is in accordance with the data available in literature \cite{Madabhushi1991,Broglia2003}. It can be noticed that in the corner region, $\langle u \rangle$ in laden flow case is approximately four times higher than that in unladen flow, and the highest value is observed closer to the corner in laden flow, whereas in unladen flow the highest value is observed at a farther distance from the corner. It can also be seen that $x$-velocity is nearly similar in both laden and unladen flows in the regions away from the corner. This higher value of $\langle u \rangle$ near the corner region is associated with the presence of bubbles.
\begin{figure}[H]
	\begin{center}
		\begin{subfigure}[b]{0.48\textwidth}
			\includegraphics[width=1\textwidth]{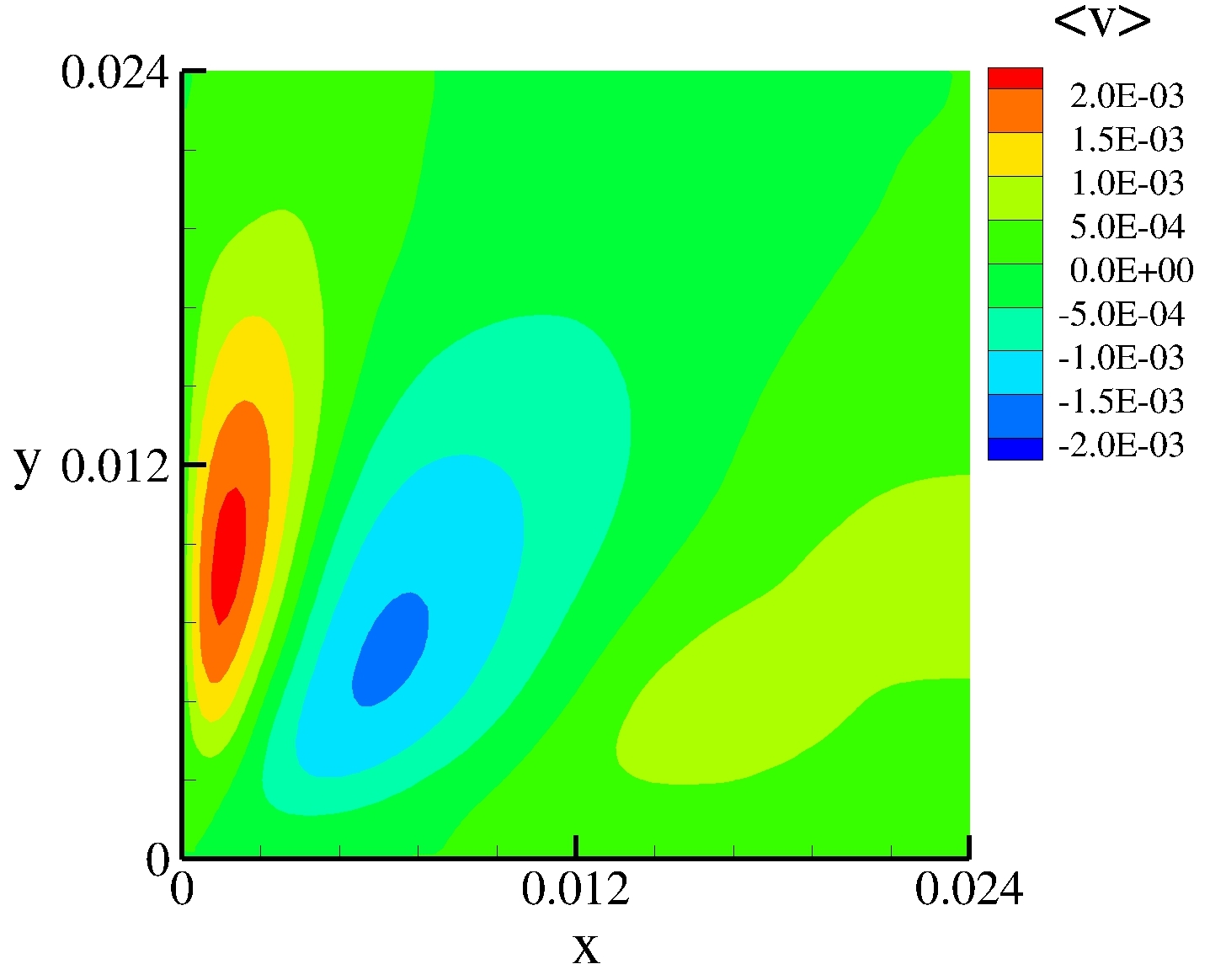}
			\caption{}
			\label{fig:mean_v_vel_unladen:ch:bubbly_flow:sec:results}
		\end{subfigure} %
		\hspace{3mm}
		\begin{subfigure}[b]{0.48\textwidth}
			\includegraphics[width=1\textwidth]{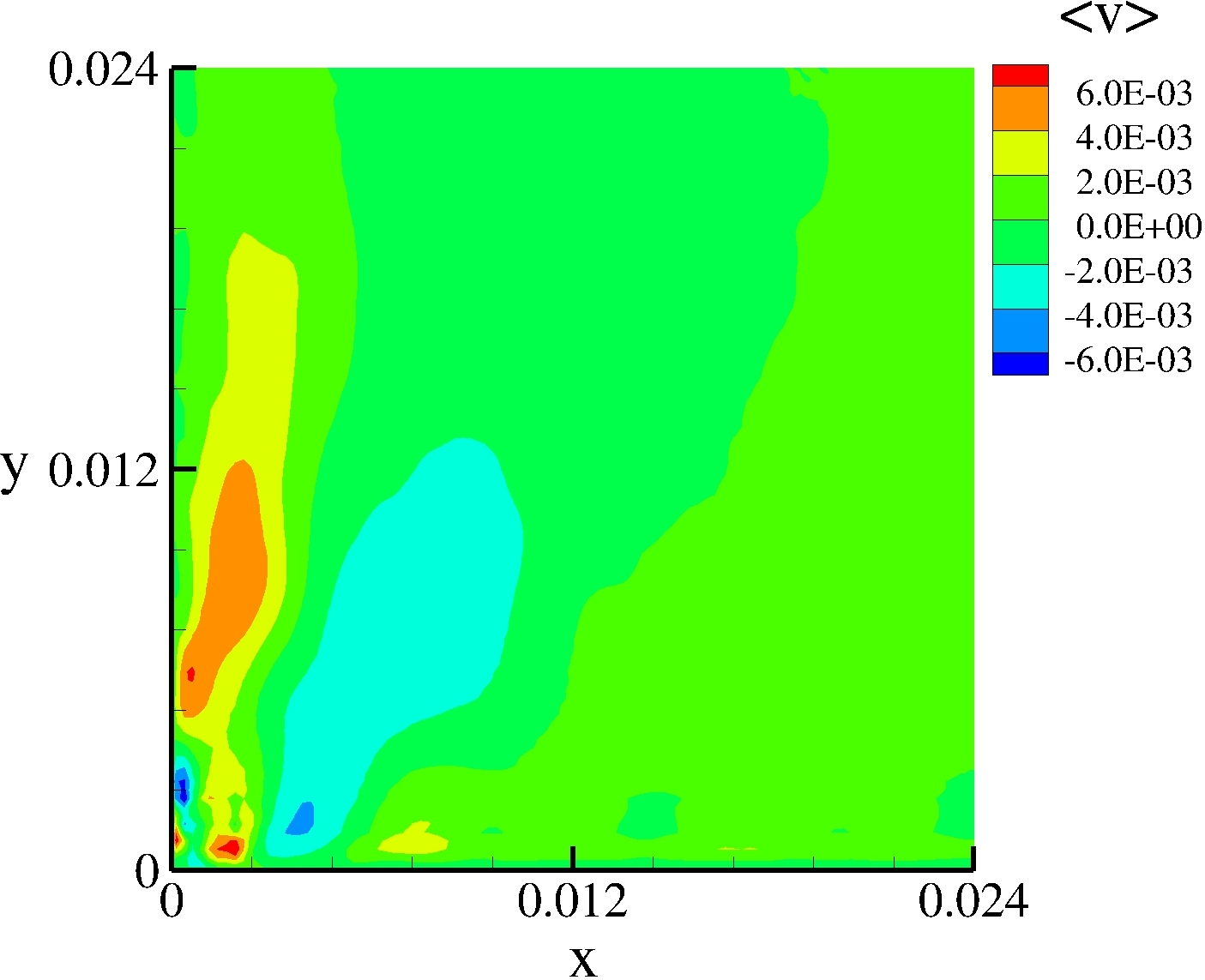}
			\caption{}
			\label{fig:mean_v_vel_laden:ch:bubbly_flow:sec:results}
		\end{subfigure} %
		\caption{Contours of mean $y$-direction velocity in the lower left cross-section of the duct: (a) unladen flow and (b) bubble laden flow}
		\label{fig:mean_v_vel:ch:bubbly_flow:sec:results}
	\end{center}
\end{figure}
Contours of $y$-direction spanwise velocity are shown in \cref{fig:mean_v_vel_unladen:ch:bubbly_flow:sec:results,fig:mean_v_vel_laden:ch:bubbly_flow:sec:results}. In the case of unladen flow, the maximum magnitude of $\langle v \rangle$ is close to 2\% of the streamwise bulk velocity\footnote{The bulk velocity of the unladen flow is 0.110 m/s} $(w_b)$. This observation is similar to that for $\langle u \rangle$. In the case of laden flow, the maximum magnitude of $\langle v \rangle$ is smaller than that of $\langle u \rangle$. This is somewhat surprising as we would expect both $\langle u \rangle$ and $\langle v \rangle$ to be similar as there is no source of directional preference and can be attributed to a lower sampling time of the simulation. The highest value of $\langle v \rangle$ in laden flow is three times higher than that in the unladen flow. Both location of the highest value and distribution of $\langle v \rangle$ in the bulk region is similar in unladen and laden flows. \par 
It can also be noticed from \cref{fig:mean_u_vel_unladen:ch:bubbly_flow:sec:results,fig:mean_v_vel_unladen:ch:bubbly_flow:sec:results} that $\langle u \rangle$ is an approximate reflection of $\langle v \rangle$ about corresponding diagonal. The slight asymmetry between $\langle u \rangle$ and $\langle v \rangle$ is a mere indication of the quadrant, not octagonal, averaging. It should also be noted that the range of $x$ and $y$ velocities are nearly identical and in accordance with physics of the problem. However, in the case of laden flow, no such parallels can be drawn between \cref{fig:mean_u_vel_laden:ch:bubbly_flow:sec:results,fig:mean_v_vel_laden:ch:bubbly_flow:sec:results}. \par 
Although the mean secondary flow velocities are two orders of magnitude smaller than streamwise velocity, it has a significant impact on the distribution of streamwise velocity. As, it can be observed from \cref{fig:mean_w_vel_unladen_vectors:ch:bubbly_flow:sec:results} that the $w$-contour lines near the corner are pointed and bulged away from the duct corner. The mechanism is related to the two counter-rotating vortices present near the corner.

\subsubsection{Reynolds stresses}
\label{ch:unladen_flow:sec:results:sub:Reynolds_stresses}
The Reynolds stresses are presented along all three locations, namely wall bisector, wall and corner bisector as mentioned in \cref{fig:presentation_direction:ch:bubbly_flow:sec:results}. The data points are extracted using a visualization software; thus we expect a small interpolation errors in the results. Although the laden flow has not reached a stationary-state, we can analyze the trends of the Reynolds stresses and compare them with unladen flow to understand the effects of bubbles.
\subsubsection{Along Wall Bisector}
\begin{figure}[H]
	\begin{center}
		\begin{subfigure}[b]{0.49\textwidth}
			\includegraphics[width=1\textwidth]{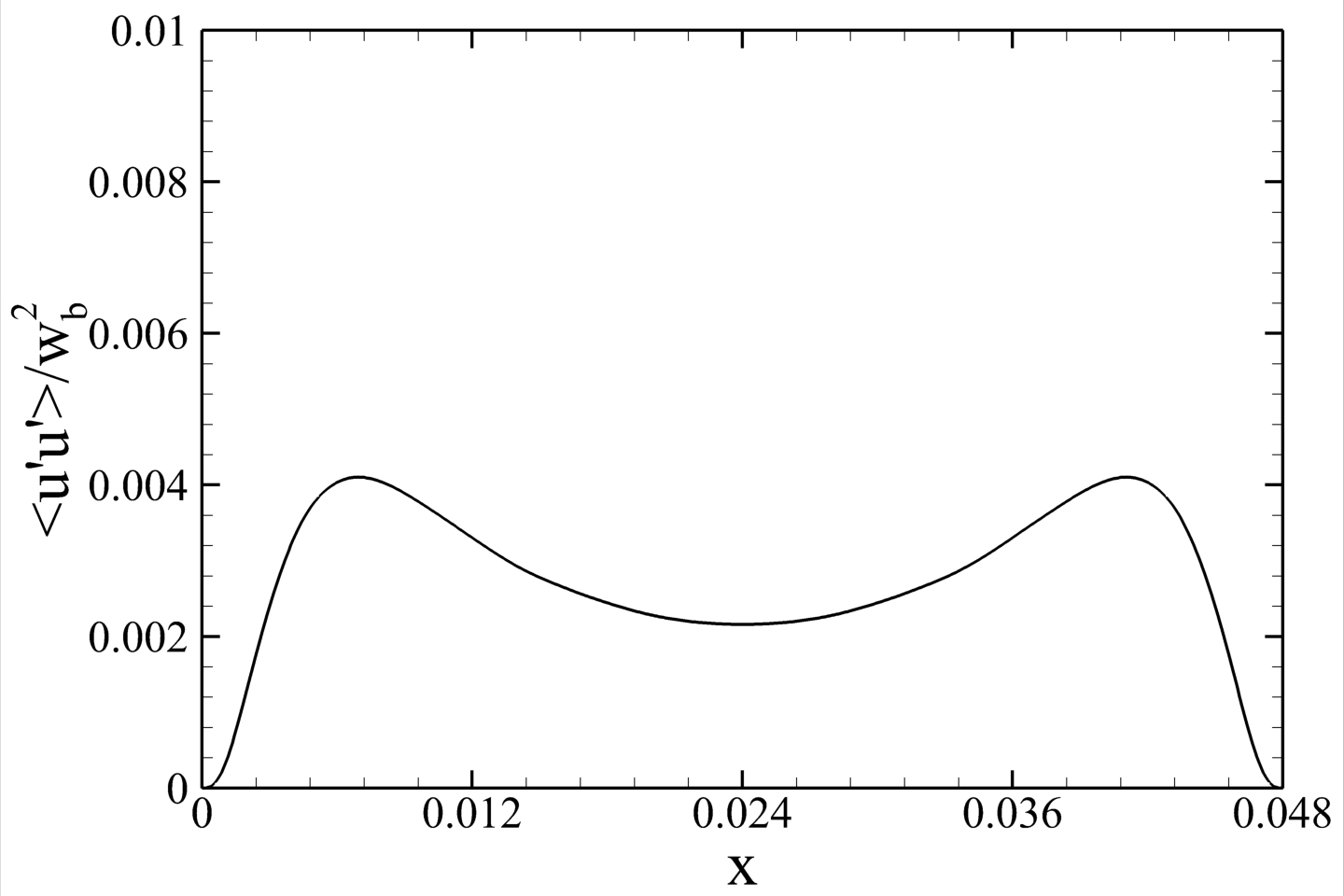}
			\caption{$\langle u'u' \rangle$ component for unladen flow}
			\label{fig:uprime_uprime_mean_unladen_wall_bisector:ch:bubbly_flow:sec:results}
		\end{subfigure} %
		\hspace{1mm}
		\begin{subfigure}[b]{0.49\textwidth}
			\includegraphics[width=1\textwidth]{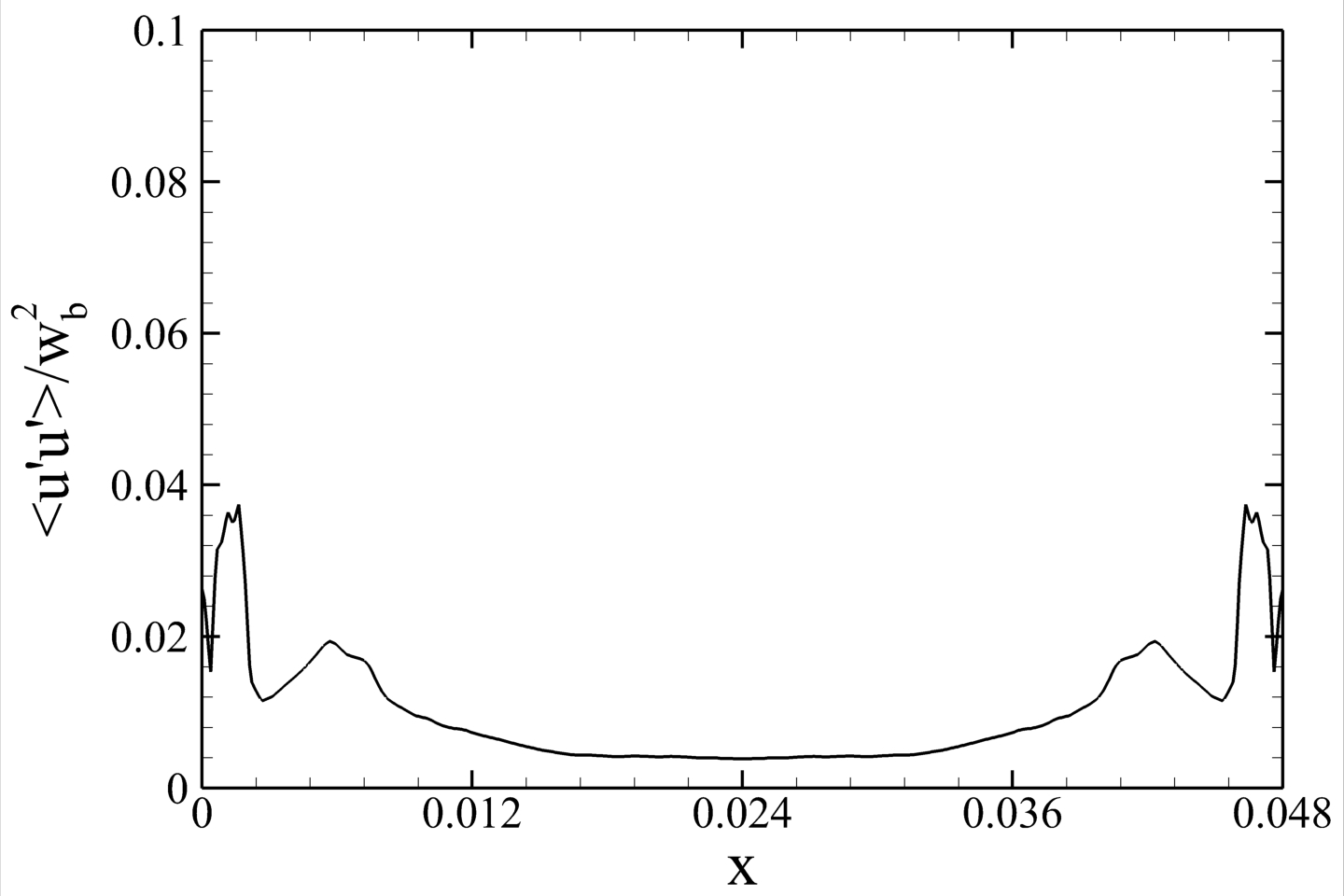}
			\caption{$\langle u'u' \rangle$ component for bubble laden flow}
			\label{fig:uprime_uprime_mean_laden_wall_bisector:ch:bubbly_flow:sec:results}
		\end{subfigure} %
		\hspace{1mm}
		\begin{subfigure}[b]{0.49\textwidth}
			\includegraphics[width=1\textwidth]{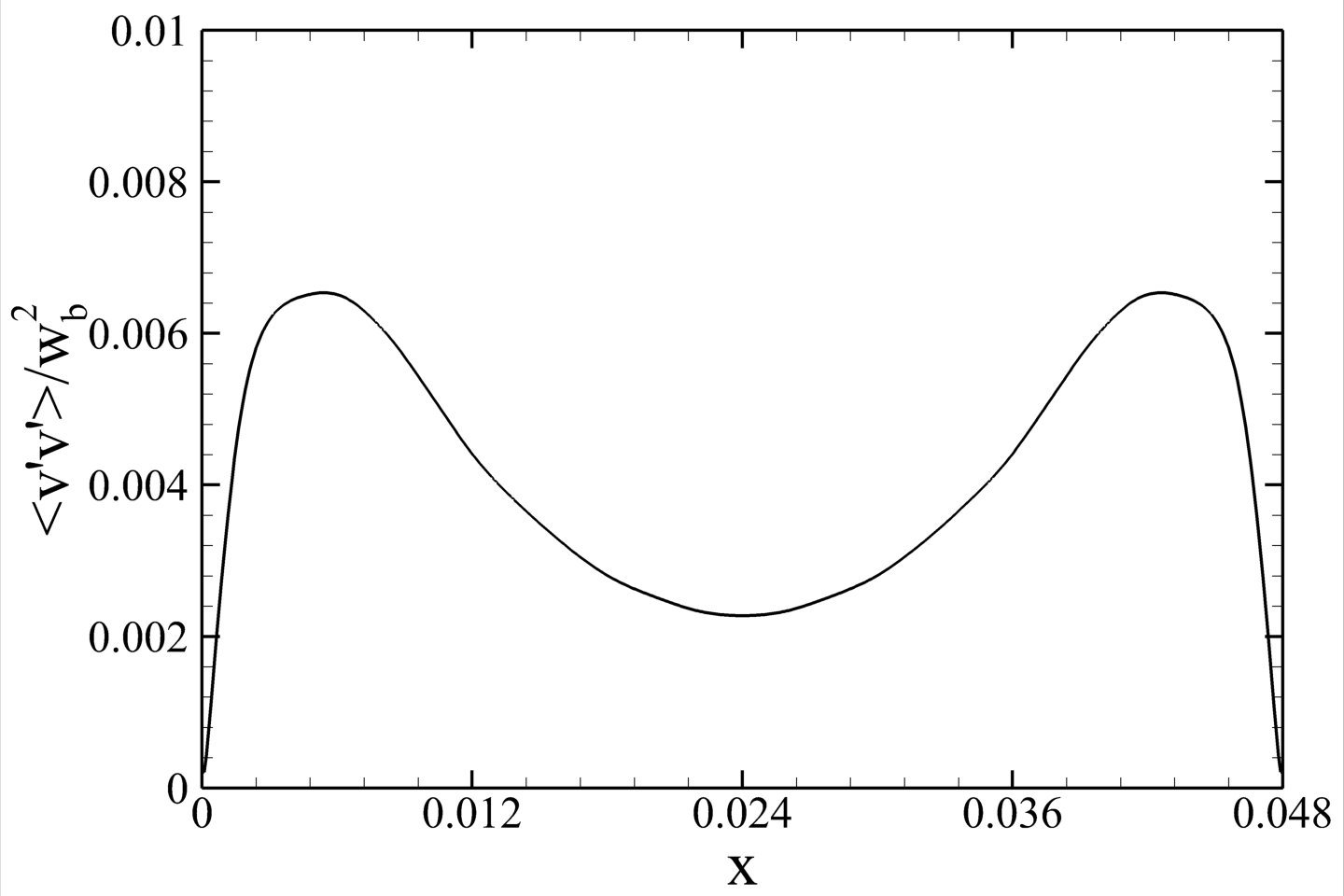}
			\caption{$\langle v'v' \rangle$ component for unladen flow}
			\label{fig:vprime_vprime_mean_unladen_wall_bisector:ch:bubbly_flow:sec:results}
		\end{subfigure} %
		\hspace{1mm}
		\begin{subfigure}[b]{0.49\textwidth}
			\includegraphics[width=1\textwidth]{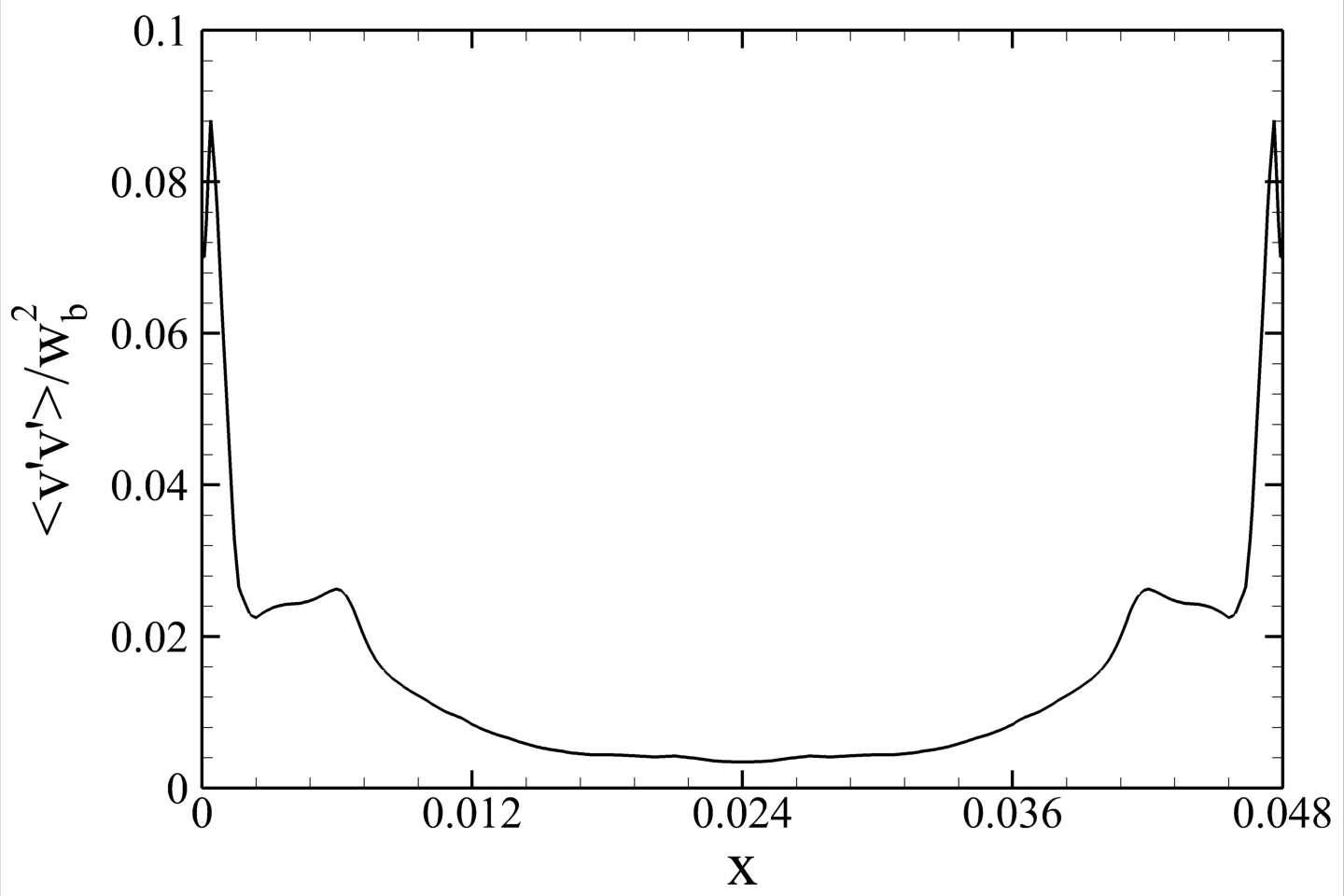}
			\caption{$\langle v'v' \rangle$ component for bubble laden flow}
			\label{fig:vprime_vprime_mean_laden_wall_bisector:ch:bubbly_flow:sec:results}
		\end{subfigure} %
		\hspace{1mm}
		\begin{subfigure}[b]{0.49\textwidth}
			\includegraphics[width=1\textwidth]{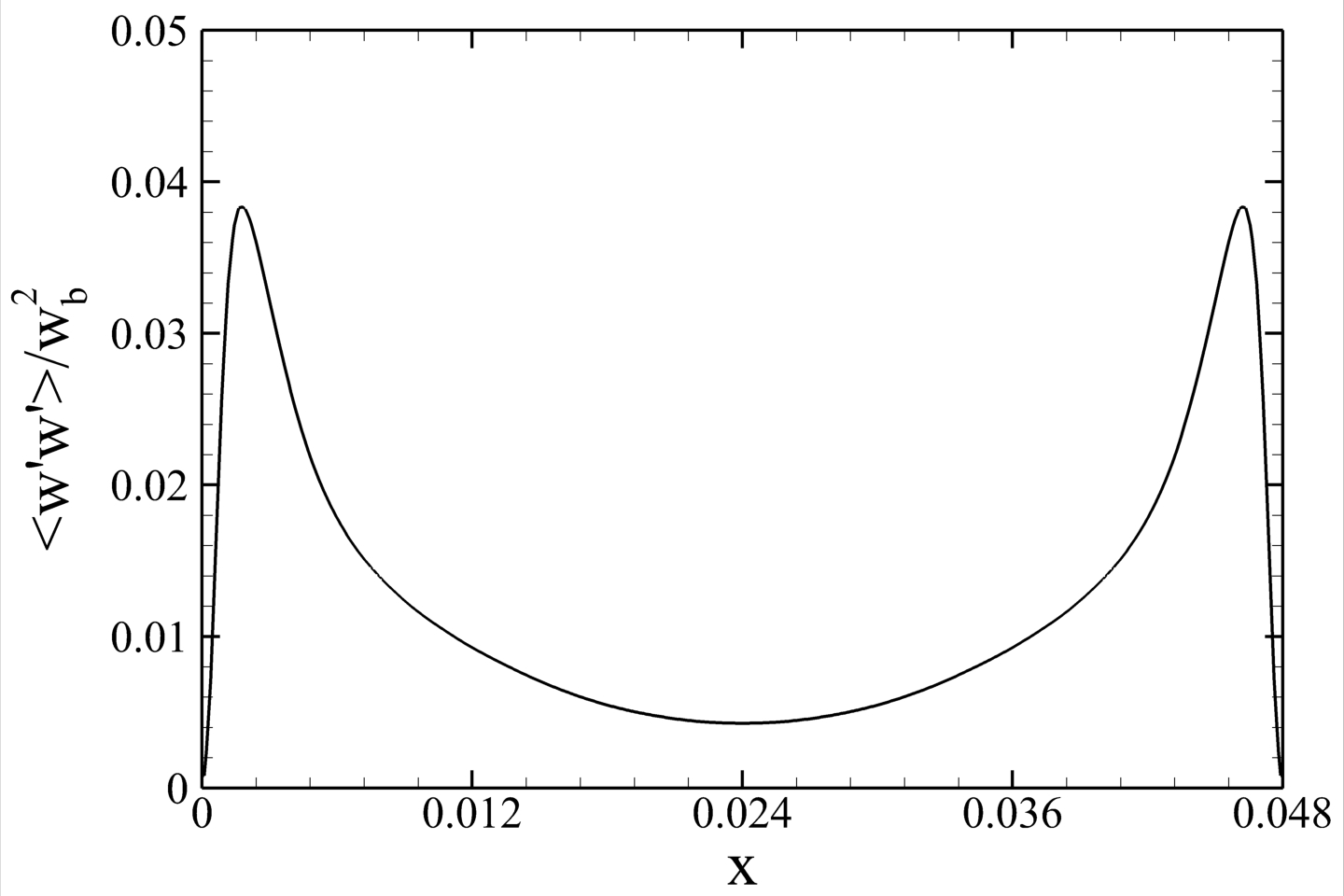}
			\caption{$\langle w'w' \rangle$ component for unladen flow}
			\label{fig:wprime_wprime_mean_unladen_wall_bisector:ch:bubbly_flow:sec:results}
		\end{subfigure} %
		\hspace{1mm}
		\begin{subfigure}[b]{0.49\textwidth}
			\includegraphics[width=1\textwidth]{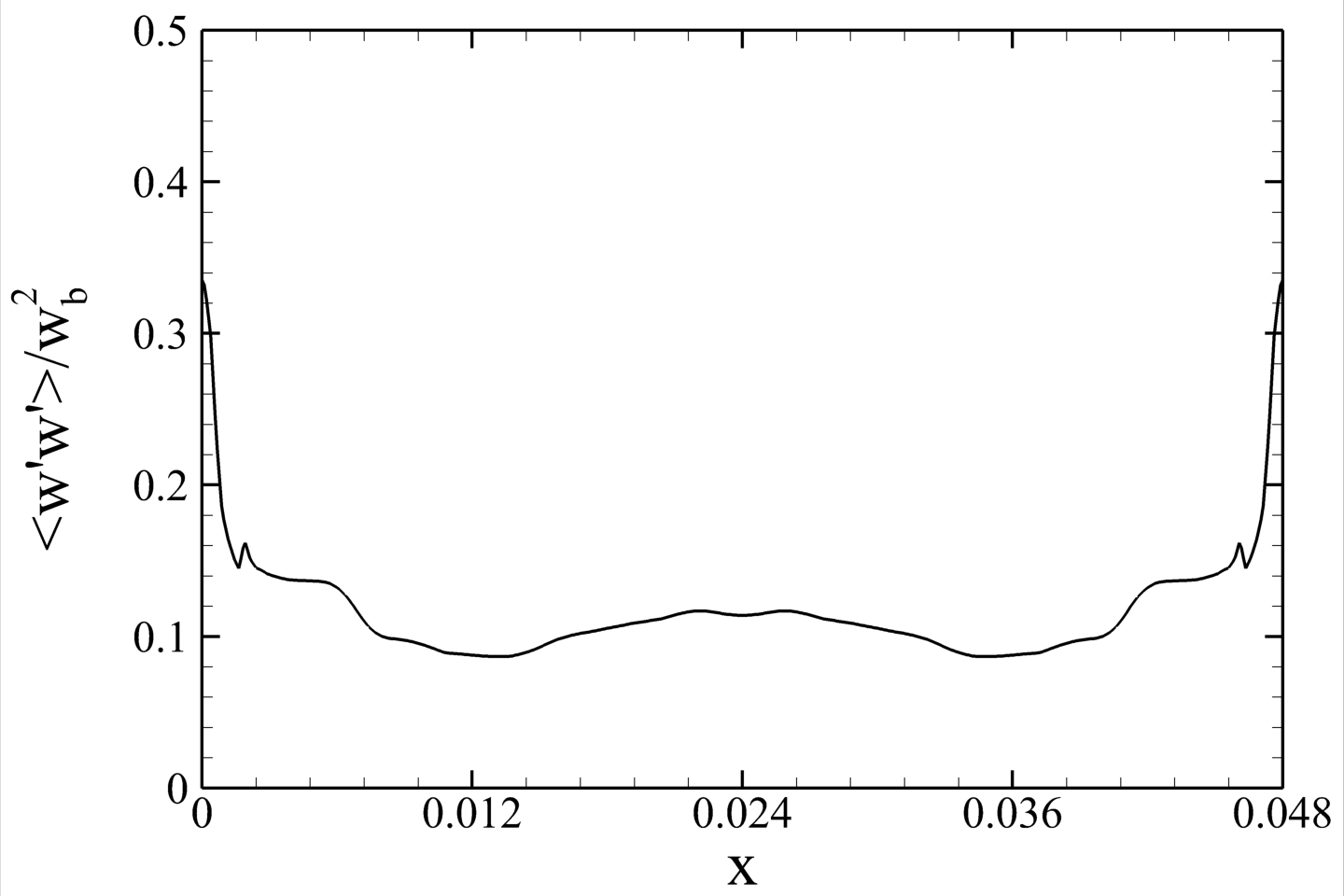}
			\caption{$\langle w'w' \rangle$ component for bubble laden flow}
			\label{fig:wprime_wprime_mean_laden_wall_bisector:ch:bubbly_flow:sec:results}
		\end{subfigure} %
		\caption{Reynold stress components: (a-b) $\langle u'u' \rangle$, (c-d) $\langle v'v' \rangle$ and (e-f)  $\langle w'w' \rangle$}
		\label{fig:reynolds_stresses_uu_vv_ww_wall_bisector:ch:bubbly_flow:sec:results}
	\end{center}
\end{figure}

The normal components $\langle u'u' \rangle$, $\langle v'v' \rangle$,  and $\langle w'w' \rangle$ of Reynolds stress along the wall bisector for both unladen and bubble laden flows are given in \cref{fig:reynolds_stresses_uu_vv_ww_wall_bisector:ch:bubbly_flow:sec:results}. All the components of Reynolds stress are scaled with the square of unladen flow bulk velocity. It can be noticed from the \cref{fig:uprime_uprime_mean_unladen_wall_bisector:ch:bubbly_flow:sec:results,fig:uprime_uprime_mean_laden_wall_bisector:ch:bubbly_flow:sec:results,fig:vprime_vprime_mean_unladen_wall_bisector:ch:bubbly_flow:sec:results,fig:vprime_vprime_mean_laden_wall_bisector:ch:bubbly_flow:sec:results,fig:wprime_wprime_mean_unladen_wall_bisector:ch:bubbly_flow:sec:results,fig:wprime_wprime_mean_laden_wall_bisector:ch:bubbly_flow:sec:results} that $\langle u'u' \rangle / w_b^2$, $\langle v'v' \rangle / w_b^2$, $\langle w'w' \rangle / w_b^2$ are zero at the wall and symmetric about $x = 0.024$ or $W/2$ which is expected as the quadrant symmetry has been enforced. 

In the case of unladen flow, all three normal components initially increase with distance from the wall to reach a maximum and then decrease to reach a minimum. The location of the maximum is closer to the wall, and the minimum is at center. It can also be noticed that the rate of increase of these quantities is higher than the rate of their decrease. The corresponding maximums of $\langle u'u' \rangle / w_b^2$, $\langle v'v' \rangle / w_b^2$, $\langle w'w' \rangle / w_b^2$ are observed at $x = 0.0041$, 0.0054 and 0.0018 or $x = 0.085W$, $0.113W$ and $0.037W$, respectively. The maximum values of $\langle w'w' \rangle / w_b^2$ and $\langle v'v' \rangle / w_b^2$ are roughly 10 and 1.5 times higher than that of $\langle u'u' \rangle / w_b^2$. It can also be noticed that the ratio between maximum and centerline values of $\langle u'u' \rangle$, $\langle v'v' \rangle$,  and $\langle w'w' \rangle$ are 2, 3 and 11 respectively. These values have also been predicted earlier in Madabhushi and Vanka \cite{Madabhushi1991}. \par 
In the case of laden flow, there are multiple local maximum and minimum for all three normal components. It can be seen from \cref{fig:uprime_uprime_mean_laden_wall_bisector:ch:bubbly_flow:sec:results,fig:vprime_vprime_mean_laden_wall_bisector:ch:bubbly_flow:sec:results,fig:wprime_wprime_mean_laden_wall_bisector:ch:bubbly_flow:sec:results} that the maximum is always observed near the wall, and the minimum of $\langle u'u' \rangle$ and $\langle v'v' \rangle$ are observed at the center. The minimum of $\langle w'w' \rangle$ is observed between the wall and duct center. It can also be noticed that the values of all three quantities at all locations are considerably higher in the presence of bubbles . For example, at the center $\langle u'u' \rangle/ w_b^2$ in unladen case is 0.0024, whereas in laden flow, it is approximately 0.0065. 

\begin{figure}[H]
	\begin{center}
		\begin{subfigure}[b]{0.49\textwidth}
			\includegraphics[width=1\textwidth]{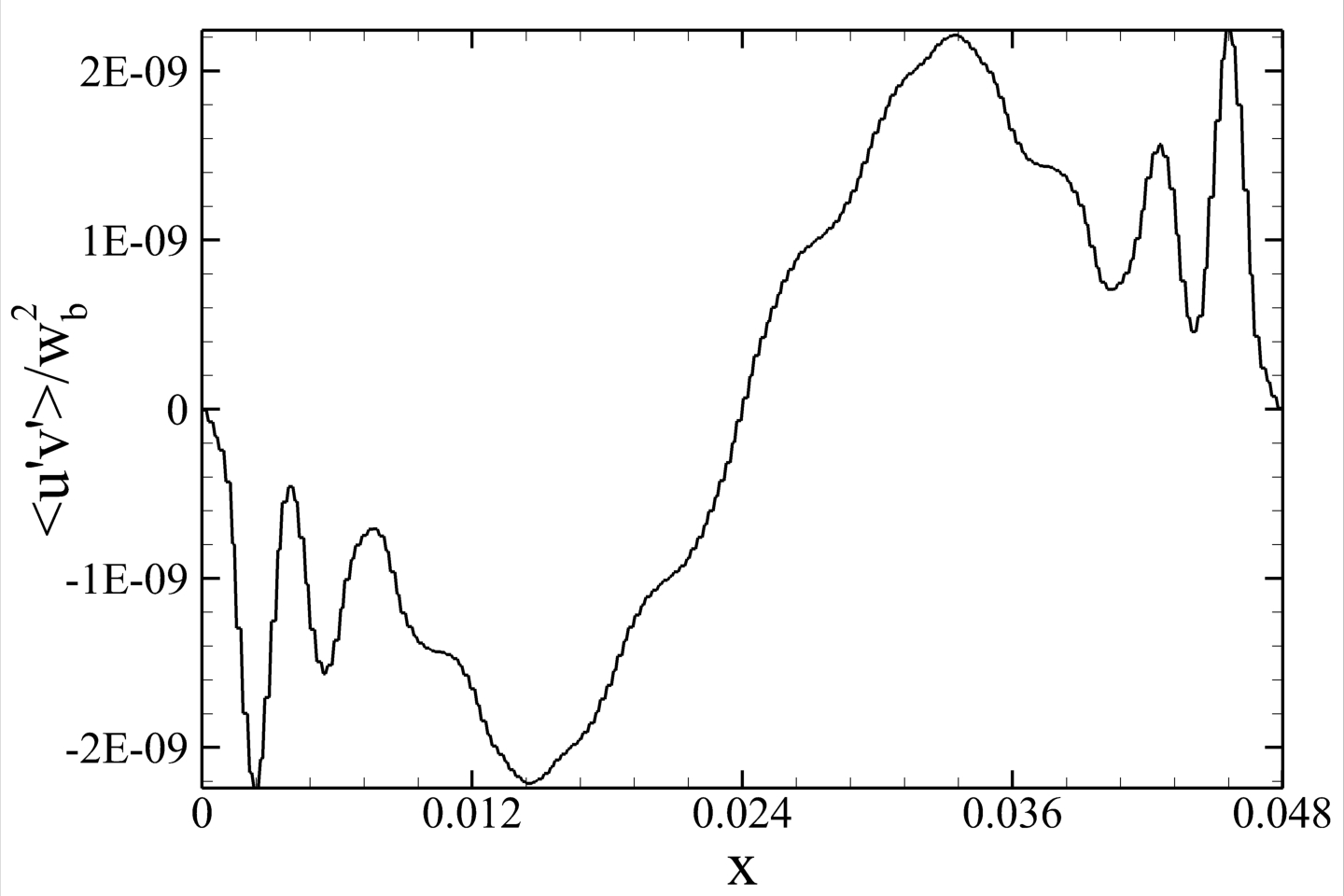}
			\caption{$\langle u'v' \rangle$ component for unladen flow}
			\label{fig:uprime_vprime_mean_unladen_wall_bisector:ch:bubbly_flow:sec:results}
		\end{subfigure} %
		\hspace{1mm}
		\begin{subfigure}[b]{0.49\textwidth}
			\includegraphics[width=1\textwidth]{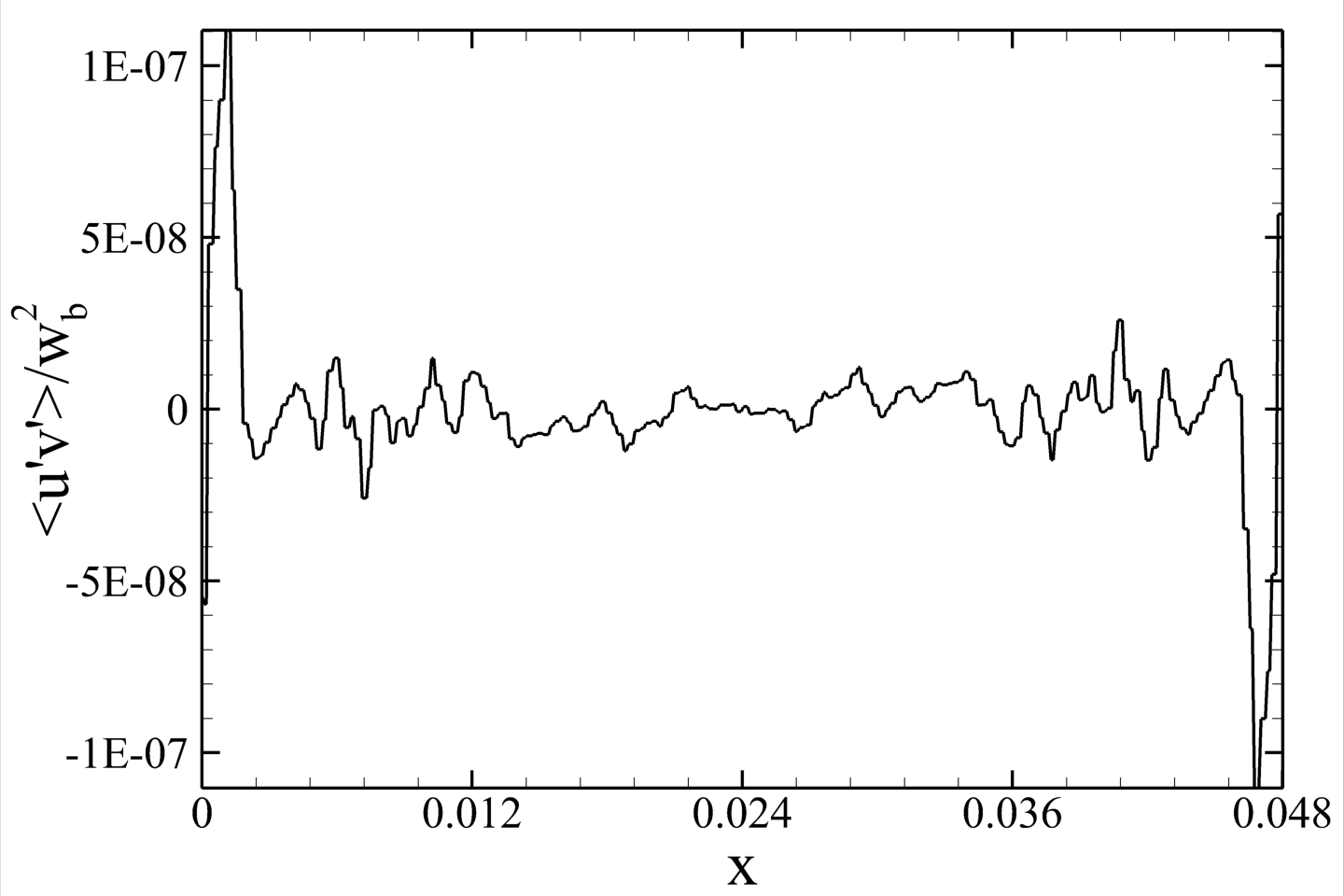}
			\caption{$\langle u'v' \rangle$ component for bubble laden flow}
			\label{fig:uprime_vprime_mean_laden_wall_bisector:ch:bubbly_flow:sec:results}
		\end{subfigure} %
		\hspace{1mm}
		\begin{subfigure}[b]{0.49\textwidth}
			\includegraphics[width=1\textwidth]{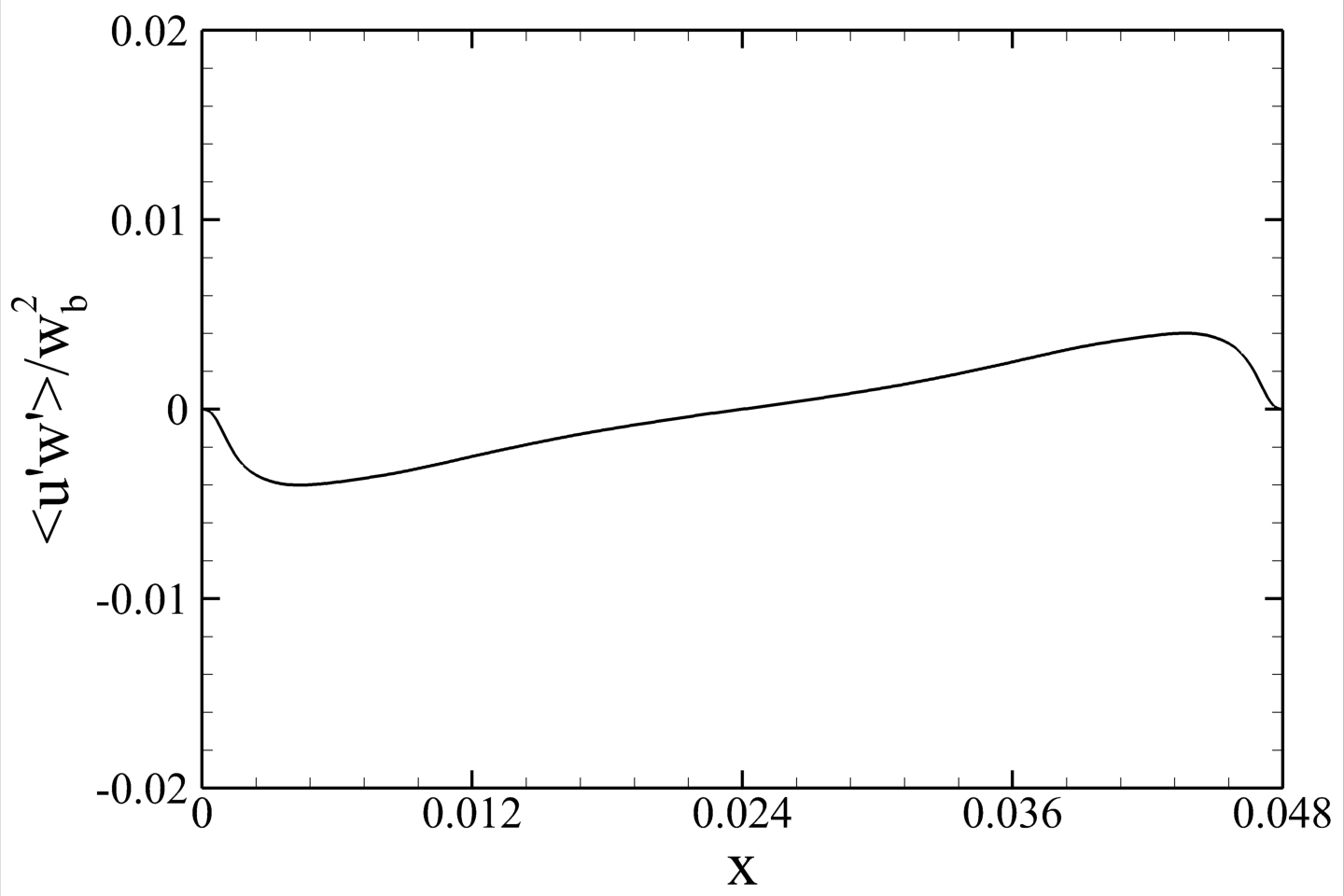}
			\caption{$\langle u'w' \rangle$ component for unladen flow}
			\label{fig:uprime_wprime_mean_unladen_wall_bisector:ch:bubbly_flow:sec:results}
		\end{subfigure} %
		\hspace{1mm}
		\begin{subfigure}[b]{0.49\textwidth}
			\includegraphics[width=1\textwidth]{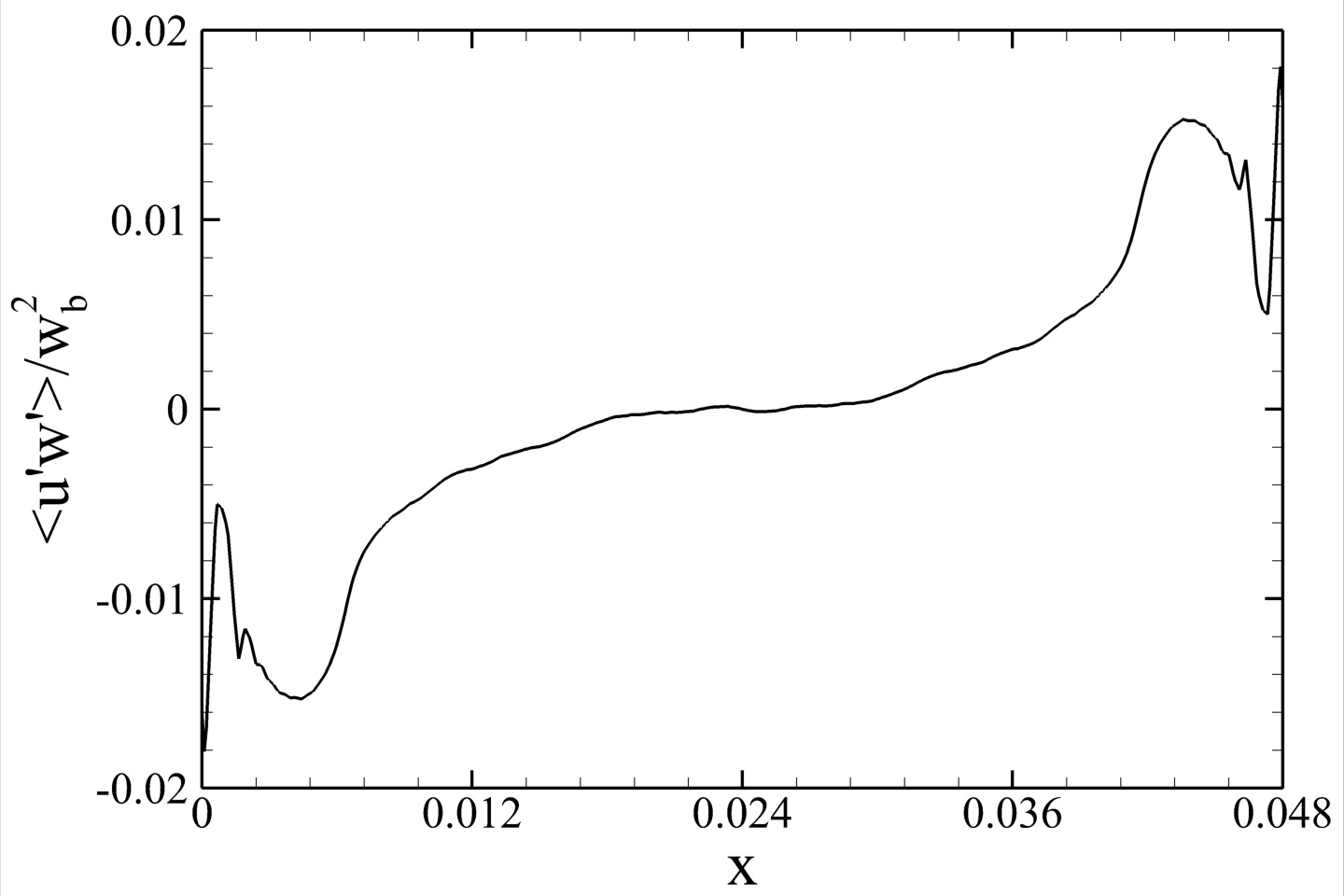}
			\caption{$\langle u'w' \rangle$ component for bubble laden flow}
			\label{fig:uprime_wprime_mean_laden_wall_bisector:ch:bubbly_flow:sec:results}
		\end{subfigure} %
		\caption{Reynold stress components: $\langle u'v' \rangle$ and $\langle u'w' \rangle$}
		\label{fig:reynolds_stresses_uv_uw_wall_bisector:ch:bubbly_flow:sec:results}
	\end{center}
\end{figure}

Similar to unladen flow, the maximum value of $\langle w'w' \rangle/ w_b^2$ and $\langle v'v' \rangle/ w_b^2$ is approximately 8 and 3 times higher than $\langle u'u' \rangle/ w_b^2$ for bubble laden flow as well. Trend that the first peak of $\langle w'w' \rangle/ w_b^2$ is closer to the wall than that of $\langle v'v' \rangle/ w_b^2$ and the peak of $\langle v'v' \rangle/ w_b^2$ is closer to the wall than that of $\langle u'u' \rangle/ w_b^2$ is consistent between unladen and bubble laden flows. However, there is one key difference in shape of the $\langle w'w' \rangle/ w_b^2$. In the case of unladen flow, the curve is `M' shaped, whereas in the case of bubble laden flow, the curve is `W' shaped. 
\Cref{fig:uprime_vprime_mean_unladen_wall_bisector:ch:bubbly_flow:sec:results,fig:uprime_vprime_mean_laden_wall_bisector:ch:bubbly_flow:sec:results,fig:uprime_wprime_mean_unladen_wall_bisector:ch:bubbly_flow:sec:results,fig:uprime_wprime_mean_laden_wall_bisector:ch:bubbly_flow:sec:results} show the $\langle u'v' \rangle$ and $\langle u'w' \rangle$ components of Reynolds stress along the wall bisector for both unladen and bubble laden flows. It can be noticed that these two quantities are asymmetric about the center or $x = W/2 = 0.024$. For both unladen and laden flows, $\langle u'v' \rangle$ is nearly zero along the wall bisectors. This signifies the very small fluctuations of $x$ and $y$ velocities along the wall bisector.
%
%
%
%
It can be noticed from \cref{fig:uprime_wprime_mean_unladen_wall_bisector:ch:bubbly_flow:sec:results,fig:uprime_wprime_mean_laden_wall_bisector:ch:bubbly_flow:sec:results} that the $\langle u'w' \rangle$ initially decreases with the distance to reach a minimum and then increases to reach a maximum at the same location on the other side of the duct. In the unladen flow, the minimum and maximum are observed at $x = 0.0044$ or $0.091W$ and $x = 0.909W$ respectively. The minimum and maximum values are -0.004 and 0.004 respectively, which are approximately same as $\langle u'u' \rangle/w_b^2$. In the case of laden flow, the minimum and maximum are observed at $x = 0.0044$ and $x = 0.0435$ respectively which are also at the same locations where they were observed in the unladen flow. The shape of $\langle u'w' \rangle$ in the central region is also similar to each other as their approximate slopes are 0.2042 and 0.2837 respectively in unladen and laden flows.

From figures of unladen and laden flow Reynolds stresses, we can make two qualitative conclusions: (1) Reynolds stresses increase when bubbles are introduced in the flow and (2) their maximum modification are observed near the wall. Due to the added effects of the buoyancy forces, bubbles travel with higher speed than the surrounding liquid in the upward flow, hence it introduces additional fluctuations in the flow; thus, we observe an increase in velocity fluctuations everywhere in the domain. The second observation is related to the migration of bubbles toward the wall region. As they concentrate near the wall, they increase the velocity fluctuations. 
%
%
%
%
%
\section{Conclusions}
\label{ch:bubbly_flow:sec:summary}
In the present study, the dynamics of a bubble swarm in a turbulent flow in a vertical square duct was addressed, and comparisons between unladen and bubble laden flows were conducted. A DNS of unladen flow for a frictional Reynolds number of $360$ was conducted, and unladen flow turbulence quantities were validated with the literature. The influence of bubbles on liquid phase was investigated, and the following overall conclusions can be drawn.

\begin{itemize}
	\item The introduction of bubbles strongly modifies the elongated flow structure usually observed in the unladen flow, and the longest structures are seen to occupy half of the duct. In the laden flow, instantaneous flow structures are oscillatory, and flatter. The longer structures are concentrated near the wall, and the shorter ones are distributed randomly away from the wall. The instantaneous laden flow is dominated by high and low values of positive and negative fluctuations respectively. 
	\item Starting from a uniform configuration, bubbles migrate toward the wall. They form a layer near the wall and a cluster near the corner. The lateral migration of bubbles, which leads to preferential concentration, is a result of balance between shear-induced lift force and turbophoresis effects. The secondary flows, which transport high energy fluids from the central region to the wall region along the corner bisector, are also responsible for the migration of bubbles in the corner region.
	\item Turbulence quantities, including velocity fluctuations and Reynolds stress, are enhanced by the introduction of bubbles. In the unladen flow, all turbulence quantities have one peak and one valley. Along the wall bisector they achieved their peaks near the wall and valleys at the center, but in the laden flow, multiple peaks and valleys were observed. The first peak was observed near the wall, and the second peak was observed between the wall and half distance. 
\end{itemize}

%
%
\section*{Acknowledgment}
The authors thank the financial support from the Air Conditioning and Refrigeration Center (ACRC), University of Illinois at Urbana-Champaign, USA. This research is also part of the Blue Waters sustained-petascale computing project, which is supported by the National Science Foundation (awards OCI-0725070 and ACI-1238993) and the state of Illinois. Blue Waters is a joint effort of the University of Illinois at Urbana-Champaign and its National Center for Supercomputing Applications. The authors further thank NVIDIA Hardware Grant Program for providing the GPUs for an in-house workstation.

\bibliographystyle{unsrtnat}	
\bibliography{ms.bib}

\begin{thebibliography}{61}
\providecommand{\natexlab}[1]{#1}
\providecommand{\url}[1]{\texttt{#1}}
\expandafter\ifx\csname urlstyle\endcsname\relax
  \providecommand{\doi}[1]{doi: #1}\else
  \providecommand{\doi}{doi: \begingroup \urlstyle{rm}\Url}\fi

\bibitem[Tomiyama et~al.(2002)Tomiyama, Tamai, Zun, and
  Hosokawa]{Tomiyama2002_bubble_migration}
A.~Tomiyama, H.~Tamai, I.~Zun, and S.~Hosokawa.
\newblock {Transverse migration of single bubbles in simple shear flows}.
\newblock \emph{Chemical Engineering Science}, 57:\penalty0 1849--1858, 2002.
\newblock ISSN 0066-4189.
\newblock \doi{10.1146/annurev.fluid.37.061903.175803}.

\bibitem[Afghan(1988)]{Zun1988_transition_wall_peaking}
N.~H. Afghan, editor.
\newblock \emph{{Transient Phenomena in Multiphase Flow}}.
\newblock Hemisphere Publishing, 1988.

\bibitem[Zun et~al.(1993)Zun, Kljenak, Serizawa, and
  Moze]{Zun1993_space_time_evolution}
I.~Zun, I.~Kljenak, A.~Serizawa, and S.~Moze.
\newblock {Space time evolution of non homogeneous bubble distribution in
  upward flow}.
\newblock \emph{International Journal of Multiphase Flow}, 19:\penalty0
  151--172, 1993.
\newblock ISSN 0066-4189.
\newblock \doi{10.1146/annurev.fluid.37.061903.175803}.

\bibitem[Liu(1989)]{Liu1989_phd_thesis}
T.J. Liu.
\newblock \emph{{Experimental investigation of turbulence structure in
  two-phase bubbly flow}}.
\newblock Phd thesis, Northwest University, Evanston, Illinois, 1989.

\bibitem[Liu and Bankoff(1993a)]{Liu1993a}
T.~J. Liu and S.~G. Bankoff.
\newblock {Structure of air-water bubbly flow in a vertical pipe-1. Liquid
  velocity and turbulence measurements}.
\newblock \emph{International Journal of Heat and Mass Transfer}, 36:\penalty0
  1049--1060, 1993a.
\newblock ISSN 0066-4189.
\newblock \doi{10.1146/annurev.fluid.37.061903.175803}.

\bibitem[Liu and Bankoff(1993b)]{Liu1993b}
T.~J. Liu and S.~G. Bankoff.
\newblock {Structure of air-water bubbly flow in a vertical pipe-1. Void
  fraction, bubble velocity and bubble size distribution}.
\newblock \emph{International Journal of Heat and Mass Transfer}, 36:\penalty0
  1061--1072, 1993b.
\newblock ISSN 0066-4189.
\newblock \doi{10.1146/annurev.fluid.37.061903.175803}.

\bibitem[Colin et~al.(2012)Colin, Fabre, and Kamp]{Colin2012}
Catherine Colin, Jean Fabre, and Arjan Kamp.
\newblock {Turbulent bubbly flow in pipe under gravity and microgravity
  conditions}.
\newblock \emph{Journal of Fluid Mechanics}, 711:\penalty0 469--515, 2012.

\bibitem[Lance and Bataille(1991)]{Lance1991}
M.~Lance and J.~Bataille.
\newblock {Turbulence in the liquid phase of a uniform bubbly air–water
  flow}.
\newblock \emph{Journal of Fluid Mechanics}, 222:\penalty0 95--118, 1991.
\newblock ISSN 0022-1120.
\newblock \doi{10.1017/S0022112091001015}.

\bibitem[Yuan and Michaelides(1992)]{Yuan1992}
Z.~Yuan and E.~E. Michaelides.
\newblock {Turbulence modulation in particulate flows-A theoretical approach}.
\newblock \emph{International Journal of Multiphase Flow}, 18\penalty0
  (5):\penalty0 779--785, 1992.
\newblock ISSN 03019322.
\newblock \doi{10.1016/0301-9322(92)90045-I}.

\bibitem[Hosokawa and Tomiyama(2013)]{Hosokawa2013}
Shigeo Hosokawa and Akio Tomiyama.
\newblock {Bubble-induced pseudo turbulence in laminar pipe flows}.
\newblock \emph{International Journal of Heat and Fluid Flow}, 40:\penalty0
  97--105, 2013.

\bibitem[Alm{\'{e}}ras et~al.(2015)Alm{\'{e}}ras, Risso, Roig, Cazin, Plais,
  and Augier]{Almeras2015}
Elise Alm{\'{e}}ras, Fr{\'{e}}d{\'{e}}ric Risso, V{\'{e}}ronique Roig,
  S{\'{e}}bastien Cazin, C{\'{e}}cile Plais, and Fr{\'{e}}d{\'{e}}ric Augier.
\newblock {Mixing by bubble-induced turbulence}.
\newblock \emph{Journal of Fluid Mechanics}, 776:\penalty0 458--474, 2015.

\bibitem[Hirt and Nichols(1981)]{Hirt1981}
C.~W Hirt and B.~D Nichols.
\newblock {Volume of Fluid (VOF) method for the dynamics of free boundaries}.
\newblock \emph{Journal of Computational Physics}, 39\penalty0 (1):\penalty0
  201--225, January 1981.

\bibitem[Rudman(1998)]{Rudman1998}
Murray Rudman.
\newblock {A volume-tracking method for incompressible multifluid flows with
  large density variations}.
\newblock \emph{International Journal for Numerical Methods in Fluids},
  28\penalty0 (2):\penalty0 357--378, August 1998.
\newblock ISSN 0271-2091.

\bibitem[Cummins et~al.(2005)Cummins, Francois, and Kothe]{Cummins2005}
Sharen~J. Cummins, Marianne~M. Francois, and Douglas~B. Kothe.
\newblock Estimating curvature from volume fractions.
\newblock \emph{Computers \& Structures}, 83\penalty0 (6-7):\penalty0 425 --
  434, 2005.
\newblock ISSN 0045-7949.
\newblock \doi{http://dx.doi.org/10.1016/j.compstruc.2004.08.017}.
\newblock URL
  \url{http://www.sciencedirect.com/science/article/pii/S0045794904004110}.

\bibitem[Francois et~al.(2006)Francois, Cummins, Dendy, Kothe, Sicilian, and
  Williams]{Francois2006}
Marianne~M. Francois, Sharen~J. Cummins, Edward~D. Dendy, Douglas~B. Kothe,
  James~M. Sicilian, and Matthew~W. Williams.
\newblock {A balanced-force algorithm for continuous and sharp interfacial
  surface tension models within a volume tracking framework}.
\newblock \emph{Journal of Computational Physics}, 213\penalty0 (1):\penalty0
  141--173, March 2006.
\newblock ISSN 00219991.

\bibitem[Wang and Tong(2010)]{Wang2010}
Zhaoyuan Wang and Albert~Y. Tong.
\newblock {A sharp surface tension modeling method for two-phase incompressible
  interfacial flows}.
\newblock \emph{International Journal for Numerical Methods in Fluids},
  64\penalty0 (7):\penalty0 709--732, September 2010.
\newblock ISSN 02712091.

\bibitem[Kumar and Vanka(2015)]{Kumar2015confinement}
P.~Kumar and S.P. Vanka.
\newblock Effects of confinement on bubble dynamics in a square duct.
\newblock \emph{International Journal of Multiphase Flow}, 77:\penalty0 32 --
  47, 2015.
\newblock ISSN 0301-9322.

\bibitem[Kumar et~al.(2015{\natexlab{a}})Kumar, Jin, and
  Vanka]{Kumar2015numerical}
Purushotam Kumar, Kai Jin, and S.~Pratap Vanka.
\newblock A multi-{GPU} based accurate algorithm for simulations of gas-liquid
  flows.
\newblock \emph{Computational Methods and Tools in Thermal Fluids Sciences},
  pages 1--17, 2015{\natexlab{a}}.

\bibitem[Kumar et~al.(2015{\natexlab{b}})Kumar, Jin, and
  Vanka]{Kumar2015non_newtonian}
Purushotam Kumar, Kai Jin, and S.~Pratap Vanka.
\newblock Bubble rise and deformation in a non-newtonian fluid in a square
  duct.
\newblock \emph{Computational Methods and Tools in Thermal Fluids Sciences},
  pages 1--15, 2015{\natexlab{b}}.

\bibitem[Kumar et~al.(2019)Kumar, Jin, and Vanka]{Kumar2019AJKFluids}
Purushotam Kumar, Kai Jin, and Surya~Pratap Vanka.
\newblock {Numerical Simulation of a Gas Bubble Rising in Power-Law Fluids
  Using a Sharp Surface Force Implementation}.
\newblock \textit{Proceedings of the ASME-JSME-KSME 2019 8th Joint Fluids
  Engineering Conference. Volume 5: Multiphase Flow}. San Francisco,
  California, USA, July 28-August 01, 2019.

\bibitem[Rider and Kothe(1998)]{Rider1998}
William~J. Rider and Douglas~B. Kothe.
\newblock Reconstructing volume tracking.
\newblock \emph{Journal of Computational Physics}, 141\penalty0 (2):\penalty0
  112--152, April 1998.
\newblock ISSN 00219991.

\bibitem[Noh and Woodward(1976)]{Noh1976}
WF~Noh and P~Woodward.
\newblock {SLIC (simple line interface calculation)}.
\newblock In \emph{Proceedings of the Fifth International Conference}, pages
  330--340, Berlin, 1976. Springer-Verlag.

\bibitem[Li(1995)]{Li1995}
Jie Li.
\newblock {Calcul d'Interface Affine par Morceaux}(piecewise linear interface
  calculation).
\newblock \emph{C. R. Acad. Sci. Paris}, 320\penalty0 (8):\penalty0 391--396,
  1995.

\bibitem[Ashgriz and Poo(1991)]{Ashgriz1991}
N~Ashgriz and JY~Poo.
\newblock {FLAIR:} flux line-segment model for advection and interface
  reconstruction.
\newblock \emph{Journal of Computational Physics}, 93\penalty0 (2):\penalty0
  449--468, April 1991.
\newblock ISSN 00219991.

\bibitem[Renardy and Renardy(2002)]{Renardy2002}
Yuriko Renardy and Michael Renardy.
\newblock {PROST:} a parabolic reconstruction of surface tension for the
  volume-of-fluid method.
\newblock \emph{Journal of Computational Physics}, 183\penalty0 (2):\penalty0
  400--421, December 2002.
\newblock ISSN 00219991.

\bibitem[Sussman(2003)]{Sussman2003}
Mark Sussman.
\newblock {A second order coupled level set and volume-of-fluid method for
  computing growth and collapse of vapor bubbles}.
\newblock \emph{Journal of Computational Physics}, 187\penalty0 (1):\penalty0
  110--136, May 2003.
\newblock ISSN 00219991.

\bibitem[Sussman et~al.(2007)Sussman, Smith, Hussaini, Ohta, and
  Zhi-Wei]{Sussman2007}
M.~Sussman, K.M. Smith, M.Y. Hussaini, M.~Ohta, and R.~Zhi-Wei.
\newblock {A sharp interface method for incompressible two-phase flows}.
\newblock \emph{Journal of Computational Physics}, 221\penalty0 (2):\penalty0
  469--505, February 2007.
\newblock ISSN 00219991.

\bibitem[Gueyffier et~al.(1999)Gueyffier, Li, Nadim, Scardovelli, and
  Zaleski]{Gueyffier1999}
Denis Gueyffier, Jie Li, Ali Nadim, Ruben Scardovelli, and St\'{e}phane
  Zaleski.
\newblock {Volume-of-Fluid} interface tracking with smoothed surface stress
  methods for three-dimensional flows.
\newblock \emph{Journal of Computational Physics}, 152\penalty0 (2):\penalty0
  423--456, July 1999.
\newblock ISSN 00219991.

\bibitem[Vanka et~al.(2016)Vanka, Kumar, Jin, and Thomas]{Vanka2016Single}
{S. P.} Vanka, P.~Kumar, K.~Jin, and {B. G.} Thomas.
\newblock Single and multiphase flow computations on graphics processing units.
\newblock \emph{International Conference on Computational Methods for Thermal
  Problems}, \penalty0 (217349), 2016.
\newblock ISSN 2305-5995.

\bibitem[Kumar(2016)]{Kumar2016thesis}
Purushotam Kumar.
\newblock \emph{Development and application of Lattice Boltzmann and accurate
  volume of fluid numerical techniques on graphics processing units}.
\newblock PhD thesis, University of Illinois at Urbana-Champaign, 2016.

\bibitem[Horwitz et~al.(2012)Horwitz, Kumar, and Vanka]{horwitz2012simulations}
Jeremy Horwitz, Purushotam Kumar, and Pratap Vanka.
\newblock Simulations of multiphase flow in a t-junction and distributor
  header.
\newblock \emph{Bulletin of the American Physical Society}, 57, 2012.

\bibitem[Horwitz et~al.(2013)Horwitz, Kumar, and Vanka]{horwitz2013simulations}
Jeremy Horwitz, Purushotam Kumar, and Pratap Vanka.
\newblock Simulations of three-dimensional droplet deformation in a square-duct
  at moderate reynolds numbers.
\newblock \emph{Bulletin of the American Physical Society}, 58, 2013.

\bibitem[Kumar et~al.(2013)Kumar, Horwitz, and Vanka]{kumar2013three}
Purushotam Kumar, Jeremy Horwitz, and Surya Vanka.
\newblock A three-dimensional numerical study of immiscible droplet deformation
  in a right angle bend.
\newblock \emph{Bulletin of the American Physical Society}, 58, 2013.

\bibitem[Horwitz et~al.(2014)Horwitz, Kumar, and Vanka]{Horwitz2014_lbm}
JAK Horwitz, P~Kumar, and SP~Vanka.
\newblock Three-dimensional deformation of a spherical droplet in a square duct
  flow at moderate reynolds numbers.
\newblock \emph{International journal of multiphase flow}, 67:\penalty0 10--24,
  2014.

\bibitem[Horwitz et~al.(2019)Horwitz, Vanka, and Kumar]{Horwitz2019AJKFluids}
Jeremy A.~K. Horwitz, S.~P. Vanka, and P.~Kumar.
\newblock Lbm simulations of dispersed multiphase flows in a channel: Role of a
  pressure poisson equation.
\newblock \textit{Proceedings of the ASME-JSME-KSME 2019 8th Joint Fluids
  Engineering Conference. Volume 5: Multiphase Flow}. San Francisco,
  California, USA, July 28-August 01, 2019.

\bibitem[{Vanka} et~al.(2015){Vanka}, {Jin}, {Kumar}, and
  {Thomas}]{Vanka2015APS_DFD}
Surya~Pratap {Vanka}, Kai {Jin}, Purushotam {Kumar}, and Brian {Thomas}.
\newblock {Rise of an argon bubble in liquid steel in the presence of a
  transverse magnetic field}.
\newblock \emph{Bulletin of the American Physical Society}, 60, 2015.

\bibitem[Jin et~al.(2016)Jin, Kumar, Vanka, and Thomas]{Jin2016mhd_bubble}
K.~Jin, P.~Kumar, S.~P. Vanka, and B.~G. Thomas.
\newblock {Rise of an argon bubble in liquid steel in the presence of a
  transverse magnetic field}.
\newblock \emph{Physics of Fluids}, 28\penalty0 (9), 2016.
\newblock ISSN 10897666.
\newblock \doi{10.1063/1.4961561}.

\bibitem[{Vanka} et~al.(2016){Vanka}, {Kumar}, and {Jin}]{Vanka2016APS_DFD}
Pratap {Vanka}, Purushotam {Kumar}, and Kai {Jin}.
\newblock {Numerical Simulation of Turbulent Bubbly Flow in a Vertical Square
  Duct}.
\newblock \emph{Bulletin of the American Physical Society}, 61, 2016.

\bibitem[Nickolls and Dally(2010)]{Nickolls2010_gpu_era}
J.~Nickolls and W.~J. Dally.
\newblock The gpu computing era.
\newblock \emph{IEEE Micro}, 30\penalty0 (2):\penalty0 56--69, March 2010.
\newblock ISSN 0272-1732.
\newblock \doi{10.1109/MM.2010.41}.

\bibitem[Cud(2015)]{Cuda_programming_guide}
{CUDA programming guide}, September 2015.
\newblock URL
  \url{http://docs.nvidia.com/cuda/pdf/CUDA_C_Programming_Guide.pdf}.

\bibitem[Nickolls et~al.(2008)Nickolls, Buck, Garland, and
  Skadron]{Nickolls2008_scalable_parallel}
John Nickolls, Ian Buck, Michael Garland, and Kevin Skadron.
\newblock Scalable parallel programming with cuda.
\newblock \emph{Queue}, 6\penalty0 (2):\penalty0 40--53, March 2008.
\newblock ISSN 1542-7730.
\newblock \doi{10.1145/1365490.1365500}.
\newblock URL \url{http://doi.acm.org/10.1145/1365490.1365500}.

\bibitem[Association(2000)]{InfiniBand2000}
I.~T. Association.
\newblock \emph{{InfiniBand Architecture Specification: Release 1.0}}.
\newblock InfiniBand Trade Association, 2000.

\bibitem[Gropp et~al.(1996)Gropp, Lusk, Doss, and
  Skjellum]{Gropp1996_high_performance_MPI}
W~Gropp, E~Lusk, N~Doss, and A~Skjellum.
\newblock A high-performance, portable implementation of the mpi message
  passing interface standard.
\newblock \emph{Parallel Computing}, 22\penalty0 (6):\penalty0 789--828, 1996.

\bibitem[Gropp et~al.(1999)Gropp, Lusk, and Thakur]{Gropp1999_MPI}
W~Gropp, E~Lusk, and R~Thakur.
\newblock \emph{{Using MPI-2: Advanced Features of the Message Passing
  Interface}}.
\newblock MIT press, 1999.

\bibitem[Kim et~al.(1987)Kim, Moin, and Moser]{Kim1987_turbulence_stat}
John Kim, Parviz Moin, and Robert Moser.
\newblock Turbulence statistics in fully developed channel flow at low reynolds
  number.
\newblock \emph{Journal of Fluid Mechanics}, 177:\penalty0 133--166, 4 1987.
\newblock ISSN 1469-7645.
\newblock \doi{10.1017/S0022112087000892}.
\newblock URL \url{http://journals.cambridge.org/article_S0022112087000892}.

\bibitem[Madabhushi and Vanka(1991)]{Madabhushi1991}
Ravi~K Madabhushi and S~P Vanka.
\newblock {Large eddy simulation of turbulence-driven secondary flow in a
  square duct}.
\newblock \emph{Physics of Fluids A: Fluid Dynamics}, 3\penalty0 (11):\penalty0
  2734, 1991.
\newblock ISSN 08998213.
\newblock \doi{10.1063/1.858163}.

\bibitem[Panidis and Papailiou(2000)]{Panidis2000}
Thrasyvoulos Panidis and Demosthenes~D. Papailiou.
\newblock {The structure of two-phase grid turbulence in a rectangular channel:
  An experimental study}.
\newblock \emph{International Journal of Multiphase Flow}, 26\penalty0
  (8):\penalty0 1369--1400, 2000.
\newblock ISSN 03019322.
\newblock \doi{10.1016/S0301-9322(99)00085-3}.

\bibitem[Santarelli and Fr{\"{o}}hlich(2015)]{Santarelli2015}
C.~Santarelli and J.~Fr{\"{o}}hlich.
\newblock {Direct Numerical Simulations of spherical bubbles in vertical
  turbulent channel flow}.
\newblock \emph{International Journal of Multiphase Flow}, 75:\penalty0
  174--193, 2015.

\bibitem[Bunner and Tryggvason(2002)]{Bunner2002}
B~Bunner and G.~Tryggvason.
\newblock {Dynamics of homogeneous bubbly flows Part 2. Velocity fluctuations}.
\newblock \emph{Journal of Fluid Mechanics}, 466:\penalty0 53--84, 2002.
\newblock ISSN 0022-1120.
\newblock \doi{10.1017/S0022112002001180}.

\bibitem[Bunner and Tryggvason(2003)]{Bunner2003}
B.~Bunner and G.~Tryggvason.
\newblock {Effect of bubble deformation on the properties of bubbly flows}.
\newblock \emph{Journal of Fluid Mechanics}, 495:\penalty0 77--118, 2003.
\newblock ISSN 00221120.
\newblock \doi{10.1017/S0022112003006293}.

\bibitem[Grace(1973)]{Grace1973}
JR~Grace.
\newblock Shapes and velocities of bubbles rising in infinite liquids.
\newblock \emph{Trans. Inst. Chem. Eng.}, 51\penalty0 (2):\penalty0 116--120,
  1973.

\bibitem[Gavrilakis(1992)]{Gavrilakis1992}
S.~Gavrilakis.
\newblock {Numerical simulation of low-Reynolds-number turbulent flow through a
  straight square duct}.
\newblock \emph{Journal of Fluid Mechanics}, 244\penalty0 (-1):\penalty0 101,
  1992.
\newblock ISSN 0022-1120.
\newblock \doi{10.1017/S0022112092002982}.

\bibitem[Huser and Biringen(1993)]{Huser1993}
Asmund Huser and Sedat Biringen.
\newblock Direct numerical simulation of turbulent flow in a square duct.
\newblock \emph{Journal of Fluid Mechanics}, 257:\penalty0 65--95, 12 1993.
\newblock ISSN 1469-7645.
\newblock \doi{10.1017/S002211209300299X}.
\newblock URL \url{http://journals.cambridge.org/article_S002211209300299X}.

\bibitem[Broglia et~al.(2003)Broglia, Pascarelli, and Piomelli]{Broglia2003}
Riccardo Broglia, Andrea Pascarelli, and Ugo Piomelli.
\newblock {Large-eddy simulations of ducts with a free surface}.
\newblock \emph{Journal of Fluid Mechanics}, 484:\penalty0 223--253, 2003.
\newblock ISSN 00221120.
\newblock \doi{10.1017/S0022112003004257}.

\bibitem[Zhang et~al.(2015)Zhang, Trias, Gorobets, Tan, and Oliva]{Zhang2015}
Hao Zhang, F.~Xavier Trias, Andrey Gorobets, Yuanqiang Tan, and Assensi Oliva.
\newblock {Direct numerical simulation of a fully developed turbulent square
  duct flow up to $Re_{\tau}$ = 1200}.
\newblock \emph{International Journal of Heat and Fluid Flow}, 54:\penalty0
  258--267, 2015.

\bibitem[Winkler et~al.(2004)Winkler, Rani, and
  Vanka]{Winkler2004_preferential}
C.M. Winkler, Sarma~L. Rani, and S.P. Vanka.
\newblock Preferential concentration of particles in a fully developed
  turbulent square duct flow.
\newblock \emph{International Journal of Multiphase Flow}, 30\penalty0
  (1):\penalty0 27 -- 50, 2004.
\newblock ISSN 0301-9322.
\newblock \doi{http://dx.doi.org/10.1016/j.ijmultiphaseflow.2003.11.003}.
\newblock URL
  \url{http://www.sciencedirect.com/science/article/pii/S0301932203002039}.

\bibitem[Winkler et~al.(2006)Winkler, Rani, and
  Vanka]{Winkler2006_wall_deposition}
C.M. Winkler, Sarma~L. Rani, and S.P. Vanka.
\newblock A numerical study of particle wall-deposition in a turbulent square
  duct flow.
\newblock \emph{Powder Technology}, 170\penalty0 (1):\penalty0 12 -- 25, 2006.
\newblock ISSN 0032-5910.
\newblock \doi{http://dx.doi.org/10.1016/j.powtec.2006.08.009}.
\newblock URL
  \url{http://www.sciencedirect.com/science/article/pii/S0032591006003391}.

\bibitem[Chaudhary et~al.(2009)Chaudhary, Vanka, and Thomas]{Chaudhary2009}
R.~Chaudhary, S.~P. Vanka, and B.~G. Thomas.
\newblock Direct numerical simulations of magnetic field effects on turbulent
  duct flows.
\newblock In \emph{Proceedings of the ASME 2009 International Mechanical
  Engineering Congress and Exposition (IMECE 2009)}, pages 229--308, 2009.
\newblock \doi{10.1115/IMECE2009-11527}.
\newblock URL
  \url{http://proceedings.asmedigitalcollection.asme.org/proceeding.aspx?articleid=1640392}.

\bibitem[Chaudhary et~al.(2010)Chaudhary, Vanka, and
  Thomas]{Chaudhary2010_DNS_MHD}
R.~Chaudhary, S.~P. Vanka, and B.~G. Thomas.
\newblock Direct numerical simulations of magnetic field effects on turbulent
  flow in a square duct.
\newblock \emph{Physics of Fluids}, 22\penalty0 (7):\penalty0 075102, 2010.
\newblock \doi{http://dx.doi.org/10.1063/1.3456724}.
\newblock URL
  \url{http://scitation.aip.org/content/aip/journal/pof2/22/7/10.1063/1.3456724}.

\bibitem[Chaudhary et~al.(2011)Chaudhary, Shinn, Vanka, and
  Thomas]{Chaudhary2011_DNS_MHD}
R.~Chaudhary, A.F. Shinn, S.P. Vanka, and B.G. Thomas.
\newblock Direct numerical simulations of transverse and spanwise magnetic
  field effects on turbulent flow in a 2:1 aspect ratio rectangular duct.
\newblock \emph{Computers \& Fluids}, 51\penalty0 (1):\penalty0 100 -- 114,
  2011.
\newblock ISSN 0045-7930.
\newblock \doi{http://dx.doi.org/10.1016/j.compfluid.2011.08.002}.
\newblock URL
  \url{http://www.sciencedirect.com/science/article/pii/S004579301100243X}.

\bibitem[Shinn(2011)]{Shinn2011}
Aaron~F. Shinn.
\newblock \emph{Large eddy simulations of turbulent flows on graphics
  processing units: application to film-cooling flows}.
\newblock PhD thesis, University of Illinois at Urbana-Champaign, 2011.

\end{thebibliography}

\end{document}